# The role of charge recombination to spin-triplet excitons

# in non-fullerene acceptor organic solar cells

*Alexander J. Gillett[1*], Alberto Privitera[2], Rishat Dilmurat[3], Akchheta Karki[4], Deping Qian[5,6], Anton Pershin[3,7], Giacomo Londi[3], William K. Myers[8], Jaewon Lee[4,9], Jun Yuan[5], Seo-Jin Ko[4,10], Moritz K. Riede[2], Feng Gao[5], Guillermo C. Bazan[4], Akshay Rao[1], Thuc-Quyen Nguyen[4*], David Beljonne[3*] and Richard H. Friend[1*].*

[1]Cavendish Laboratory, University of Cambridge, JJ Thomson Avenue, Cambridge, CB3 0HE, UK.

[2]Clarendon Laboratory, University of Oxford, Parks Road, Oxford, OX1 3PU, UK.

[3]Laboratory for Chemistry of Novel Materials, Université de Mons, Place du Parc 20, 7000 Mons, Belgium.

[4]Centre for Polymers and Organic Solids, Department of Chemistry and Biochemistry, University of California at Santa Barbara, CA 93106, USA.

[5]Biomolecular and Organic Electronics, IFM, Linköping University, Linköping, 58183, Sweden.

[6]Department of Chemistry and Centre for Plastic Electronics, Imperial College London, Imperial College Road, London, SW7 2AZ, UK.

[7]Wigner Research Centre for Physics, PO Box 49, H-1525, Budapest, Hungary.

[8]Centre for Advanced ESR, Inorganic Chemistry Laboratory, University of Oxford, South Parks Road, Oxford, OX1 3QR, UK.

[9]Department of Chemical Engineering and Applied Chemistry, Chungnam National University, 99, Daehak-ro, Yuseong-gu, Daejeon, 34134, Republic of Korea.

[10]Division of Advanced Materials, Korea Research Institute of Chemical Technology, Daejeon 34114, Republic of Korea.

*Corresponding authors: Alexander J. Gillett: E-mail: ajg216@cam.ac.uk; Thuc-Quyen Nguyen: E-mail: quyen@chem.ucsb.edu; David Beljonne: E-mail: david.beljonne@umons.ac.be; Richard H. Friend: E-mail: rhf10@cam.ac.uk.

The power conversion efficiencies (PCEs) of organic solar cells (OSCs) using non-fullerene acceptors (NFAs) have now reached 18%[1]. However, this is still lower than inorganic solar cells, for which PCEs >20% are commonplace[2]. A key reason is that OSCs still show low open-circuit voltages ($V_{OC}$) relative to their optical band gaps[3], attributed to non-radiative recombination[4]. For OSCs to compete with inorganics in efficiency, all non-radiative loss pathways must be identified and where possible, removed. Here, we show that in most NFA OSCs, the majority of charge recombination at open-circuit proceeds via formation of non-emissive NFA triplet excitons ($T_1$); in the benchmark PM6:Y6 blend[5], this fraction reaches 90%, contributing 60 mV to the reduction of $V_{OC}$. We develop a new design to prevent recombination via this non-radiative channel through the engineering of significant hybridisation between the NFA $T_1$ and the spin-triplet charge transfer exciton ($^3CTE$). We model that the rate of the back charge transfer from $^3CTE$ to $T_1$ can be reduced by an order of magnitude, allowing re-dissociation of the $^3CTE$. We then demonstrate NFA systems where $T_1$ formation is suppressed. This work therefore provides a clear design pathway for improved OSC performance to 20% PCE and beyond.

Within the Shockley-Queisser model, an ideal solar cell should possess only radiative recombination, thus acting as an ideal light emitting diode with 100% electroluminescence external quantum efficiency (EQE$_{EL}$) in forward bias[4,6–8]. This sets the limit to the photon energy loss ($\Delta E_{loss}$), defined as the difference between the optical band gap ($E_g$) and the energy of the extracted charges ($qV_{OC}$)[4]. However, when the EQE$_{EL}$ falls below 1, an additional voltage loss ($\Delta V_{nr}$) is incurred[7,8]:

$$\Delta V_{nr} = \frac{-k_B T}{q} ln(EQE_{EL}) \qquad (1)$$

where $k_B$ is the Boltzmann constant, $T$ is temperature and $q$ is the elementary charge. Inorganic technologies have effectively minimised this metric; GaAs solar cells can exhibit $EQE_{ELS} > 0.9$, corresponding to $\Delta V_{nr} = 40$ mV and $\Delta E_{loss} = 320$ meV[9]. In comparison, the $EQE_{EL}$ of the best NFA OSCs are currently $\sim 10^{-4}$–$10^{-5}$, resulting in $\Delta V_{nr} \sim 230$–290 mV and $\Delta E_{loss} = 500$–600 meV[3,10–13]. For PCEs >20% to be achieved in OSCs, $\Delta E_{loss}$ must be reduced to <430 meV[14,15]; as the Shockley-Queisser model includes $\sim 250$-300 meV of unavoidable $\Delta E_{loss}$ due to radiative recombination[6], $\Delta V_{nr}$ is the obvious area for improvement[14,16]. To better understand the factors controlling the $EQE_{EL}$, it is useful to separate the different contributions[17]:

$$EQE_{EL} = \gamma \Phi_{PL} \chi \eta_{out} \qquad (2)$$

where $\gamma$ is the charge balance factor (often engineered to be $\sim 1$), $\Phi_{PL}$ is the photoluminescence quantum efficiency (the upper limit will be set by $\Phi_{PL}$ for the low $E_g$ component), $\chi$ is the fraction of recombination events that can decay radiatively (excitons in the spin-singlet configuration) and $\eta_{out}$ is the photon out-coupling efficiency (typically $\sim 0.3$). The two key factors that can be manipulated are $\Phi_{PL}$ and $\chi$. We note that $\Phi_{PL}$ is often small for low $E_g$ emitters due to efficient multiphonon decay processes[18]. Consequently, a $\Phi_{PL}$ of $\sim 0.1\%$ is typical for NFA OSCs with $E_g \sim 1.4$ eV[19]. Whilst the recent empirical advances in $EQE_{EL}$ have been achieved by raising $\Phi_{PL}$[12,14,20,21], in this work, we address the role of $\chi$ in NFA OSCs.

In OSCs, the recombination of free charges (FC) proceeds via the formation of charge transfer excitons (CTEs), with an electron on the acceptor (A) and a hole on the adjacent donor (D) material. These CTEs will be created in a 1:3 ratio of spin-singlet ($^1$CTE) and spin-triplet ($^3$CTE) states via spin-statistical non-geminate recombination (NGR)[22]. Due to the low exchange energy in CTEs, [1]CTE and [3]CTE can interconvert via spin-orbit coupling (SOC) or hyperfine interactions on ns timescales[23,24]. However, CTEs can be weakly bound at room temperature and can readily re-dissociate into FC; this is exemplified by a Langevin reduction factor of <0.2, used to empirically scale the Langevin recombination rate to match the low charge recombination rates observed in OSCs[25,26]. In the situation where [3]CTE can only decay by spin-mixing with [1]CTE, or re-dissociation, an equilibrium of CTEs and FC in the device can be anticipated[25]; in this case $\chi = 1$ since the only spin-allowed recombination pathway is via [1]CTE[27]. However, OSC systems studied to date possess molecular triplet states on either the D or A lower in energy than the [3]CTE. Thus, it is possible for back charge transfer (BCT) from [3]CTE to $T_1$ to occur[27–29]. Since the $S_1$-$T_1$ energy gap in most organic semiconductors is ~0.6–1 eV[30], $T_1$ will be too low in energy to allow thermal re-dissociation and must decay non-radiatively, for example, via triplet-charge annihilation[31]. This non-radiative pathway disrupts the CTE/FC equilibrium, resulting in $\chi < 1$ and an increased $\Delta V_{nr}$. For OSCs that use fullerenes as acceptors, some (optimised) morphologies of the D:A blend do allow kinetic suppression of this pathway[28], but in general, $T_1$ formation is observed.

The recent advances in OSC performance are due to the replacement of fullerenes with NFAs. Here, we address the role of triplet states in NFA OSCs; it has been noted that this has not yet been extensively considered[15,16,32]. As [3]CTE can be formed from both geminate and non-geminate charge carrier pairs, BCT to $T_1$ can occur through two distinct pathways; from [3]CTE formed by spin-mixing with the geminate [1]CTE[24,33] (Fig. 1a) or via [3]CTE generated through NGR[26,28,34] (Fig. 1b). However, the factor that determines whether $T_1$ forms is the competition between BCT and re-dissociation of the CTEs[28,35]. Therefore, in order to gain a full understanding of $T_1$ formation in NFA OSCs, both geminate and non-geminate $T_1$ formation pathways, as well as the rates of BCT ($k_{BCT}$) and CTE dissociation ($k_{dissociation}$), must all be considered. To achieve this, we utilise complementary techniques; transient absorption (TA) to probe non-geminate $T_1$ and transient electron paramagnetic resonance (trEPR) to investigate geminate $T_1$.

In TA, $T_1$ states often possess distinct photo-induced absorption (PIA) features. Furthermore, by investigating the fluence dependence of the $T_1$ dynamics, we can readily determine whether formation follows the bimolecular kinetics expected for NGR[28], or the monomolecular kinetics anticipated if $T_1$ is produced from geminate processes[24]. We can also quantify the $T_1$ population directly from TA through knowledge of the $T_1$ absorption cross section $(\sigma_T)$[36].

Triplet states can also be detected using EPR measurements, but signals from the population densities achieved through photoexcitation are only measurable when the triplet sublevels $T_+$, $T_0$ and $T_-$ have non-thermal occupancies[37]. In this case, we observe enhanced absorptive ($a$) and emissive ($e$) characters of the EPR transitions, from which the generation mechanism of the triplet can be determined[37]. Triplets produced via NGR will have thermal sublevel occupancies and are thus not detectable[22]. However, triplets produced from geminate processes result in sublevel occupancies far from thermal equilibrium and can be readily observed[29,37,38]. $T_1$ formation via geminate BCT can be understood in the framework of the spin-correlated radical pair mechanism[39–42], where spin-mixing first occurs between ${}^1CTE_0$ and ${}^3CTE_0$, followed by BCT to the molecular triplet sublevels. Depending on the sign of the zero-field splitting D-parameter, the overpopulation of either $T_0$ or $T_+/T_-$ results in an *aeeaae* or *eaaeea* spin-polarization pattern of the $T_1$ trEPR signal; a clear and unique fingerprint of the geminate pathway[29,37,38]. Though performed at 80 K, we expect that these measurements are of relevance to the blend behaviour at 293 K; a detailed discussion of the influence of temperature and an in-depth review of EPR theory are available in the SI.

We report here on a representative set of 13 high-performing OSC systems. The structures of the four polymer donors and nine small molecule acceptors used in this study are shown in Fig. 1c. A summary of the device performance (current density-voltage and $EQE_{EL}$ curves are available in Figs. S2-S3), $\Delta V_{nr}$ and whether the blend exhibits geminate or non-geminate $T_1$ formation are given in Table 1 (TA and trEPR for every blend is presented in the SI). From this, we note that geminate $T_1$ formation is not observed in our NFA OSCs. However, non-geminate $T_1$ formation is generally seen, occurring in all NFA blends studied, with the exception of the closely-related PTB7-Th:IEICO-0F and -2F systems. In contrast, we observe geminate $T_1$ formation in three of four fullerene acceptor blends studied here, and two also exhibit non-geminate $T_1$ formation. From this extensive study, we have selected two NFA blends to act as representative case studies; PM6:Y6 as one of the best performing OSC systems[5], despite exhibiting non-geminate $T_1$ formation, and PTB7-Th:IEICO-2F as a sample that had no detectable BCT $T_1$. A full account of all other blends is presented in the SI.

Fig. 2a shows the TA of PM6:Y6, pumped at 532 nm for preferential PM6 excitation. Here, we focus solely on the infrared probe spectral region where the PIAs of $T_1$ states are typically found (full spectral range TA is in Fig. S14). At the earliest time of 0.1-0.2 ps, we observe PIA bands at 1250 nm and 1550 nm; these are assigned to the PM6 and Y6 $S_1$ states though comparison to the spectra of the neat materials (Figs. S5-6). As charge transfer develops, these features are lost and a new PIA at 1450 nm develops beyond a few ps. We confirm this to be the Y6 $T_1$ by triplet sensitisation experiments (Fig. S4b). The kinetics taken from this spectral region (Fig. 2b) show a strong fluence dependence in $T_1$ formation, demonstrating that triplets are generated via bimolecular processes. Deviation of the $T_1$ region kinetics at different fluences begins on sub-ps timescales, confirming that NGR can occur extremely quickly when the excitation fluence is sufficiently high; this infers that the BCT rate ($k_{BCT}$) must be ~$10^{11}$-$10^{12}$ s$^{-1}$ for $T_1$ to be detected this rapidly[27]. In contrast, we find that $k_{dissociation}$ of the thermalized CTEs is between $10^{10}$-$10^{11}$ s$^{-1}$ for all NFA blends studied here (Figs. S33-S38). Thus, as $k_{BCT} \gg k_{dissociation}$ in PM6:Y6, it is clear why $T_1$ is observed. Furthermore, through knowledge of $\sigma_T$ for Y6, we can quantify the extent of triplet formation. Through the utilisation of a previously-employed kinetic model (full details in the SI), we determine that ~90% of the recombination in this blend under conditions equivalent to open-circuit proceeds non-radiatively via the NFA $T_1$ (Fig. S41). Here, the $T_1$ recombination fraction can be greater than the 75% predicted by spin-statistics as it is possible for CTEs to form and separate multiple times prior to recombining[25]. We note that the presence of non-geminate $T_1$ formation in PM6:Y6 is representative of most NFA blends studied here.

We next turn to trEPR to investigate geminate $T_1$ pathways. Fig. 2c shows the trEPR spectra of PM6:Y6 at representative time points after 532 nm excitation. At 1 μs, we observe a single, intense peak at 346 mT that can be attributed to FC (polarons)[43] and a broader weak triplet feature. However, at 5 μs there is no evidence of remaining triplets; likely due to the rapid triplet-charge annihilation in this blend (Fig. S42) and the loss of spin polarisation. The triplet detected at 1 μs can be simulated by a single *eeeaaa* component, characteristic of $T_1$ formed via direct SOC-mediated intersystem crossing (ISC)[29,37,38]. We attribute this triplet to un-dissociated $S_1$ states undergoing ISC. Importantly, the absence of any species with an *aeeaae* or *eaaeea* polarisation pattern confirms that geminate $T_1$ formation does not occur in this blend; this is a characteristic observation of all the NFA OSC systems studied.

We now focus on PTB7-Th:IEICO-2F, an NFA blend where $T_1$ generation from CTEs could not be detected. Its TA is shown in Fig. 2d, where excitation at 620 nm was used to preferentially pump PTB7-Th. In the infrared probe region (full spectral range data is in Fig. S17), two distinct PIA features at 1175 nm and 1550 nm are observed at the earliest time of 0.2-0.3 ps. Through comparison to the TA of the neat materials (Figs. S7-8), we assign the former to the edge of the IEICO-2F $S_1$ and the latter to the PTB7-Th $S_1$. As charge transfer develops, both of these PIAs are lost and only the edge of the PTB7-Th hole PIA is visible at 1175 nm. Importantly, there is no detectable formation of the IEICO-2F $T_1$ PIA, which we find to be at 1350 nm from triplet sensitisation measurements (Fig. S4c). Furthermore, there is no fluence dependence of the kinetics taken from the IEICO-2F $T_1$ region (Fig. 2e), providing additional evidence that non-geminate $T_1$ formation is not a significant recombination pathway in this blend.

In the trEPR of PTB7-Th:IEICO-2F excited at 532 nm (Fig. 2f), we observe a prominent SOC-ISC $T_1$ feature with a clear *eeeaaa* polarisation pattern that flips to *aaaeee* by 5 μs[29,37,38], as well as an *ea* polarisation CTE at 346 mT that evolves into FC[29]. The unusual $T_1$ spectral inversion is due to differing decay rates from the three high-field triplet levels[44,45], as further discussed in the SI. To explain the increased $T_1$ intensity in PTB7-Th:IEICO-2F, we note that IEICO derivatives exhibit surprisingly high $T_1$ quantum yields from ISC of ~5% (Fig. S39), indicating that $T_1$ formation from any un-dissociated $S_1$ is rapid. Importantly, geminate BCT $T_1$ states, with a characteristic *aeeaae* or *eaaeea* polarisation pattern[29,37,38], are absent.

We next evaluate the impact of $T_1$ formation on device performance. In PM6:Y6, 90% of the recombination at open-circuit proceeds non-radiatively via the NFA $T_1$; this results in $\chi$ =0.1, reducing the $EQE_{EL}$ by a factor of ten. From equation 1, this leads to an increase in $\varDelta V_{nr}$ by ~60 mV, corresponding alone to an absolute PCE loss of over 1%. Additionally, the PTB7-Th:IEICO-2F and PTB7-Th:IEICO-4F blends provide a further opportunity to assess the impact of $T_1$ formation on $\Delta V_{nr}$. Here, the NFA structures differ only by two fluorine atoms and the CTE energies are 1.29 and 1.26 eV (Fig. S68), respectively. However, only PTB7-Th:IEICO-4F exhibits non-geminate $T_1$ formation (Fig. S26); we discuss this further in Fig. S72. Therefore, these blends enable a direct appraisal of the impact of $T_1$ formation on $\Delta V_{nr}$ without a significant influence of molecular structure or CTE energy[46]. A $\Delta V_{nr}$ =280 mV is obtained for PTB7-Th:IEICO-2F, whereas $\Delta V_{nr}$ =340 mV is found for PTB7-Th:IEICO-4F (Fig. S3); this is consistent with our estimate of ~60 mV extra losses from significant $T_1$ generation. However, despite 90% of recombination proceeding via $T_1$, PM6:Y6 exhibits a smaller $\Delta V_{nr}$ =250 mV; the increased $EQE_{EL}$ must therefore come from a relatively high $\Phi_{PL}$. By analysing the neat Y6 and PM6:Y6 blend electroluminescence spectra, we find that 75% of the blend emission originates from the Y6 $S_1$ (Fig. S69). From this, we conclude that the primary radiative pathway in PM6:Y6 is via the relatively emissive NFA $S_1$[47] ($\Phi_{PL}$ =1.3% for a neat Y6 film), enabled by the small $S_1$-$^1$CTE energy gap of only 50 meV[48]. As the NFA $S_1$ will have a higher $\Phi_{PL}$ than even an $S_1$-$^1$CTE hybridised state[19,49], this can explain the improved $EQE_{EL}$. Furthermore, as our subsequent quantum chemical calculations suggest few NFA blends possess the necessary D/A electronic coupling for significant hybridisation[20], we expect the NFA $S_1$ to be a significant radiative pathway in most low-offset NFA OSCs[47].

In order to optimise PCEs, we consider it of critical importance that OSCs are designed to avoid $T_1$ formation. We have therefore explored the role of D/A intermolecular interactions in $T_1$ generation using quantum-chemical calculations. As discussed below for the two representative cases of PM6:Y6 and PTB7-Th:IEICO-2F, we find that if D/A electronic coupling is significant and energy level alignment is correct, hybridisation of the CTE with the local exciton states (LE) on the NFA can occur[12,20,21]. This then results in a substantial rearrangement of the $^1$CTE and $^3$CTE energies. However, further calculations on many of the other systems studied here suggest that whilst the D/A electronic coupling may be sufficient for charge transfer to take place, it is rarely enough to induce significant CTE:LE hybridisation (Fig. S72). Importantly, BCT $T_1$ formation is only suppressed in the blends that exhibit strong CTE:LE hybridisation (Figs. 3a, 3b, S73).

Beginning with PTB7-Th:IEICO-2F, we have calculated the $^1$CTE and $^3$CTE energies at the equilibrium geometry. We find the energy ordering of the CTEs is inverted from that expected when considering exchange interactions[30], with the $^3$CTE higher than the $^1$CTE by ~70 meV. To investigate this further, we have run a rigid scan of the $^1$CTE and $^3$CTE excitation energies as a function of D/A separation. The results displayed in Fig. 3a show that below 0.5 nm, the $^1$CTE is rapidly stabilised, whilst the $^3$CTE is destabilised. In contrast, the explored PM6:Y6 configurations display the expected CTE energy ordering (Figs. 3b and S72). By analysing the excited-state wavefunctions, we conclude that the inversion of $^1$CTE and $^3$CTE in PTB7-Th:IEICO-2F arises from CTE:LE hybridisation. This inversion occurs because the NFA $S_1$ is higher in energy than the $^1$CTE and the NFA $T_1$ is lower than the $^3$CTE; hybridisation of these states therefore stabilises the $^1$CTE and destabilises the $^3$CTE (Fig. 3c). The primary reason for hybridisation is the enhanced CTE-LE electronic coupling in the PTB7-Th:IEICO-2F complex, due to: (i) the similar bonding-antibonding pattern of the HOMOs, with the same sequence of vertical nodal planes along the main molecular axis (Fig. 3d); (ii) the almost perfect registry between the NFA and the polymer backbone, offering significant molecular overlap (Fig. S71). We note that the results presented above are sensitive to the local supramolecular organization; in the case of PTB7-Th:IEICO-2F, we also find configurations where hybridisation is weaker or absent (Fig. S72). However, these are substantially higher in energy than the configurations where hybridisation is present.

The consequence of $^3$CTE destabilisation at close D/A separations is that it causes the electron and hole to remain distant from the interface in this CTE spin configuration. This has significant implications as the D/A electronic coupling, and thus $k_{BCT}$, falls exponentially with distance[50], providing time for thermal re-dissociation of $^3$CTE. This is consistent with previous findings, where BCT $T_1$ formation was enhanced when the CTE radius was reduced[51]. Calculations of $k_{BCT}$ as a function of intermolecular distance indicate that $k_{BCT}$ is reduced by an order of magnitude through hybridisation effects (Fig. 3a); in the absence of hybridisation, $k_{BCT}$ is ~$10^{12}$ s$^{-1}$, in-line with experimental observations (when the D $T_1$ is also energetically accessible from $^3$CTE, $k_{BCT}$ to the NFA is consistently higher, Fig. S70). Critically, the observed $k_{dissociation}$ for NFA OSCs of between $10^{10}$–$10^{11}$ s$^{-1}$ is comparable to the reductions in $k_{BCT}$ possible through $^3$CTE-$T_1$ hybridisation, confirming that it is a feasible route to suppress BCT to $T_1$.

Whilst the importance of optimising $\Phi_{PL}$ in OSCs is already well known[14], the insight provided by this work demonstrates the critical role of $T_1$ states in $\Delta E_{loss}$. If $T_1$ formation can be inhibited, with $^3$CTE-$T_1$ hybridisation providing one viable pathway, $\Delta E_{loss}$ can be reduced by ~60 meV; enough to enable PCEs of 20% with the current best device performance metrics[14,15]. Therefore, future OSC development should focus on simultaneously increasing $\Phi_{PL}$ and engineering-out $T_1$ formation. To achieve this, quantum chemical calculations on the D/A electronic interactions will provide a valuable predictive tool for screening perspective D/A pairs *in silico*. Furthermore, we anticipate the unprecedented spin-control over charge recombination demonstrated here will be of great interest to the broader field of excitonic semiconductors.

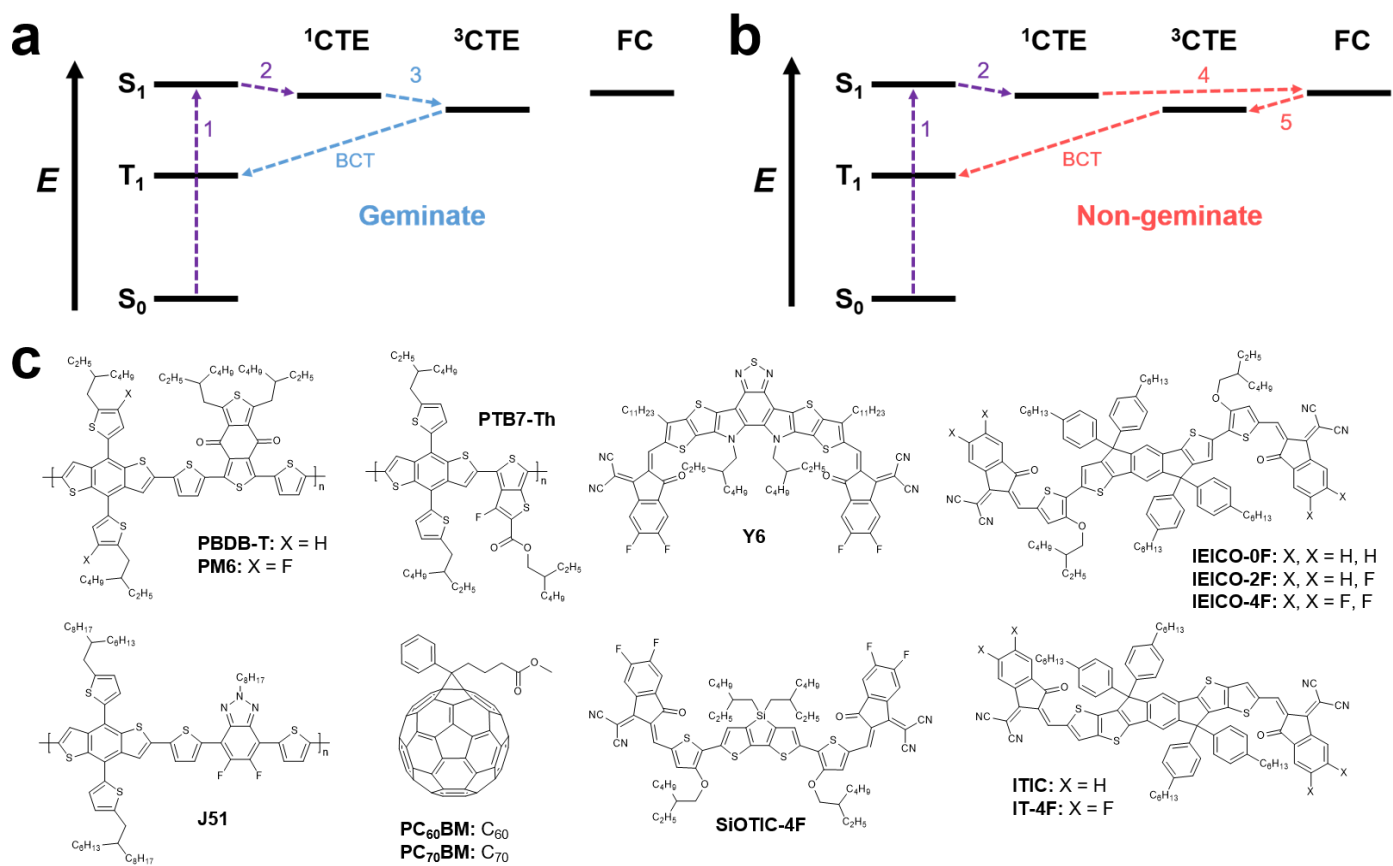

**Figure 1: (a)** A diagram to illustrate the geminate pathway for $T_1$ formation in OSCs. After optical excitation (1), charge transfer from the $S_1$ to $^1$CTE occurs (2). However, the $^1$CTE does not manage to separate into FC before spin-mixing with the $^3$CTE occurs on ns timescales (3). From the $^3$CTE, BCT to a lower energy $T_1$ on either the D or A can occur. **(b)** A diagram to illustrate the non-geminate pathway for $T_1$ formation in OSCs. After optical excitation (1), charge transfer from the $S_1$ to $^1$CTE occurs (2). The $^1$CTE then successfully dissociates in FC (4). The FC then undergo non-geminate recombination, forming a 3:1 ratio of $^3$CTE to $^1$CTE (5). From the $^3$CTE, BCT to a lower energy $T_1$ on either the D or A can occur. **(c)** The molecular structures of the four polymer donors and nine small molecule acceptor materials used in this study.

| OSC Class | Blend | PCE (%) | $EQE_{EL}$ | $\Delta V_{nr}$ (V) | Geminate $T_1$ | Non-geminate $T_1$ |
|---|---|---|---|---|---|---|
| **NFA** | PM6:Y6 | 15.2 | $4.3 \times 10^{-5}$ | 0.25 | No | Yes |
| | PM6:IT-4F | 12.0 | $9.5 \times 10^{-7}$ | 0.35 | No | Yes |
| | PM6:ITIC | 9.2 | $5.0 \times 10^{-5}$ | 0.25 | No | Yes |
| | PBDB-T:ITIC | 11.2 | $8.8 \times 10^{-7}$ | 0.35 | No | Yes |
| | J51:ITIC | 7.2 | $7.1 \times 10^{-8}$ | 0.42 | No | Yes |
| | PTB7-Th: SiOTIC-4F | 8.9 | $8.7 \times 10^{-7}$ | 0.35 | No | Yes |
| | PTB7-Th: IEICO-4F | 10.2 | $1.6 \times 10^{-6}$ | 0.34 | No | Yes |
| | <span style="color:red">PTB7-Th: IEICO-2F</span> | <span style="color:red">11.7</span> | <span style="color:red">$1.3 \times 10^{-5}$</span> | <span style="color:red">0.28</span> | <span style="color:red">No</span> | <span style="color:red">No</span> |
| | <span style="color:red">PTB7-Th: IEICO-0F</span> | <span style="color:red">7.2</span> | <span style="color:red">$1.4 \times 10^{-4}$</span> | <span style="color:red">0.22</span> | <span style="color:red">No</span> | <span style="color:red">No</span> |
| **Fullerene** | $PM6:PC_{60}BM$ | 7.4 | $1.0 \times 10^{-6}$ | 0.35 | Yes | Yes |
| | PTB7-Th: $PC_{60}BM$ | 7.5 | $3.3 \times 10^{-8}$ | 0.44 | Yes | Yes |
| | PBDB-T: $PC_{70}BM$ | 8.8 | $2.0 \times 10^{-7}$ | 0.39 | No | No |
| | $J51:PC_{70}BM$ | 4.3 | $4.1 \times 10^{-8}$ | 0.43 | Yes | No |

**Table 1:** A summary of the key device performance parameters of the OSC blends investigated in this study. The devices are split into two categories: those fabricated with NFAs and those with fullerene acceptors. For the determination of $\Delta V_{nr}$, the $EQE_{EL}$ at 293 K was taken at $-J_{SC}$ to ensure that carrier densities were relevant to device operating conditions. Additionally, it is stated whether the blend forms triplet excitons resulting from either geminate or non-geminate recombination pathways.

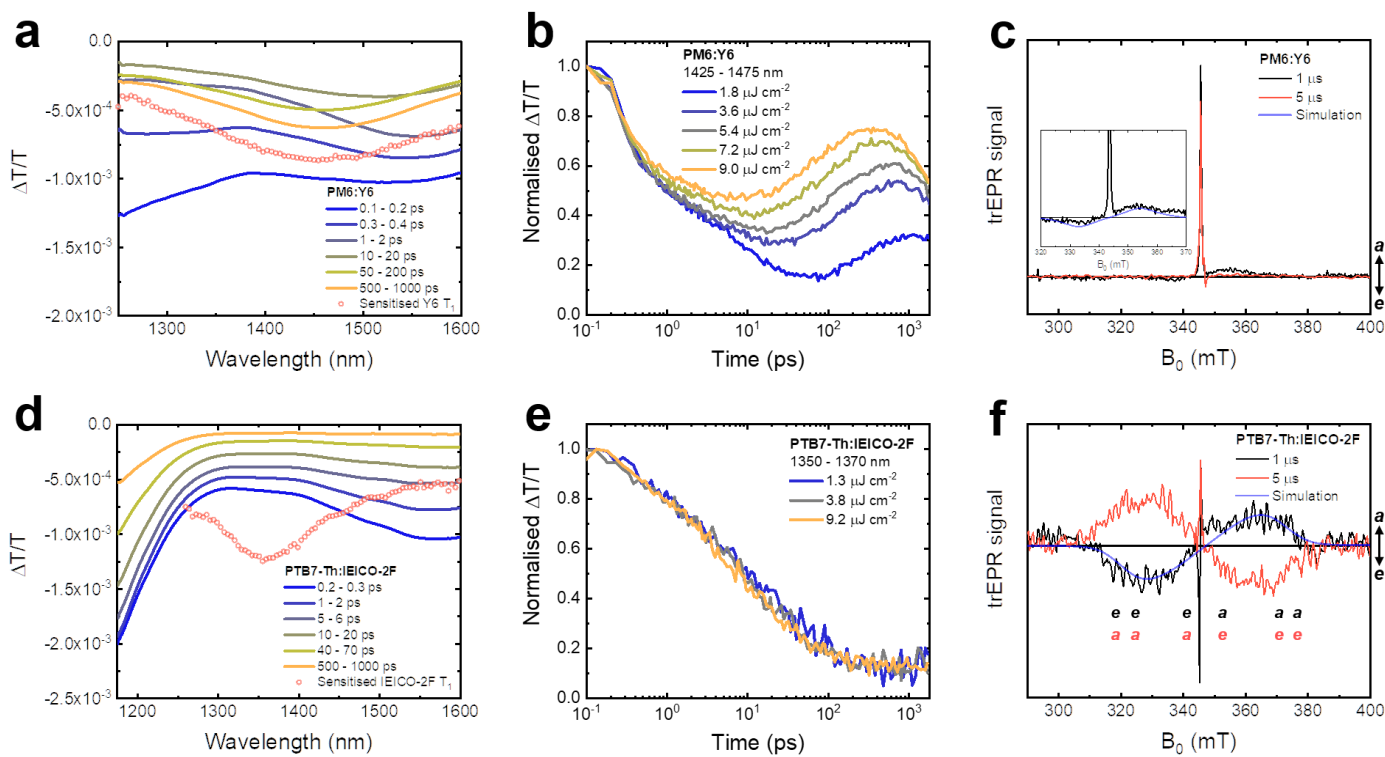

**Figure 2: (a)** The IR region TA spectra of the PM6:Y6 blend (293 K), excited with a moderate fluence of 5.4 μJ cm$^{-2}$ at 532 nm for preferential PM6 excitation. The Y6 $T_1$ PIA forms at 1450 nm, as confirmed by triplet sensitisation experiments. **(b)** The normalised TA kinetics of the PM6:Y6 blend, taken around the maximum of the Y6 $T_1$ feature between 1425 – 1475 nm. The clear fluence dependence of $T_1$ formation is indicative of a bimolecular generation pathway. **(c)** The trEPR spectra of the PM6:Y6 blend (80 K) after excitation at 532 nm, taken at 1 and 5 μs. The inset shows a magnification of the weak ISC triplet signal. **(d)** The IR region TA spectra of the PTB7-Th:IEICO-2F blend (293 K), excited with a moderate fluence of 3.8 μJ cm$^{-2}$ at 620 nm for preferential PTB7-Th excitation. The IEICO-2F $T_1$ PIA at 1350 nm does not form in the blend. **(e)** The normalised TA kinetics of the PTB7-Th:IEICO-2F blend, taken around the maximum of the IEICO-2F $T_1$ PIA at 1350 – 1370 nm. No fluence dependence in the IEICO-2F $T_1$ region is observed. **(f)** The trEPR spectra of the PTB7-Th:IEICO-2F blend (80 K) after excitation at 532 nm, taken at 1 and 5 μs. The field positions of the absorption (*a*) and emission (*e*) EPR transitions of the ISC triplet are overlaid on the plot for clarity.

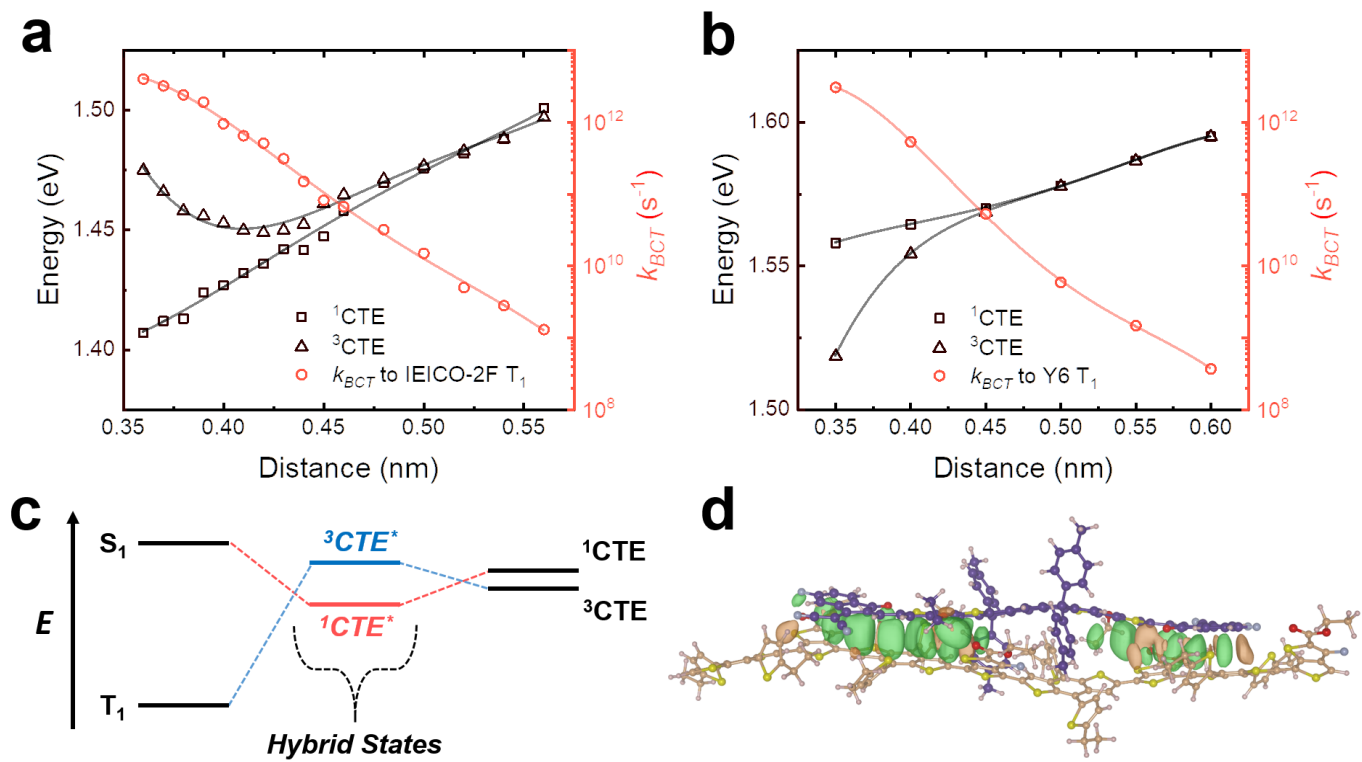

**Figure 3: (a)** The results of a rigid scan of the $^1$CTE and $^3$CTE energies for a representative PTB7-Th:IEICO-2F supramolecular configuration as a function of D/A separation. At each D/A separation, $k_{BCT}$ from $^3$CTE to $T_1$ of the NFA has also been calculated. The solid lines provide polynomial guides to the eye. **(b)** The results of a rigid scan of the $^1$CTE and $^3$CTE energies for a representative PM6:Y6 supramolecular configuration. At each D/A separation, $k_{BCT}$ from $^3$CTE to $T_1$ of the NFA has also been calculated. The solid lines provide polynomial guides to the eye. **(c)** A schematic to represent the effect of CTE-LE hybridisation on the energetic ordering of the $^1$CTE and $^3$CTE. **(d)** An image showing the same optimised supramolecular configuration between PTB7-Th (beige) IEICO-2F (purple) used for the calculations in Fig. 3a. The lobes represent regions of constructive overlap between the HOMOs of D and A: it is this HOMO overlap that mediates the mixing between the NFA LE and CTE excited states.

## Methods

### OSC device fabrication

Indium tin oxide (ITO) patterned glass substrates were cleaned by scrubbing with soapy water, followed by sonication in soapy water, deionized (DI) water, acetone, and isopropanol for 20 minutes each. The substrates were dried using compressed nitrogen and placed in an oven overnight at 100 °C. The conventional architecture devices were made by treating the ITO substrates with UV-ozone for 15 minutes and spin-coating a layer of poly(3,4-ethylenedioxythiophene):poly(styrenesulfonate) (PEDOT:PSS, Clevios P VP Al 8043) at 3000 rpm for 40 s onto the ITO substrates in air. The substrates were then annealed in air at 150 °C for 20 minutes. Active layers were spin coated on top of the PEDOT:PSS layer inside a nitrogen filled glovebox following the recipes from previous reports[10,48,52]. The substrates were then pumped down under vacuum ($<10^{-7}$ torr), and a 5 nm thick Ca interlayer followed by a 100 nm thick Al electrode were deposited on top of the active layer by thermal evaporation using the Angstrom Engineering Series EQ Thermal Evaporator. In the case of inverted architecture devices, ZnO was used as the bottom transparent electrode (replacing PEDOT:PSS), where the ZnO solution was prepared in a nitrogen glovebox by mixing tetrahydrofuran and diethylzinc (2:1). The fresh ZnO solution was then spin-coated atop the clean ITO substrates at 4000 rpm for 30 seconds and then placed on a hotplate at 110 °C for 15 minutes. Following active layer spin-coating, the inverted devices were pumped down under vacuum ($<10^{-7}$ torr), and 7 nm of MoOx and 100 nm thick Ag electrode were deposited on top of the active layer by thermal evaporation. The optimized active layer compositions used for the blend solutions were based on previously published reports[10,48,52].

**OSC device testing**

Photovoltaic characteristic measurements were carried out inside the glove box using a high-quality optical fibre to guide the light from the solar simulator equipped with a Keithley 2635A source measurement unit. *J-V* curves were measured under AM 1.5G illumination at 100 mW cm$^{-2}$ for devices with an electrode overlap area of 0.22 cm$^2$. No spectral mismatch correction was applied.

**Electroluminescence and EQE$_{EL}$ measurements**

EL measurements were performed using two setups depending on the wavelength range of interest. For measurements under 1050 nm, a home-made EL spectrometer was used. The EL emission from a sample driven by a Keithley source-measure unit (model 2602A) was collected by a lens system and focused on the entrance slit of a spectrograph (Acton Research SP-500) equipped with a Si charge-coupled detector (Princeton Instruments Pixis:400). The spectra collected by the detector were corrected for the instrument response function. The correction factors were determined by measuring the spectrum of a black body-like light source (Ocean Optics LS-1). For EL measurements in the range 900 - 1700 nm, we utilized a Photon Technology International (PTI) Quantamaster fluorimeter equipped with an Edinburgh Instruments EI-L Ge detector. The excitation monochromator of the fluorimeter was not used, and the EL emission was generated by driving the devices by a Keithley 2602 source-measure unit. An optical chopper (Thorlabs MC2000) was placed in front of the emission monochromator to make use of the fluorimeter's lock-in amplifier-based detection system. The PTI Felix fluorimeter software was used for the data collection and correction of the instrumental artefacts. The efficiency of EL was obtained by applying a bias from -1 to 2V with a dual-channel Keithley 2602 to the solar cell and placing a silicon or germanium photodiode directly in front of it to collect the emission as a function of applied bias. The current running through the device and the photodiode were simultaneously measured.

**TA spectroscopy**

TA samples were fabricated by spin-coating solutions onto quartz substrates using identical conditions to the optimised devices. The samples were encapsulated in a nitrogen glovebox environment to ensure oxygen-free measurements.

TA was performed on either one of two experimental setups. The broadband probe (525 – 1650 nm) TA was performed on a setup powered using a commercially available Ti:sapphire amplifier (Spectra Physics Solstice Ace). The amplifier operates at 1 kHz and generates 100 fs pulses centred at 800 nm with an output of 7 W. A TOPAS optical parametric amplifier (OPA) was used to provide the tuneable ~100 fs pump pulses for the "short-time" (100 fs – 1.8 ns) TA measurements, whilst the second harmonic (532 nm) of an electronically triggered, Q-switched Nd:YVO$_4$ laser (Advanced Optical Technologies Ltd AOT-YVO-25QSPX) provided the ~1 ns pump pulses for the "long-time" (1 ns – 100 μs) TA measurements. The probe was provided by a broadband visible (525 – 775 nm), NIR (800 – 1200 nm) and IR (1250 – 1650 nm) NOPAs. The probe pulses are collected with an InGaAs dual-line array detector (Hamamatsu G11608-512DA), driven and read out by a custom-built board from Stresing Entwicklungsbüro. The probe beam was split into two identical beams by a 50/50 beamsplitter. This allowed for the use of a second reference beam which also passes through the sample, but does not interact with the pump. The role of the reference was to correct for any shot-to-shot fluctuations in the probe that would otherwise greatly increase the structured noise in our experiments. Through this arrangement, very small signals with a $\frac{\Delta T}{T} = 1 \times 10^{-5}$ could be measured.

For the 500 – 950 nm continuous probe region TA, a Yb amplifier (PHAROS, Light Conversion), operating at 38 kHz and generating 200 fs pulses centred at 1030 nm with an output of 14.5 W was used. The ~200 fs pump pulse was provided by a TOPAS OPA. The probe is provided by a white light supercontinuum generated in a YAG crystal from a small amount of the 1030 nm fundamental. After passing through the sample, the probe is imaged using a Si photodiode array (Stresing S11490). This setup provided additional flexibility by allowing for broadband spectrum acquisition in one measurement, as well good signal to noise in the 750 – 850 nm region, which is difficult to obtain on the other setup due to large fluctuations in the NOPA probe around the 800 nm fundamental of the Ti:sapphire laser.

**trEPR spectroscopy**

EPR samples were fabricated by spin-coating solutions under identical conditions to the optimised devices onto Mylar substrates, which were subsequently cut into strips with a width of 3 mm. The strips were placed in quartz EPR tubes which were sealed in a nitrogen glovebox with a bi-component resin (Devcon 5-Minute Epoxy), such that all EPR measurements were performed without air exposure.

All trEPR spectra were recorded on a Bruker Elexsys E680 X-band spectrometer, equipped with a nitrogen gas-flow cryostat for sample temperature control. The sample temperature was maintained with an Oxford Instruments CF935O cryostat and controlled with an Oxford Instruments ITC503. Laser pulses for trEPR were collimated into the cryostat and resonator windows from a multi-mode optical fibre, ThorLabs FT600UMT, output and the pulses consisted of a 7 ns duration and 2 mJ energy per pulse at a wavelength of 532 nm produced by a GWU VersaScan Optical Parametric Oscillator (OPA) that is pumped by a Newport/Spectra Physics Lab 170 Quanta Ray Nd:YAG pulsed laser operating at 20 Hz, λ = 355 nm. The trEPR signal was recorded through a Bruker SpecJet II transient recorder with timing synchronisation by a Stanford Research Systems DG645 delay generator. The spectra were acquired with 2 mW microwave power and averaging 400 transient signals at each field position. The simulation of the trEPR spectra was performed using the function *pepper* of Easyspin, a MATLAB toolbox for simulating powder EPR spectra.

# Acknowledgements

A.J.G. and R.H.F. acknowledge support from the Simons Foundation (grant no. 601946) and the EPSRC (EP/M01083X/1 and EP/M005143/1). A.K. and T.-Q.N. were supported by the Department of the Navy, Office of Naval Research Award No. N00014-14-1-0580. A.K. acknowledges funding by the Schlumberger foundation. A.Privitera, R.D., A.Pershin, G.L., M.K.R. and D.B. were supported by the European Union's Horizon 2020 research and innovation programme under Marie Sklodowska Curie Grant agreement No.722651 (SEPOMO project). Computational resources in Mons were provided by the Consortium des Équipements de Calcul Intensif (CÉCI), funded by the Fonds de la Recherche Scientifiques de Belgique (F.R.S.-FNRS) under Grant No. 2.5020.11, as well as the Tier-1 supercomputer of the Fedération Wallonie-Bruxelles, infrastructure funded by the Walloon Region under Grant Agreement No. 1117545. D.B. is a FNRS Research Director. F.G. acknowledges the Stiftelsen för Strategisk Forskning through a Future Research Leader program (FFL18-0322). trEPR measurements were performed in the Centre for Advanced ESR (CAESR) located in the Department of Chemistry of the University of Oxford, and this work was supported by the EPSRC (EP/L011972/1).

# Author contributions

A.J.G, T.-Q.N. and R.H.F. conceived the work. A.J.G. performed the TA measurements. A.Privitera and W.K.M. conducted the trEPR measurements. R.D., A.Pershin and G.L. carried out the quantum chemical calculations. A.K., D.Q., J.Y. and S.-J.K. fabricated and tested the OSC devices. J.Y. performed the PLQE measurements. J.L. synthesised SiOTIC-4F and the IEICO derivatives. M.K.R., F.G., G.C.B., T.-Q.N, D.B. and R.H.F. supervised their group members involved in the project. A.J.G., A.R. and R.H.F. wrote the manuscript with input from all authors.

## Competing financial interests

The authors declare no competing interests.

## Additional information

Supplementary information accompanies this paper at [to be completed in proofs].

Correspondence and requests for materials should be addressed to R.H.F. (rhf10@cam.ac.uk), A.J.G. (ajg216@cam.ac.uk), D.B. (david.beljonne@umons.ac.be) and T.-Q.N. (quyen@chem.ucsb.edu).

Reprints and permissions information is available at www.nature.com/reprints.

## Data availability

The data that supports the findings of this study are available from the corresponding authors upon reasonable request. Source data for Figs. 2a-f and Figs. 3a and b are provided with the paper.

# Supplementary Information for

# The role of charge recombination to spin-triplet excitons in non-fullerene acceptor organic solar cells

*Alexander J. Gillett[1]\*, Alberto Privitera[2], Rishat Dilmurat[3], Akchheta Karki[4], Deping Qian[5,6], Anton Pershin[3,7], Giacomo Londi[3], William K. Myers[8], Jaewon Lee[4,9], Jun Yuan[5], Seo-Jin Ko[4,10], Moritz K. Riede[2], Feng Gao[5], Guillermo C. Bazan[4], Akshay Rao[1], Thuc-Quyen Nguyen[4]\*, David Beljonne[3]\* and Richard H. Friend[1]\*.*

[1]Cavendish Laboratory, University of Cambridge, JJ Thomson Avenue, Cambridge, CB3 0HE, UK.

[2]Clarendon Laboratory, University of Oxford, Parks Road, Oxford, OX1 3PU, UK.

[3]Laboratory for Chemistry of Novel Materials, Université de Mons, Place du Parc 20, 7000 Mons, Belgium.

[4]Centre for Polymers and Organic Solids, Department of Chemistry and Biochemistry, University of California at Santa Barbara, CA 93106, USA.

[5]Biomolecular and Organic Electronics, IFM, Linköping University, Linköping, 58183, Sweden.

[6]Department of Chemistry and Centre for Plastic Electronics, Imperial College London, Imperial College Road, London, SW7 2AZ, UK.

[7]Wigner Research Centre for Physics, PO Box 49, H-1525, Budapest, Hungary.

[8]Centre for Advanced ESR, Inorganic Chemistry Laboratory, University of Oxford, South Parks Road, Oxford, OX1 3QR, UK.

[9]Department of Chemical Engineering and Applied Chemistry, Chungnam National University, 99, Daehak-ro, Yuseong-gu, Daejeon, 34134, Republic of Korea.

[10]Division of Advanced Materials, Korea Research Institute of Chemical Technology, Daejeon 34114, Republic of Korea.

*Corresponding authors: Alexander J. Gillett: E-mail: ajg216@cam.ac.uk; Thuc-Quyen Nguyen: E-mail: quyen@chem.ucsb.edu; David Beljonne: E-mail: david.beljonne@umons.ac.be; Richard H. Friend: E-mail: rhf10@cam.ac.uk.

# Table of Contents

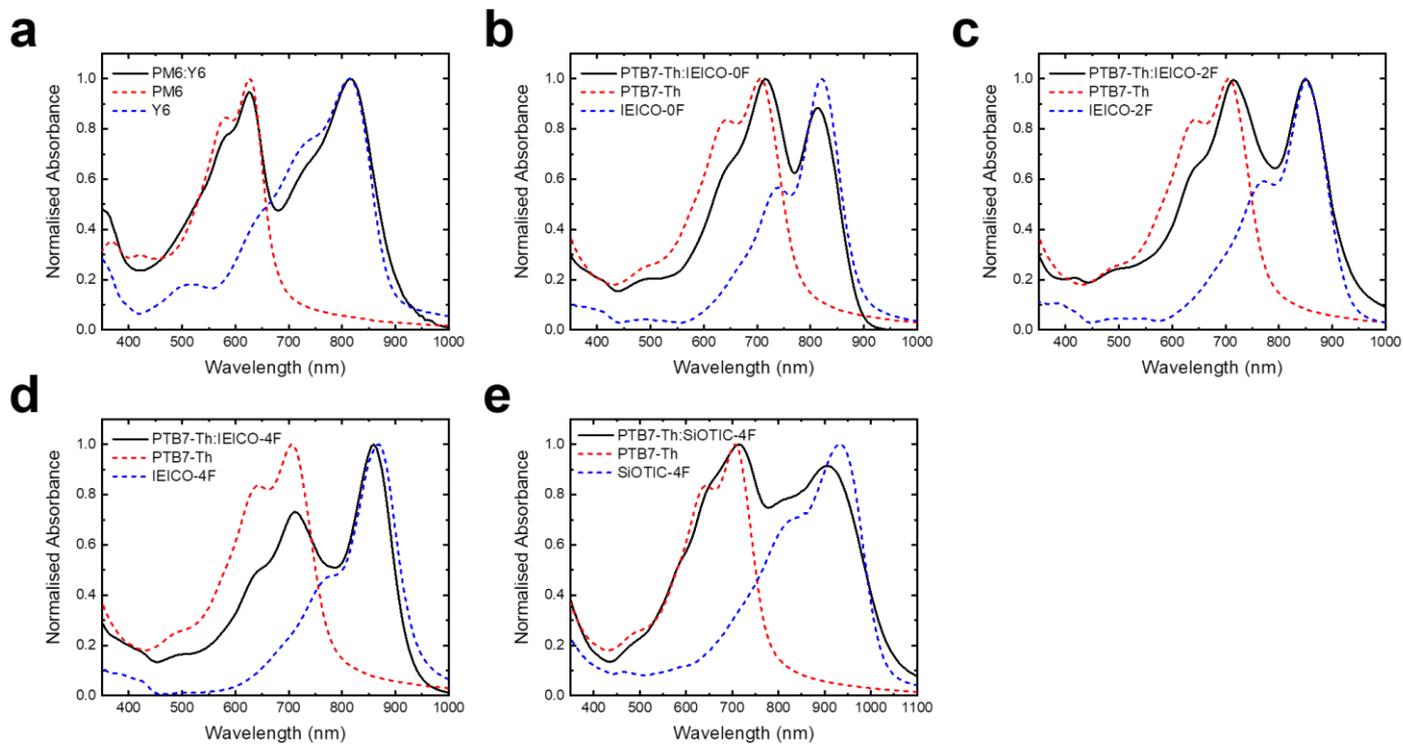

**Figure S1:** The normalised absorption spectra of the key blends examined in this study. The normalised absorption spectra of the neat materials are also overlaid for reference.

| Material | HOMO (ev) | LUMO (ev) |
|----------|-----------|-----------|
| PM6 | -5.56 | -3.50 |
| PTB7-Th | -5.20 | -3.46 |
| PBDB-T | -5.33 | -3.29 |
| J51 | -5.29 | -3.30 |
| Y6 | -5.65 | -4.10 |
| ITIC | -5.60 | -3.85 |
| IT-4F | -5.71 | -4.15 |
| IEICO-0F | -5.24 | -3.80 |
| IEICO-2F | -5.34 | -4.10 |
| IEICO-4F | -5.44 | -4.25 |
| SiOTIC-4F | -5.28 | -4.12 |
| $PC_{60}BM$ | -6.10 | -3.70 |
| $PC_{70}BM$ | -5.90 | -3.90 |

**Table S1:** The tabulated highest occupied molecular orbital (HOMO) and lowest unoccupied molecular orbital (LUMO) energy levels of the materials used in this study, as determined by cyclic voltammetry.

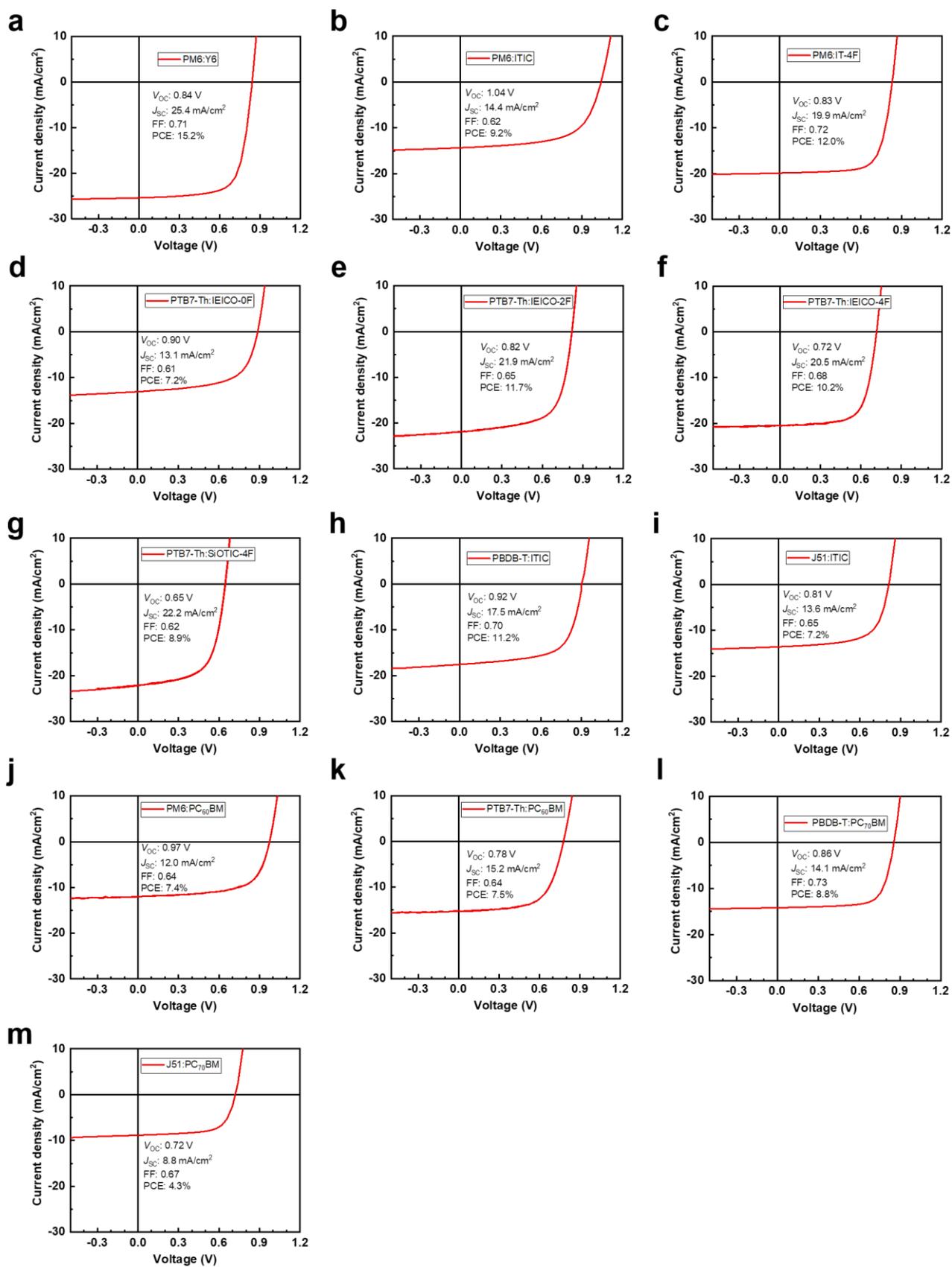

**Figure S2:** The current density-voltage (JV) curves of the OSCs investigated in this study.

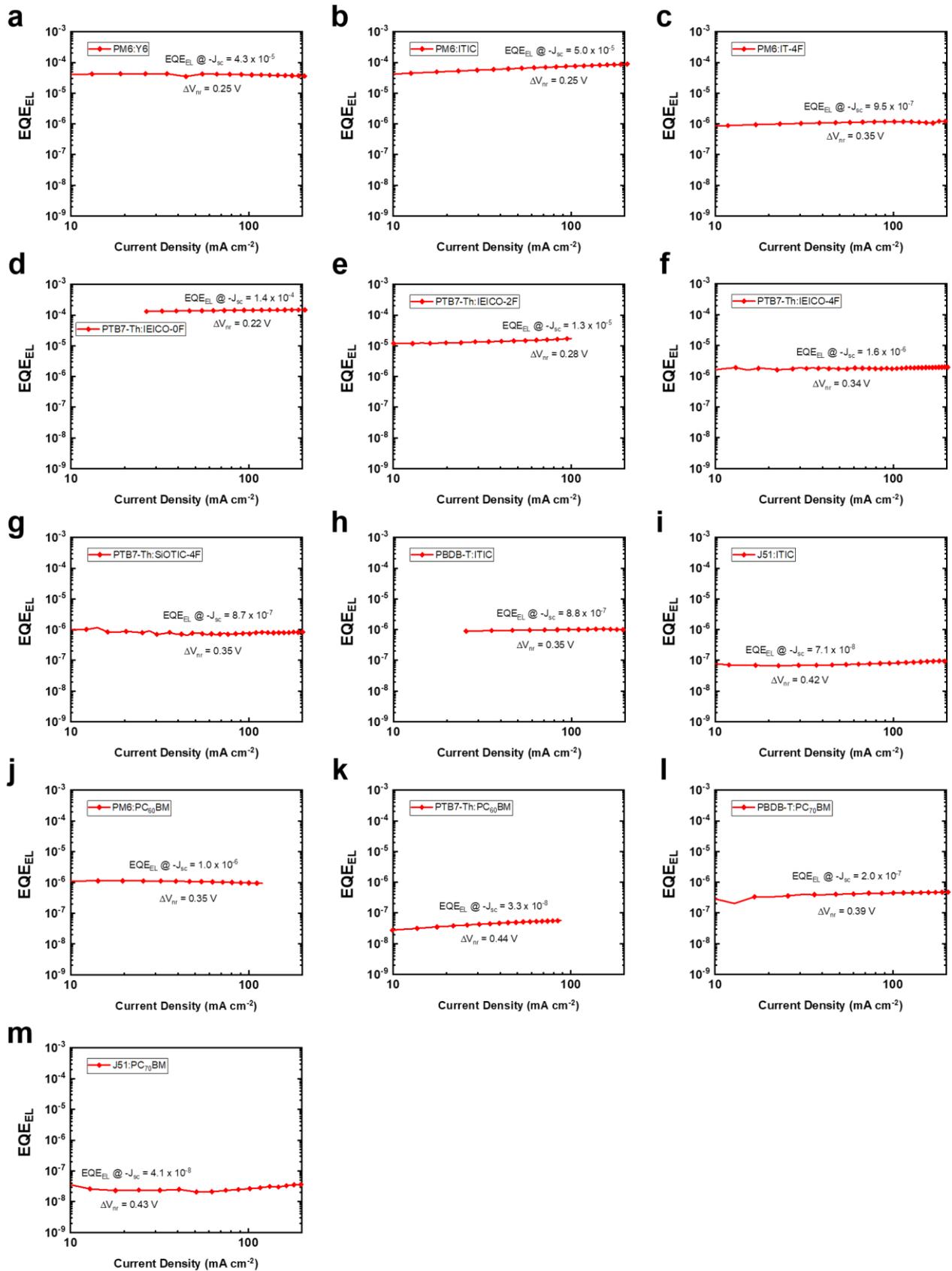

**Figure S3:** The electroluminescence external quantum efficiency (EQE$_{EL}$) of the OSCs investigated in this study.

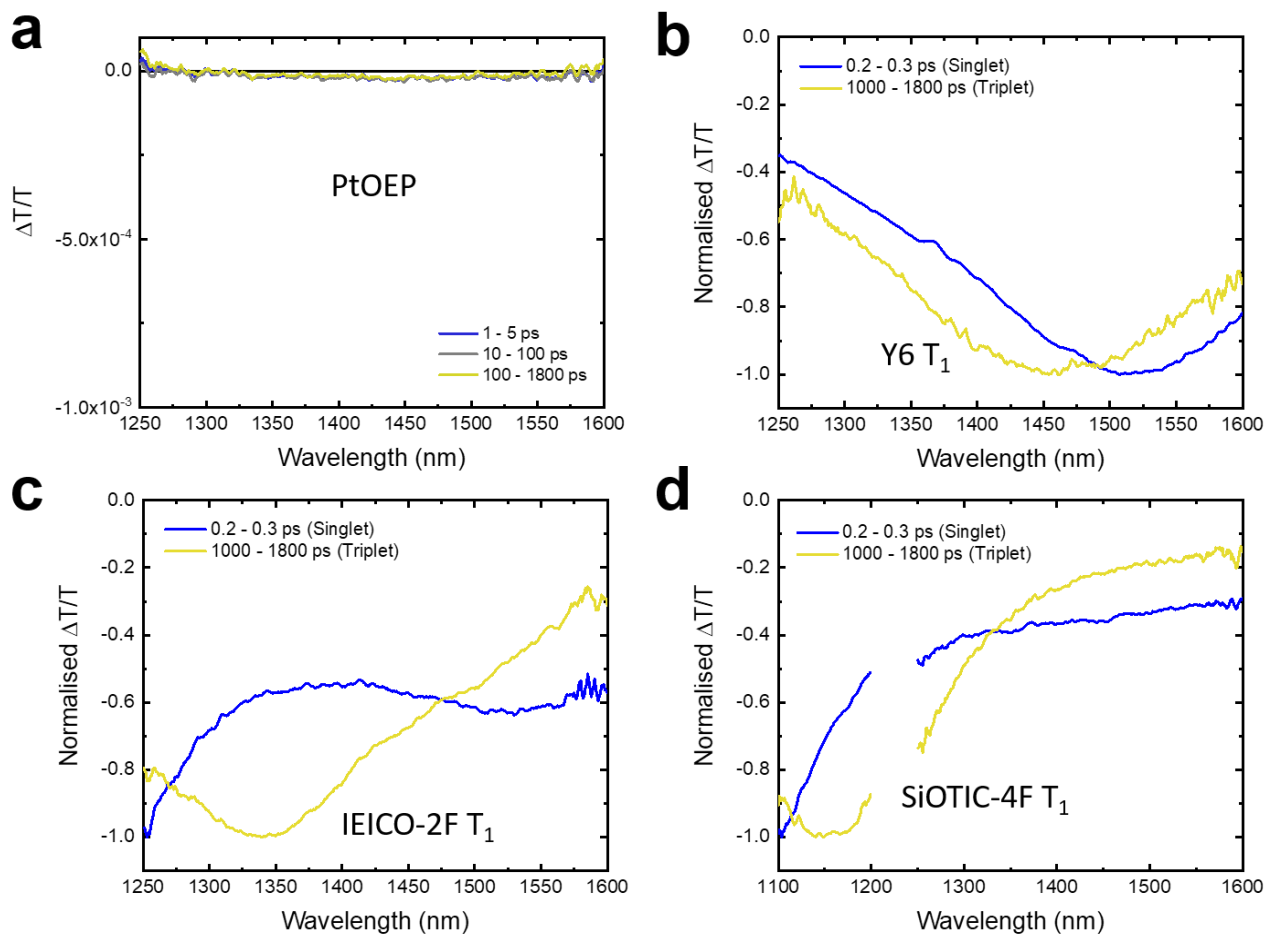

**Figure S4:** Triplet sensitisation experiments of the non-fullerene acceptors (NFA) used in this study. PtOEP was used as the triplet sensitizer. Dilute blends comprised of polystyrene (PS):PtOEP:NFA 0.94:0.03:0.03 were used to ensure that intersystem crossing (ISC) could occur on PtOEP, followed by triplet energy transfer to the NFA, before charge transfer to the NFA. All films were excited at 532 nm for preferential PtOEP excitation, though some unavoidable excitation of the NFA occurred in all blends, as evidenced by the presence of the singlet ($S_1$) PIAs at 0.2 – 0.3 ps. By 1000 – 1800 ps, triplet energy transfer from PtOEP to the NFA had taken place, leaving behind a long-lived photo-induced absorption (PIA), belonging to the $T_1$ of the NFA. **(a)** The TA spectra of a PS:PtOEP 0.94:0.06 film. There are no significant PtOEP PIA features in the IR region probed, confirming that any new PIAs in this region must belong to excited states on the NFAs. **(b)** The TA spectra of a PS:PtOEP:Y6 0.94:0.03:0.03 film. A new PIA belonging to the Y6 $T_1$ is peaked at 1450 nm. **(c)** The TA spectra of a PS:PtOEP:IEICO-2F 0.94:0.03:0.03 film. A new PIA belonging to the IEICO-2F $T_1$ is peaked at 1350 nm. **(d)** The TA spectra of a PS:PtOEP:SiOTIC-4F 0.94:0.03:0.03 film. A new PIA belonging to the SiOTIC-4F $T_1$ is peaked at 1150 nm. For ITIC and IT-4F, we note that previous works have already reported in detail the sensitisation of ITIC derivatives by PtOEP[1], confirming that the $T_1$ PIA of ITIC derivatives is peaked at 1220 nm.

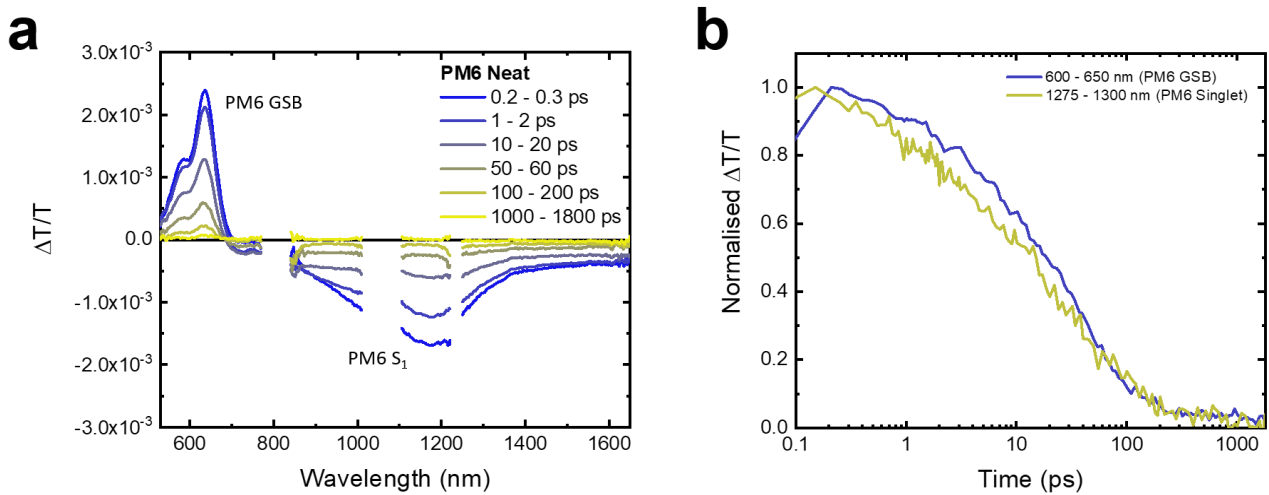

**Figure S5: (a)** The TA spectra of a neat PM6 film, excited at 532 nm with a fluence of 3.1 μJ cm$^{-2}$. The PM6 ground state bleach (GSB) is visible between 530 – 670 nm. The PM6 $S_1$ PIA is broad and spans the near infrared (NIR) region, peaked at 1150 nm. **(b)** The kinetics of the PM6 GSB and $S_1$ regions. As expected, the decay of the GSB and $S_1$ mirror each other, with most excited states decayed after a few hundred ps.

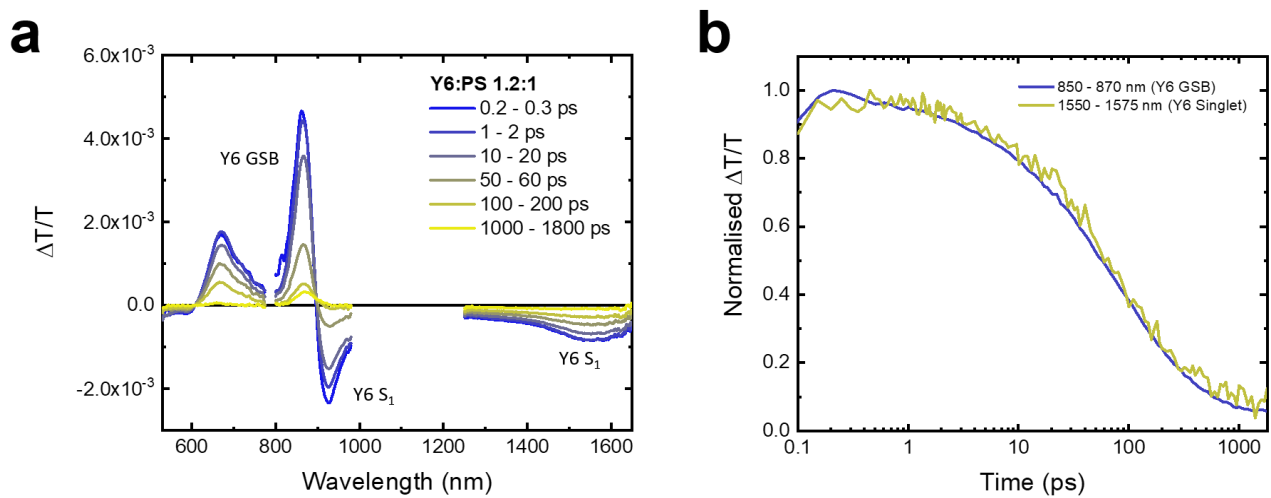

**Figure S6: (a)** The TA spectra of a PS:Y6 1:1.2 film, excited at 800 nm with a fluence of 1.8 μJ cm$^{-2}$. The Y6 GSB is visible between 600 – 900 nm, with two distinct vibronic peaks. There are two Y6 $S_1$ PIAs in the NIR region: one sharp peak adjacent to the Y6 GSB at 910 nm and a weaker, broad feature peaked at 1550 nm. **(b)** The kinetics of the Y6 GSB and $S_1$ regions. As expected, the decay of the GSB and $S_1$ mirror each other, with most excited states decayed after a few hundred ps.

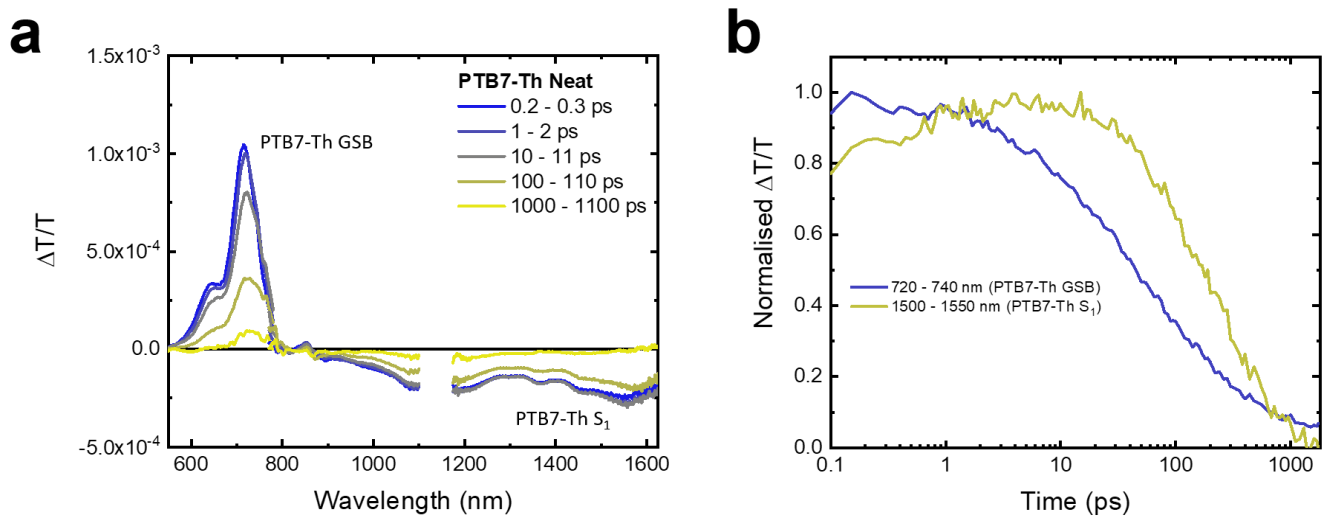

**Figure S7: (a)** The TA spectra of a neat PTB7-Th film, excited at 620 nm with a fluence of 2.1 μJ cm⁻². The PTB7-Th GSB is visible between 600 – 770 nm, with two distinct vibronic peaks. There are two PTB7-Th $S_1$ PIAs in the NIR region: one at 1150 nm and the other at 1550 nm. **(b)** The kinetics of the PTB7-Th GSB and $S_1$ regions. Interestingly, the GSB appears to decay more quickly than the $S_1$ PIA. The reasons for this are unclear and beyond the scope of this work, where we are simply interested in the spectral features of PTB7-Th.

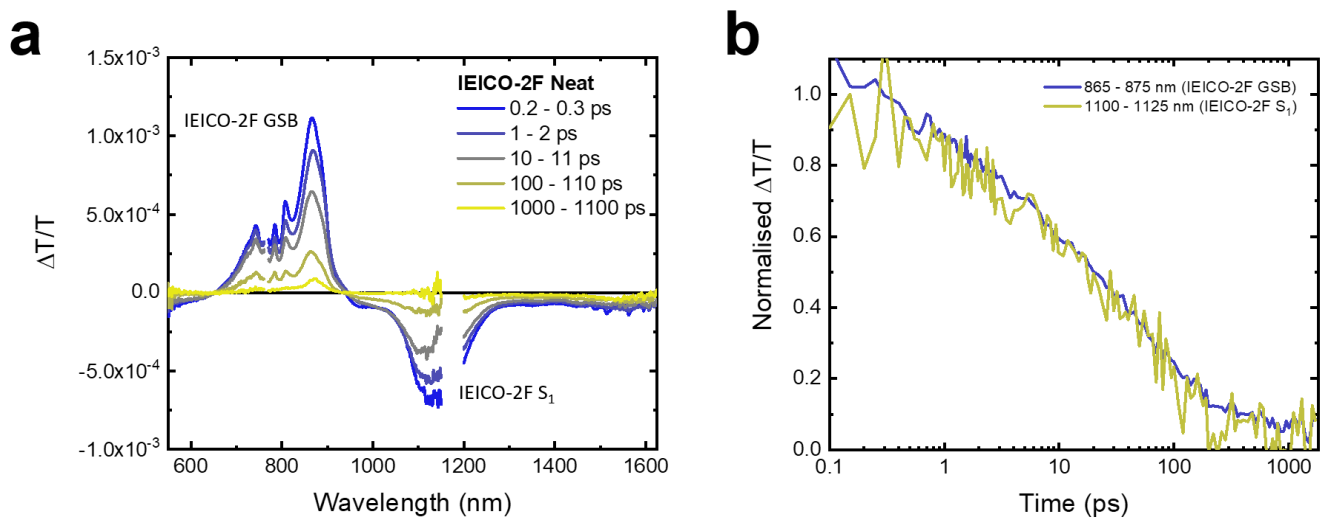

**Figure S8: (a)** The TA spectra of a neat IEICO-2F film, excited at 850 nm with a fluence of 2.1 μJ cm⁻². The IEICO-2F GSB is visible between 650 – 900 nm, with two distinct vibronic peaks. A singlet IEICO-2F $S_1$ PIAs in apparent in the NIR region, peaked at 1120 nm. **(b)** The kinetics of the IEICO-2F GSB and $S_1$ regions. As expected, the decay of the GSB and $S_1$ mirror each other, with most excited states decayed after a few hundred ps.

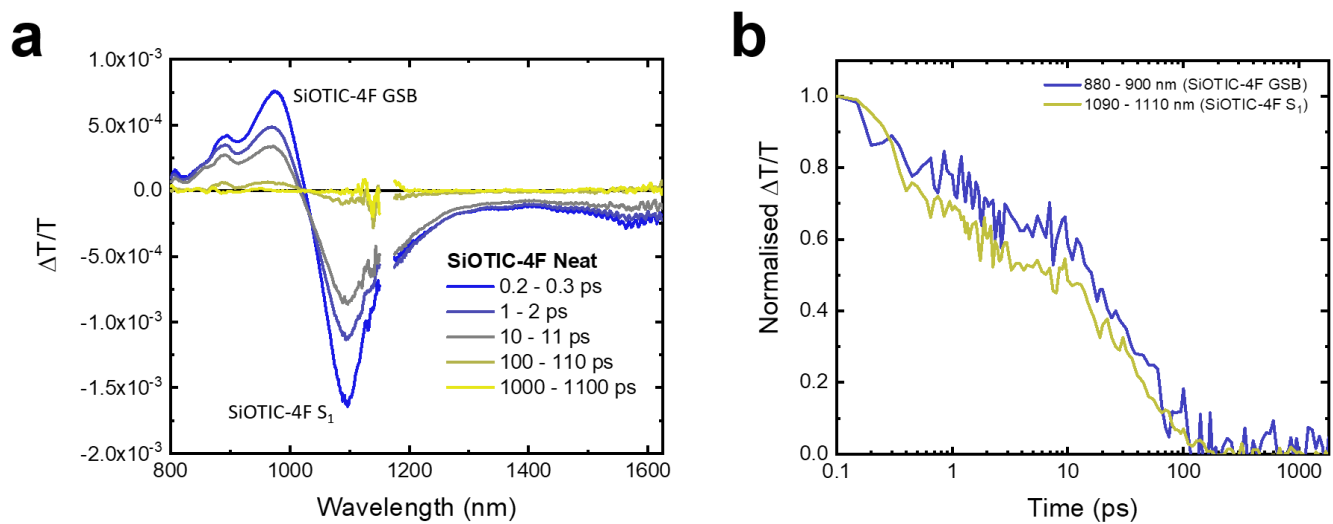

**Figure S9: (a)** The TA spectra of a neat SiOTIC-4F film, excited at 975 nm with a fluence of 3.8 μJ cm$^{-2}$. The PTB7-Th GSB is visible between 800 – 1030 nm, with two distinct vibronic peaks. There are two SiOTIC-4F $S_1$ PIAs in the NIR region: an intense peak at 1090 nm and weak band at 1550 nm. **(b)** The kinetics of the SiOTIC-4F GSB and $S_1$ regions. Interestingly, the GSB appears to decay more quickly than the $S_1$ PIA. As expected, the decay of the GSB and $S_1$ mirror each other, with most excited states decayed after 100 ps.

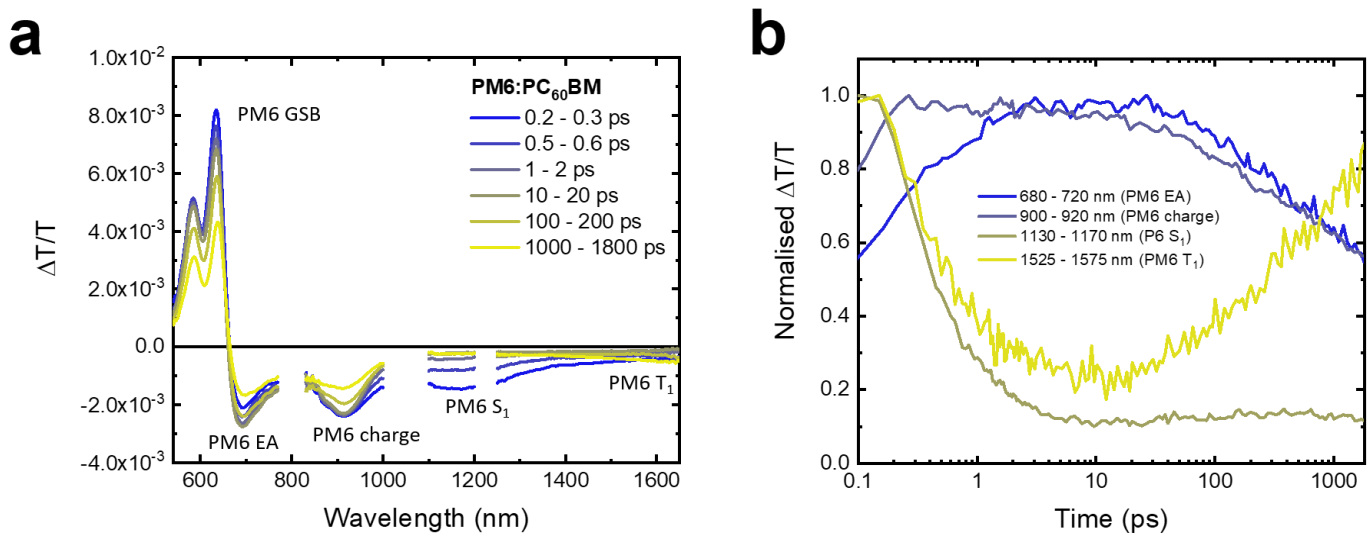

**Figure S10: (a)** The TA spectra of a PM6:PC$_{60}$BM film, excited at 532 nm with a fluence of 5.1 μJ cm$^{-2}$. The purpose of this experiment is to identify features associated with the PM6 after electron transfer to PC$_{60}$BM. As well as the PM6 GSB between 540 – 650 nm and the S$_1$ at 1150 nm, we also notice a negative features present at 690 and 920 nm, which are not present in the neat PM6 film. The feature at 920 nm is peaked almost immediately, indicating it is the PM6 hole PIA formed after ultrafast electron transfer to PC$_{60}$BM. Ultrafast electron transfer is confirmed from the loss of the PM6 S$_1$ PIA within a few ps. Interestingly, the feature at 690 nm takes until 3 ps to reach its maximum intensity. Because of this time evolution and the spectral position right at the absorption edge of PM6, we assign this band to the electro-absorption (EA) of PM6: this represents the Stark-shift of the PM6 absorption spectrum by the electric field of the separating charges[2]. The maximum EA intensity is reached when the CT states have dissociated into free charges (FC)[2–5]. Thus, tracking the EA response provides an insight into the charge separation process. At longer times, a new PIA band at 1600 nm begins to grow in. We note that this is associated with the loss of the PM6 charge PIA. Thus, we assign this new band to the PM6 T$_1$: this is confirmed by a fluence series later in the SI (Fig. S29).

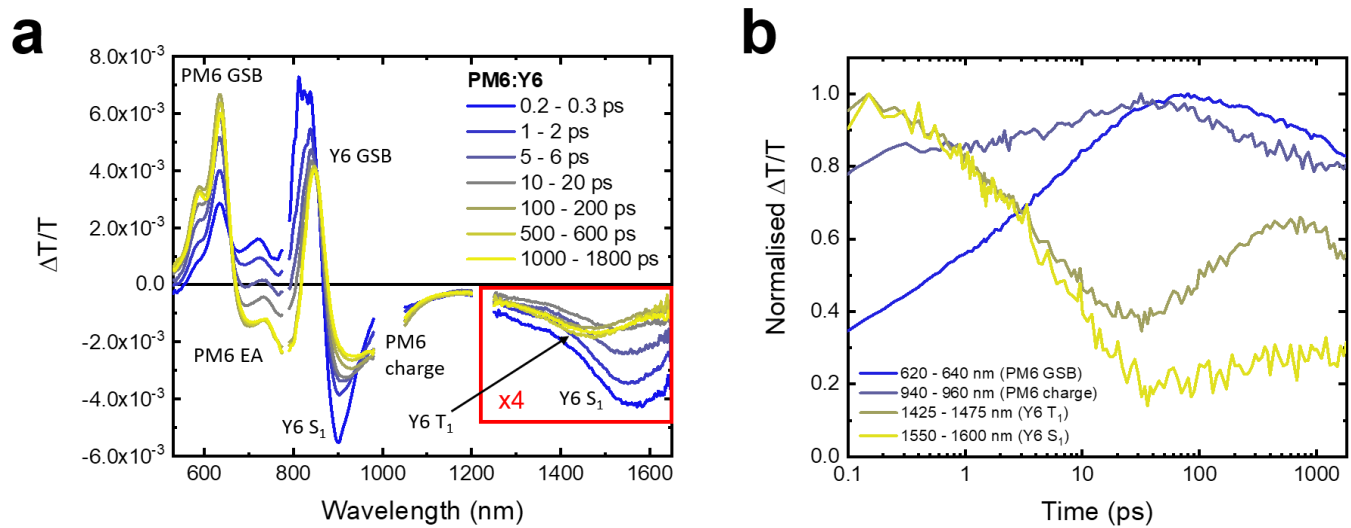

**Figure S11: (a)** The TA spectra of a PM6:Y6 film, pumped at 800 nm for selective Y6 excitation with a moderate fluence of 3.0 μJ cm$^{-2}$. At early times, features associated with Y6, including the GSB at 840 nm and $S_1$ PIAs at 900 and 1550 nm can be seen. Additionally, the PM6 GSB is already present by 200 fs, suggesting that some of the hole transfer in this blend can occur on ultrafast timescales. As time progresses, the PM6 GSB grows more intense and the Y6 $S_1$ PIAs are lost, indicating that additional hole transfer is occurring. We also notice new negative bands forming at 750 and 930 nm on the same timescales. Fig. S9 allows for the assignment of the band around 750 nm to the EA of PM6 and the PIA at 930 nm to charges on PM6. After 50 – 100 ps, the PM6 GSB and PM6 charge PIA begin to decrease in intensity, with a new PIA at 1450 nm forming. From triplet sensitisation measurements with PtOEP (Fig. S4b), we know this is the Y6 $T_1$ PIA. Therefore, it is clear that the loss of charges is associated with the formation of $T_1$ on Y6. **(b)** The kinetics of the PM6:Y6 film in relevant spectral regions. To clarify the discussion of the spectra, we clearly see the growth of the PM6 GSB and charge PIA on ps timescales, followed by their fall from ~50 ps onwards. The kinetic of the Y6 $T_1$ region exhibits an obvious growth on the same timescales the PM6 GSB and charge PIAs are lost.

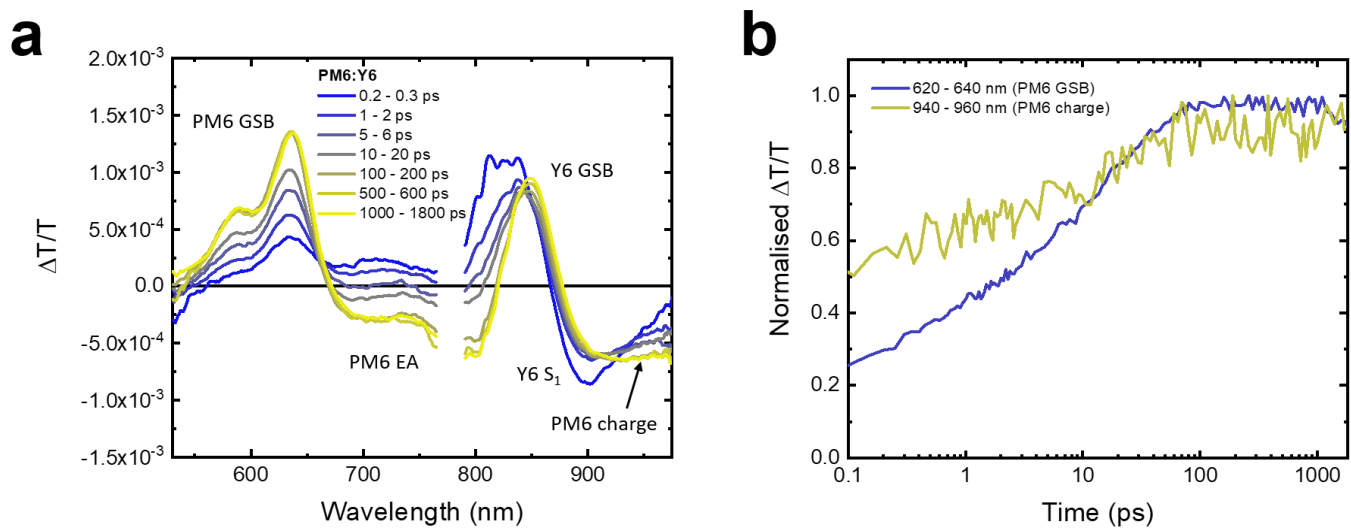

**Figure S12: (a)** The TA spectra of a PM6:Y6 film, pumped at 800 nm for selective Y6 excitation. A very low fluence of 0.5 μJ cm⁻² was used to minimise any non-geminate recombination processes on the timescales of the experiment. The evolution of the TA spectra with time follows the same path as the previous discussions in Fig. S11. **(b)** The kinetics of the PM6 GSB and charge PIA regions. An increase in intensity of these features indicates that hole transfer from Y6 to PM6 is occurring. By 100 ps, there is no further change in magnitude of the signal in these regions, signifying hole transfer has been completed. Importantly, there is no decrease in intensity of the GSB or PIA on the timescales of the TA up to 1.8 ns. This confirms that there is no excess non-geminate recombination taking place, meaning our assertion that hole transfer is completed by 100 ps is accurate. Additionally, the lack of excited state decay by 1.8 ns also suggests that geminate recombination is not a significant loss pathway, as this form of recombination is expected to be fluence-independent and typically takes place on timescales <2 ns[4,6].

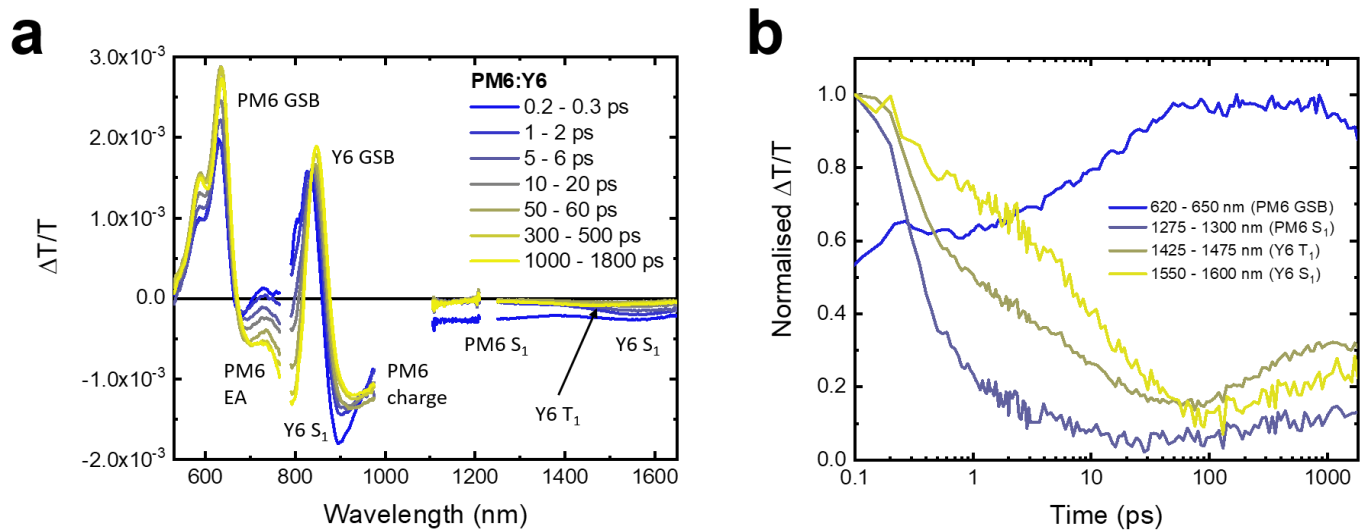

**Figure S13: (a)** The TA spectra of a PM6:Y6 film, pumped at 532 nm for preferential PM6 excitation with a low fluence of 1.8 μJ cm$^{-2}$. At the earliest times, the PM6 S$_1$ around 1150 nm is already heavily quenched and has largely disappeared by 1 ps. This confirms that the electron transfer in this blend takes place on ultrafast timescales, comparable to that of polymer:fullerene blends. From 1 ps onwards, the behaviour mimics that of the PM6:Y6 film excited at 800 nm for selective Y6 excitation. We note that whilst PM6 is preferentially excited at 532 nm, some inadvertent excitation of Y6 also occurs. This is evidenced by the presence of the Y6 S$_1$ PIAs at 900 and 1550 nm at 0.2 ps. These Y6 excitons then follow the previously observed hole transfer dynamics from Fig. S12. Additionally, the Y6 T$_1$ PIA is noticeable on the timescales of 100's ps, formed by back charge transfer (BCT) from the triplet charge transfer state ($^3$CT). **(b)** The kinetics of the PM6:Y6 film in relevant spectral regions. Immediately obvious is the rapid rate at which the PM6 S$_1$ PIA decays, confirming the presence of ultrafast electron transfer. Additionally, the PM6 GSB region clearly grows towards 100 ps, consistent with the timescale of the previously-observed hole transfer (Fig. S12).

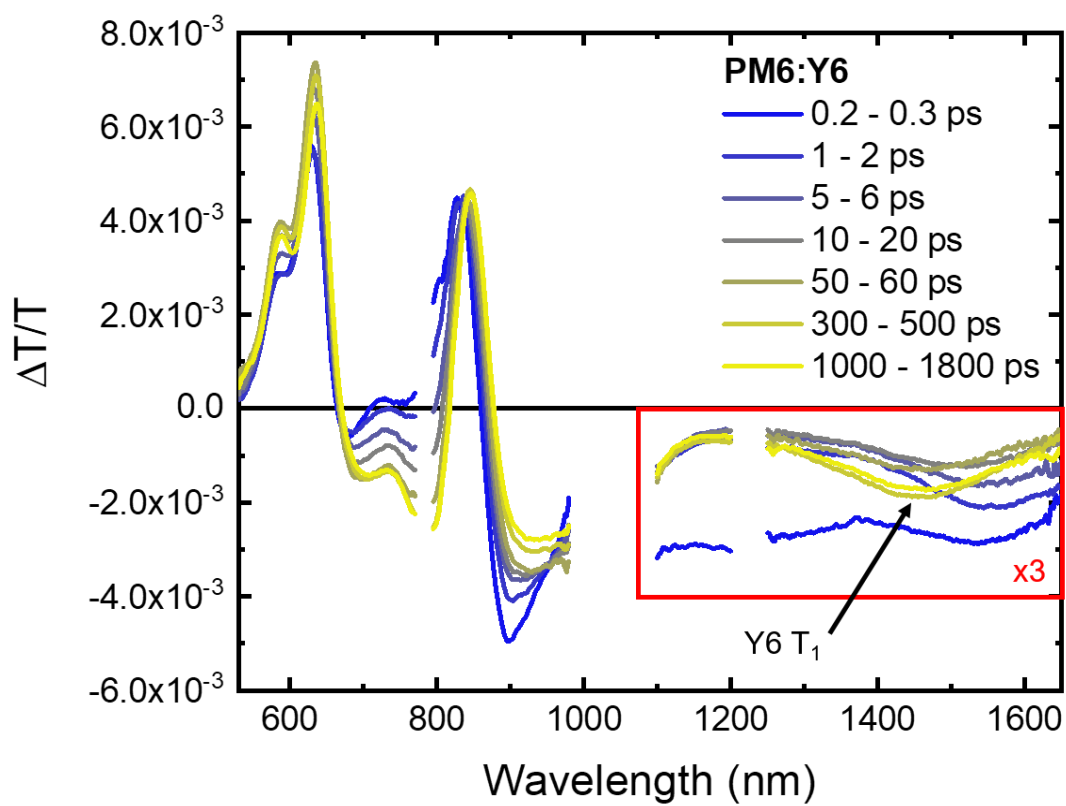

**Figure S14:** The full TA spectra of Fig. 2a: a PM6:Y6 film, pumped at 532 nm for preferential PM6 excitation with a moderate fluence of 5.4 μJ cm$^{-2}$. The rise of the Y6 $T_1$ PIA at 1450 nm is clearly correlated with the decrease in the PM6 charge PIA around 950 nm.

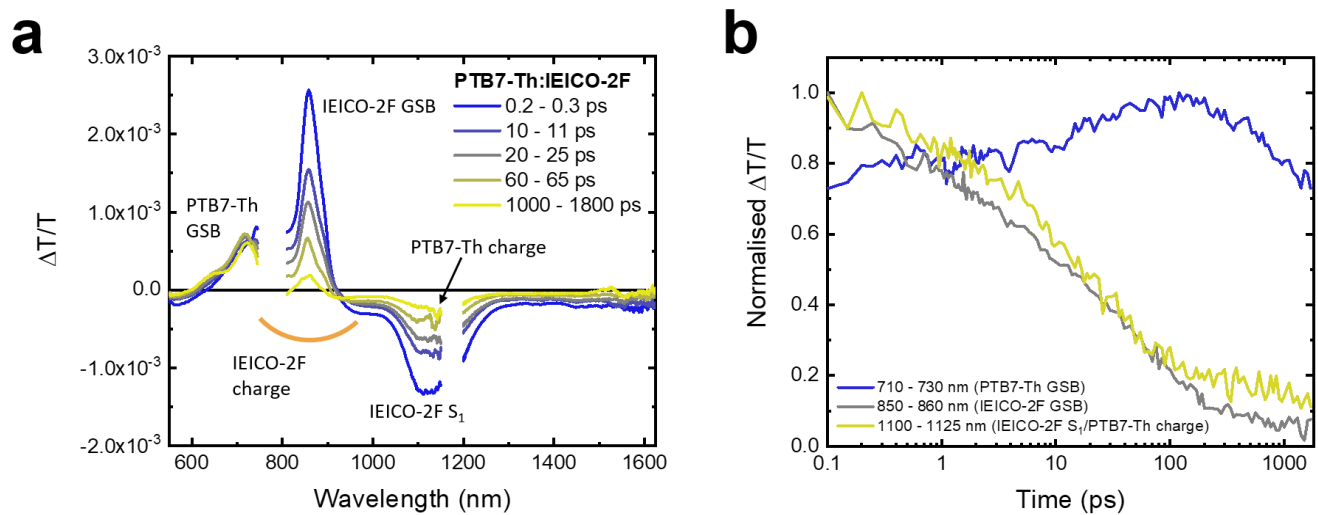

**Figure S15: (a)** The TA spectra of a PTB7-Th:IEICO-2F film, pumped at 850 nm for selective IEICO-2F excitation with a low fluence of 2.1 μJ cm$^{-2}$. At the earliest times, the spectrum resembles that of the neat IEICO-2F film, with the IEICO-2F GSB and $S_1$ PIA visible between 600 – 920 nm and at 1120 nm, respectively. As time progresses, we begin to notice the PTB7-Th GSB appearing, with a characteristic vibronic peak at 650 nm. This is the result of hole transfer from IEICO-2F. Interestingly, the IEICO-2F GSB also falls rapidly on the timescales of hole transfer. Furthermore, there is only a muted increase in the intensity of the PTB7-Th GSB region. This is unusual as if hole transfer is efficient, as suggested by the good OSC device performance, one may expect the NFA GSB to remain at roughly the same intensity and the polymer GSB to rise markedly. Indeed, this is the case in PM6:Y6. Thus, we expect that there is a new PIA forming underneath the GSB region as a result of the hole transfer process that is dragging the IEICO-2F GSB down and counteracting the expect rise in the PTB7-Th GSB. As the PTB7-Th hole PIA is widely reported to lie at 1150 nm[7–9], we assign this new PIA to the charge on IEICO-2F. Importantly, there is no new PIA formed around 1350 nm, where the IEICO-2F $T_1$ is found. **(b)** The kinetics of the PTB7-Th:IEICO-2F film in relevant spectral regions. From the rise in the PTB7-Th GSB region, we can see that hole transfer is completed by around 100 ps, which seems to be a common timescale for almost all low-offset NFA blends. The kinetics of the IEICO-2F GSB region and the IEICO-2F $S_1$ PIA almost perfectly mirror each other, suggesting that the process that is quenching singlets (hole transfer), is also responsible for the formation of the new PIA underneath the IEICO-2F GSB that is pulling it down. This provides more evidence for the formation of a new PIA band, corresponding to the IEICO-2F charge, underneath the GSB.

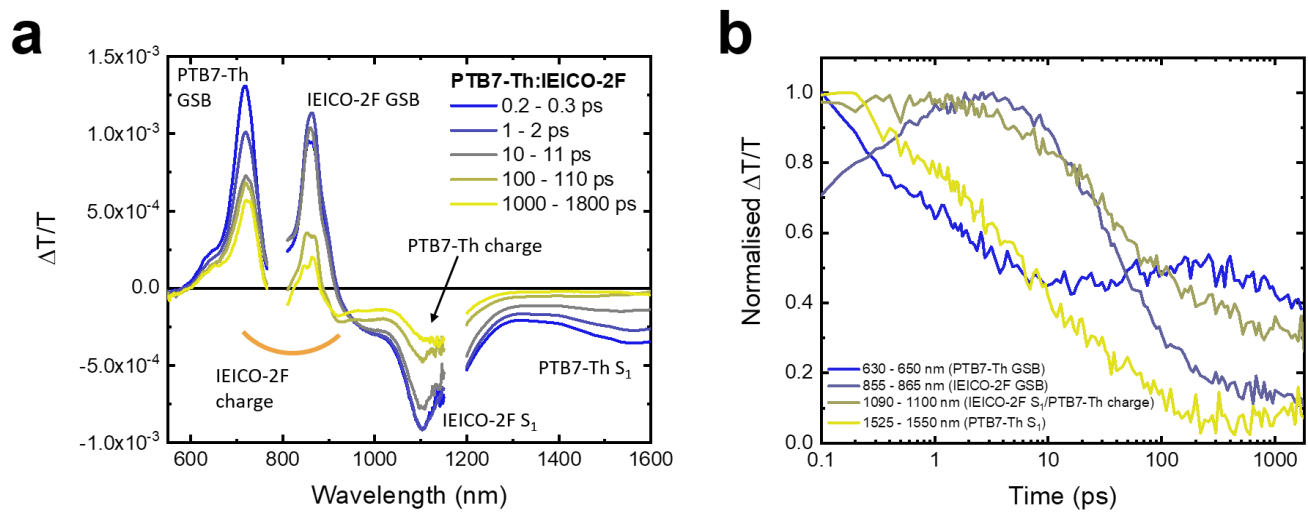

**Figure S16: (a)** The TA spectra of a PTB7-Th:IEICO-2F film, pumped at 620 nm for preferential PTB7-Th excitation with a low fluence of 1.3 μJ cm$^{-2}$. At the earliest times, the PTB7-Th GSB and S$_1$ PIA are present, as expected. However, the IEICO-2F GSB and S$_1$ PIA are also observable, suggesting that a significant amount of direct NFA excitation has occurred. Interestingly, the PTB7-Th S$_1$ PIA then decays on ps timescales, with a corresponding rise of the IOTIC-2F GSB. What is interesting is the simultaneous decrease in the PTB7-Th GSB in-line with the S$_1$ decay: if solely electron transfer was occurring, such a decrease would not be expected. Therefore, we suggest that energy transfer is occurring simultaneously with charge transfer from D to A in this blend. After ~10 ps, the IEICO-2F GSB begins to fall again, with a corresponding decrease in the PTB7-Th GSB peak and the IEICO-2F S$_1$ PIA. This is in-line with the timescales observed for the hole transfer from IEICO-2F to PTB7-Th in Fig. S12. As before, there is no evidence for the formation of the IEICO-2F T$_1$ PIA at 1350 nm. **(b)** The kinetics of the PTB7-Th:IEICO-2F film in relevant spectral regions. To support the assignments discussed above, we note that there is a clear rise in the IEICO-2F GSB towards 3 ps, suggesting a population transfer from PTB7-Th, before the back hole transfer takes place. Additionally, in the kinetic taken from the edge of the PTB7-Th GSB shows a re-bleaching of PTB7-Th chains occurs due to this back hole transfer process.

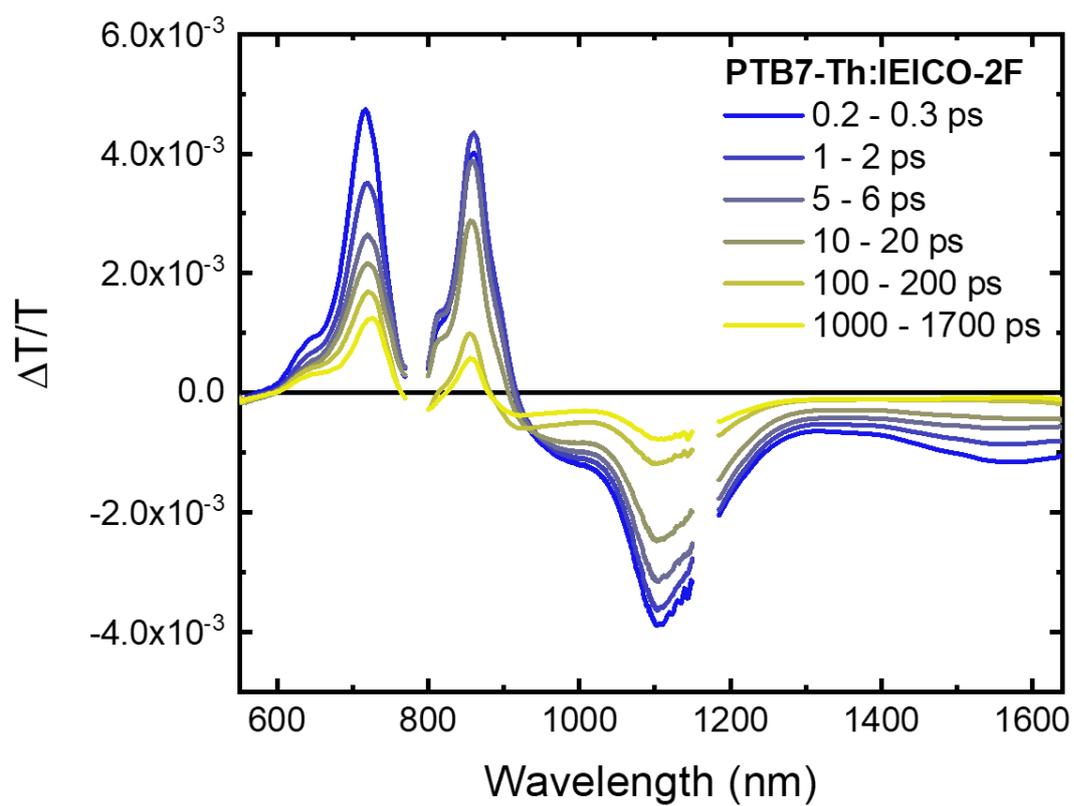

**Figure S17:** The full TA spectra of Fig. 2d: a PTB7-Th:IEICO-2F film, pumped at 620 nm for preferential PTB7-Th excitation with a moderate fluence of 3.8 μJ cm$^{-2}$.

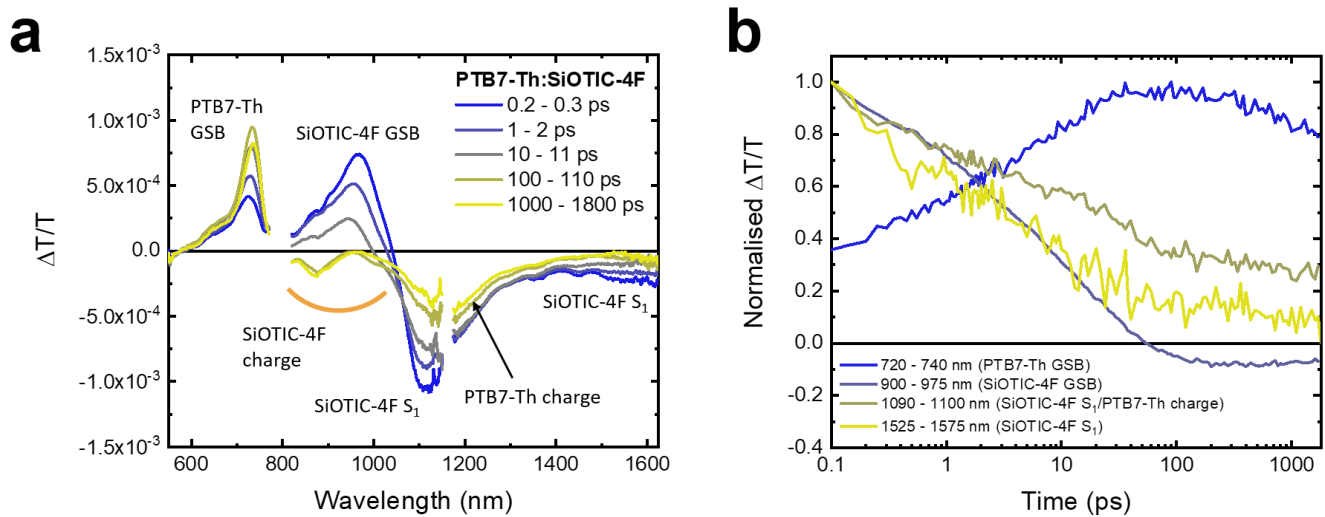

**Figure S18: (a)** The TA spectra of a PTB7-Th:SiOTIC-4F film, pumped at 975 nm for selective SiOTIC-4F excitation with a low fluence of 1.6 µJ cm⁻². At the earliest times, we can clearly see the SiOTIC-4F GSB and sharp $S_1$ PIA at 1100 nm. Additionally, the PTB7-Th GSB is also present at 0.2 ps, suggesting that some of the hole transfer occurs on ultrafast timescales. As time progresses, the PTB7-Th GSB continues to rise, with a concomitant fall in the SiOTIC-4F $S_1$ PIA at 1100 nm. The PTB7-Th hole PIA at 1150 nm is also clearly visible by 100 ps, more obvious than in the PTB7-Th:IEICO-2F blend as it is a little red-shifted from the NFA $S_1$ PIA. Additionally, this blend also exhibits similar behaviour to PTB7-Th:IEICO-2F, where upon charge transfer occurring, the GSB of the NFA falls sharply. Thus, as no other PIAs that could be attributed to charges on SiOTIC-4F are visible elsewhere, we suggest that the SiOTIC-4F charge PIA lies under the GSB. Unfortunately, the SiOTIC-4F $T_1$ PIA overlaps almost perfectly with the PTB7-Th hole PIA (Fig. S4d), meaning it is not immediately obvious to determine whether triplet formation occurs. This will be revisited shortly. **(b)** The kinetics of the PTB7-Th:SiOTIC-4F film in relevant spectral regions. We note that the PTB7-Th GSB peaks at around 30 ps, suggesting hole transfer is completed by this time.

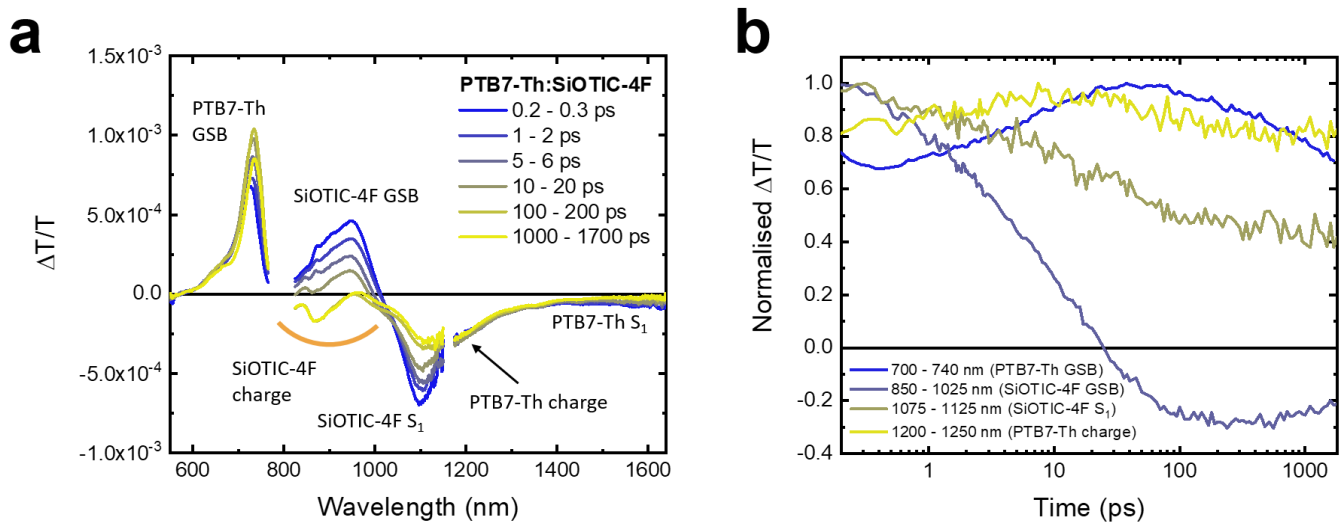

**Figure S19: (a)** The TA spectra of a PTB7-Th:SiOTIC-4F film, pumped at 620 nm for preferential PTB7-Th excitation with a low fluence of 2.1 μJ cm$^{-2}$. At the earliest times, the PTB7-Th GSB is present, as expected. However, the PTB7-Th S$_1$ PIA is not obvious and the SiOTIC-4F GSB, which would be expected to increase in intensity following electron transfer from the D, is also at a maximum at the earliest times resolvable. This suggests that the initial electron transfer process from PTB7-Th is ultrafast, occurring on sub-100 fs timescales. The SiOTIC-4F S1 PIA is also clearly visible at 1100 nm, suggesting some unintentional NFA excitation has also occurred. The spectrum then evolves in a very similar fashion to Fig. S17, when the SiOTIC-4F was selectively excited, confirming that what occurs over longer timescales is the hole transfer from SiOTIC-4F. **(b)** The kinetics of the PTB7-Th:SiOTIC-4F film in relevant spectral regions. We note that the PTB7-Th GSB peaks at around 30 ps, consistent with the hole transfer timescales observed previously when the blend was excited at 975 nm.

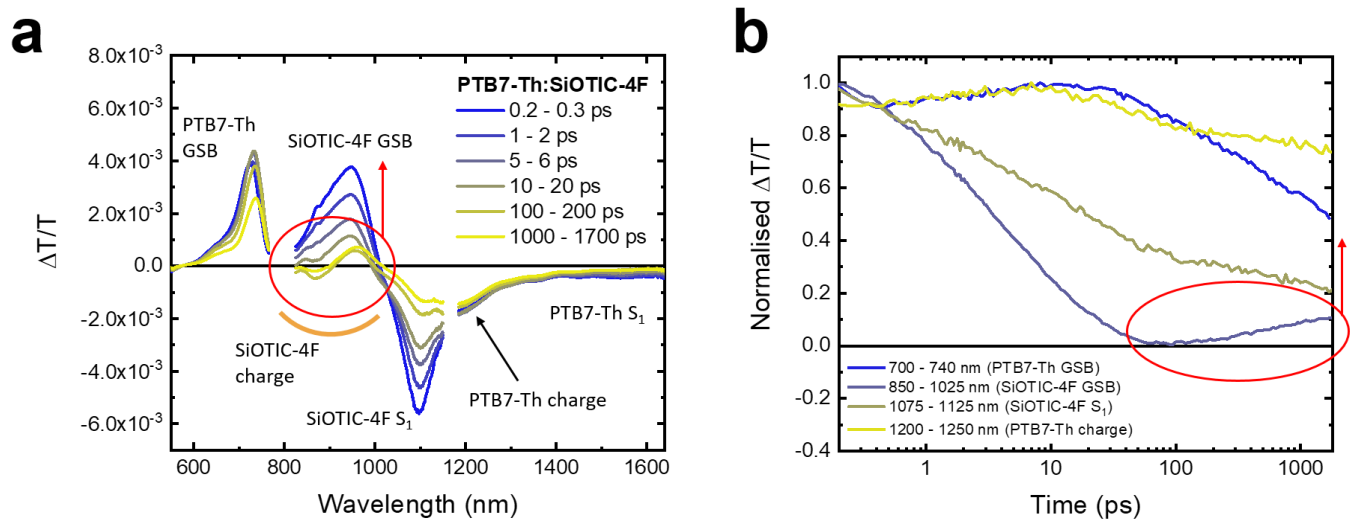

**Figure S20:** **(a)** The TA spectra of a PTB7-Th:SiOTIC-4F film, pumped at 620 nm for preferential PTB7-Th excitation with a high fluence of 10.5 µJ cm$^{-2}$. Compared to the low fluence measurement in Fig. S19, we note that the SiOTIC-4F GSB is substantially above zero by 100 ps, highlighted by the red circle. **(b)** The kinetics of the PTB7-Th:SiOTIC-4F film in relevant spectral regions. There are increased rates of non-geminate recombination in this higher fluence measurement, most obvious in the more rapid loss of the PTB7-Th GSB. The kinetic of the SiOTIC-4F GSB region in this measurement clearly differs from the low fluence measurement, where it doesn't dip below zero and actually increases from 100 ps onwards.

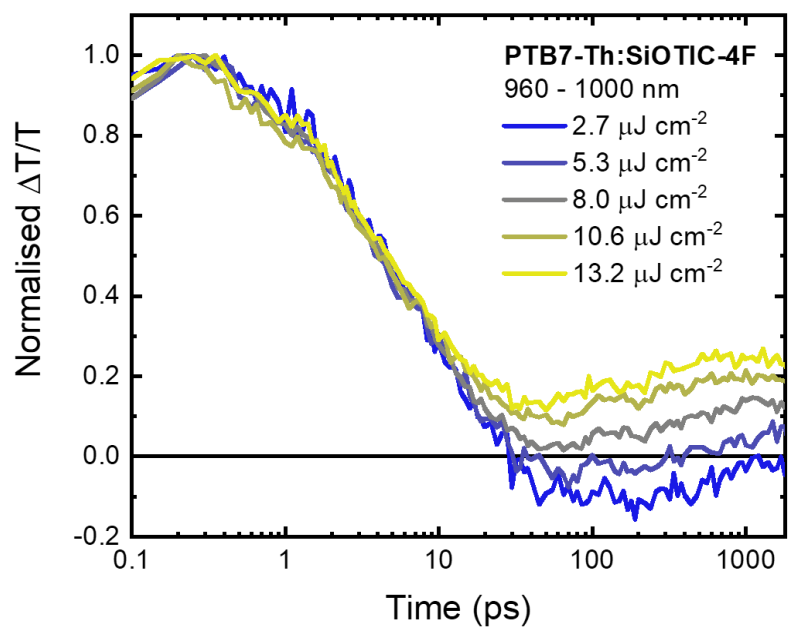

**Figure S21:** The kinetics taken from the SiOTIC-4F GSB region of a PTB7-Th:SiOTIC-4F film, pumped at 532 nm for preferential PTB7-Th excitation. In order to elucidate the dynamics of this unusual behaviour of the SiOTIC-4F, a detailed fluence series was performed. We note that with increasing fluence, the intensity of the region associated with the SiOTIC-4F GSB (960 – 1000 nm) increases more and more rapidly, reaching a higher proportion of the initial intensity at ever earlier times. This, combined with the loss of charges on PTB7-Th and the recovery of its GSB (Fig. S20), implies that a new species is being created on SiOTIC-4F from charge carriers. Given the strong fluence dependence of the creation of this new species, we assign this process to the formation of triplet excitons on SiOTIC-4F via a BCT from the $^3$CT, which is formed more rapidly via increased levels of non-geminate recombination at higher fluences[10,11]. Importantly, the rise of the SiOTIC-4F GSB is not actually due to the presence of the triplets themselves increasing the number of NFA molecules being bleached. Rather, it is due to the loss of the charge PIA underneath the SiOTIC-4F GSB, coupled with a minimal change in the number of NFA molecules not in the ground state which results in the rise; the SiOTIC-4F molecules previously bleached by an electron will continue to be bleached by the presence of a triplet. Thus, as we cannot use the $T_1$ PIA to reliably determine whether triplets form due to its overlap with the PTB7-Th charge PIA, this detailed study of the GSB dynamics provides us an alternative route to investigate triplet formation.

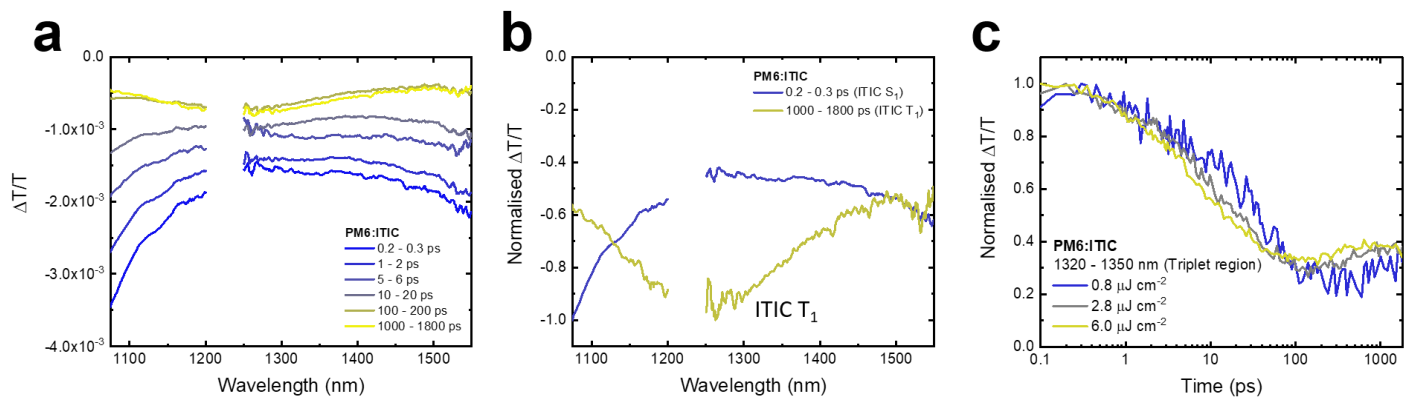

**Figure S22: (a)** The TA spectra in the NIR region of a PM6:ITIC film, pumped at 700 nm for selective ITIC excitation with a high fluence of 6.0 μJ cm$^{-2}$. Features associated with the ITIC $S_1$ at 1100 nm and 1550 nm decay away on ps timescales due to hole transfer to PM6. By 100 ps, a new PIA band centred at 1250 nm is visible. This feature is assigned to the ITIC $T_1$, due to the perfect spectral match with previous reports[1]. **(b)** The normalised spectra at 0.2 – 0.3 ps and 1000 – 1800 ps to more clearly show the ITIC $T_1$ PIA band. **(c)** TA kinetics of a fluence series taken from the spectral region associated with the ITIC $T_1$ PIA. More rapid decay can initially be seen at higher fluences, indicating bimolecular recombination processes are taking place (e.g. exciton-exciton annihilation or non-geminate recombination). However, from ~100 ps onwards, the region associated with the ITIC $T_1$ becomes increasingly more intense at earlier times with higher fluence. This is consistent with triplet formation via non-geminate recombination (NGR), as the greater charge density in the film increases the probability of NGR events that form the $^3$CT feeder state[10–12]. Therefore, the fluence dependence of $T_1$ formation confirms PM6:ITIC forms triplets via NGR.

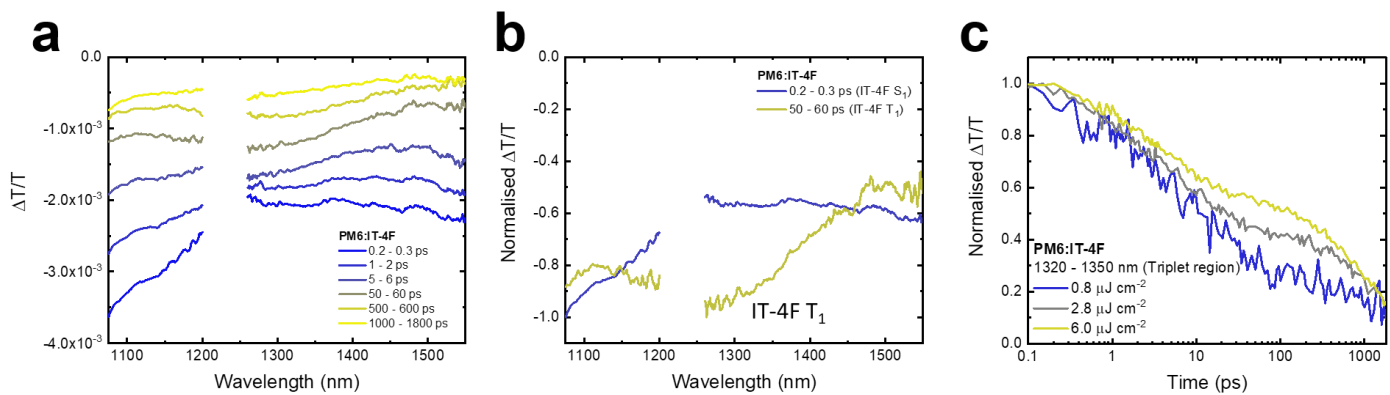

**Figure S23: (a)** The TA spectra in the NIR region of a PM6:IT-4F film, pumped at 700 nm for selective IT-4F excitation with a high fluence of 6.0 μJ cm$^{-2}$. Features associated with the IT-4F $S_1$ at 1100 nm and 1550 nm decay away on ps timescales due to hole transfer to PM6. By 100 ps, a new PIA band centred at 1250 nm is visible. This feature is assigned to the IT-4F $T_1$, due to the perfect spectral match with previous reports[1]. **(b)** The normalised spectra at 0.2 – 0.3 ps and 1000 – 1800 ps to more clearly show the IT-4F $T_1$ PIA band. **(c)** TA kinetics of a fluence series taken from the spectral region associated with the IT-4F $T_1$ PIA. From ~10 ps onwards, the region associated with the IT-4F $T_1$ becomes increasingly more intense at earlier times with higher fluence. This is consistent with triplet formation via NGR, as the greater charge density in the film increases the probability of NGR events that form the $^3$CT feeder state[10–12]. Therefore, the fluence dependence of $T_1$ formation confirms - PM6:IT-4F forms triplets via NGR.

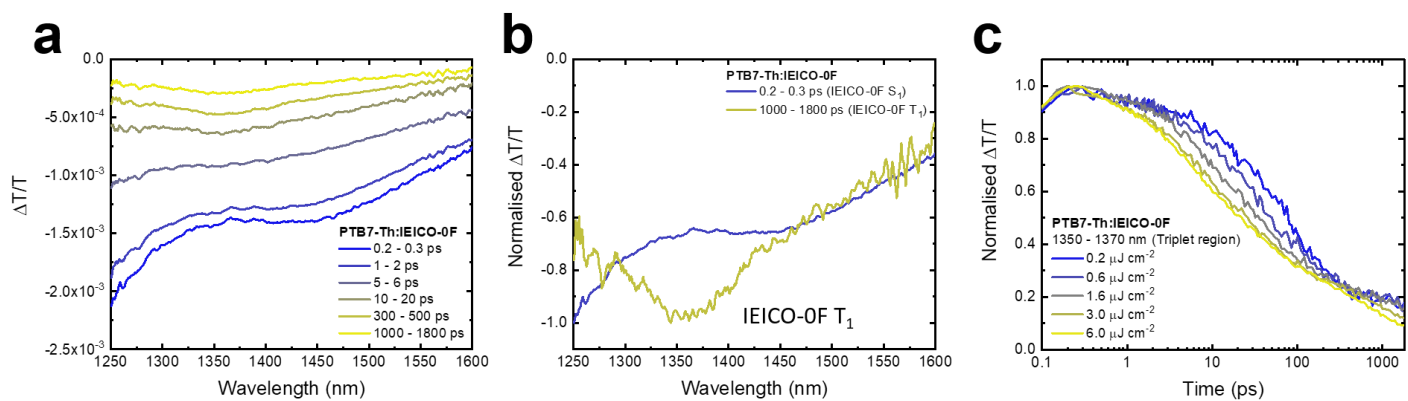

**Figure S24: (a)** The TA spectra in the NIR region of a PTB7-Th:IEICO-0F film, pumped at 800 nm for selective IEICO-0F excitation with a high fluence of 6.0 μJ cm$^{-2}$. The PIA associated with the IEICO-0F $S_1$ at 1250 nm decays away on ps timescales due to both decay back to the ground state and hole transfer to PTB7-Th. By 300 ps, a new PIA band centred at 1370 nm is visible. This feature is assigned to the IEICO-0F $T_1$, due to the perfect spectral match with Fig. S37b. **(b)** The normalised spectra at 0.2 – 0.3 ps and 1000 – 1800 ps to more clearly show the IEICO-0F $T_1$ PIA band. **(c)** TA kinetics of a fluence series taken from the 1350 – 1370 nm spectral region associated with the IEICO-0F $T_1$ PIA. Whilst triplets do indeed form in this blend, their dynamics are not what would be expected for triplets formed via NGR. A detailed fluence series reveals that <100 ps, there is a clear increase in the excited state decay rate. We attribute this to increased singlet exciton-exciton annihilation (the $S_1$ PIA also has significant intensity around 1350 – 1370 nm), not non-geminate recombination, as minimal hole transfer has occurred by 100 ps (Fig. S25). After 100 ps, when the $T_1$ PIA begins to become clearly visible, there is no fluence dependence in this region for lower fluences between 0.2 – 1.6 μJ cm$^{-2}$. A fluence dependence in the $T_1$ region only becomes apparent for higher fluences of 3.0 – 6.0 μJ cm$^{-2}$. However, the decay rate in the 1350 – 1370 nm region actually increases, which is not what would be expected if triplets were being formed via NGR. Therefore, this rules out NGR as a significant triplet formation mechanism in this blend, with the increased $T_1$ decay rate likely attributable to triplet-charge annihilation[11,13]. This leaves two possible routes for the triplet formation: ISC from the geminate $^1$CT, or direct ISC from un-dissociated IEICO-0F singlets. From the trEPR (Fig. S59), we rule out the former as only ISC triplets are present. Further, we note that the kinetics of the $T_1$ region perfectly matches the kinetics taken from the same wavelength region in the PS:IEICO-0F film, which exhibits a relatively high (~5%) yield of ISC triplets (Fig. S39b). This confirms that the primary triplet formation route in PTB7-Th:IEICO-0F is the ISC of un-dissociated singlet excitons, facilitated by the slow hole transfer rate. This is consistent with the poor device performance observed in this blend, which is much worse than the PTB7-Th:IEICO-2F and -4F devices.

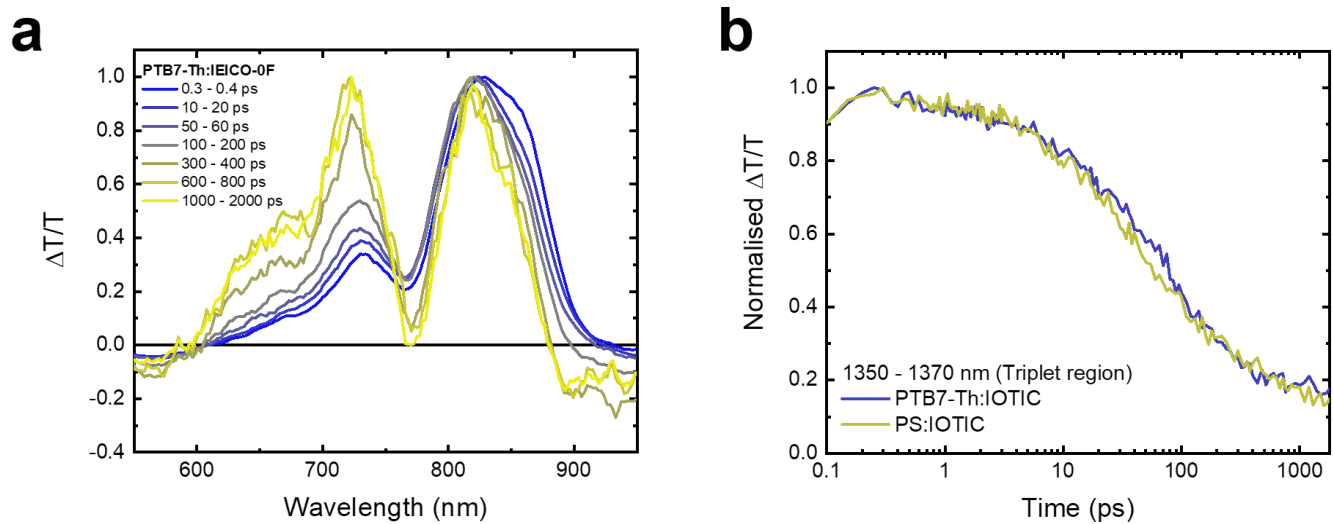

**Figure S25:** Additional data to assist with the understanding of the PTB7-Th:IEICO-0F triplet formation discussion. **(a)** The normalised TA spectra of a PTB7-Th:IEICO-0F film, excited at 860 nm for selective IEICO-0F excitation with a fluence of 0.5 μJ cm$^{-2}$. Initially, only the IEICO-0F GSB is visible, as expected. However, we note that the PTB7-Th GSB does not become readily apparent until 300 ps, by which time significant decay of IEICO-0F singlets to the ground state and to $T_1$ via ISC will have occurred. This can explain the inferior performance of the PTB7-Th:IEICO-0F blend compared to the -2F and -4F equivalents. **(b)** The normalised TA kinetics of PTB7-Th:IEICO-0F and PS:IEICO-0F 1:1.5 films taken around the maximum of the IEICO-0F $T_1$ PIA at 1350 – 1370 nm. The dynamics of this region overlap almost perfectly, strongly suggesting that the triplet formation mechanism in PTB7-Th:IEICO-0F is primarily via direct ISC of un-dissociated singlet excitons; this is consistent with the observation of an intense ISC triplet signal in the trEPR (Fig. S59).

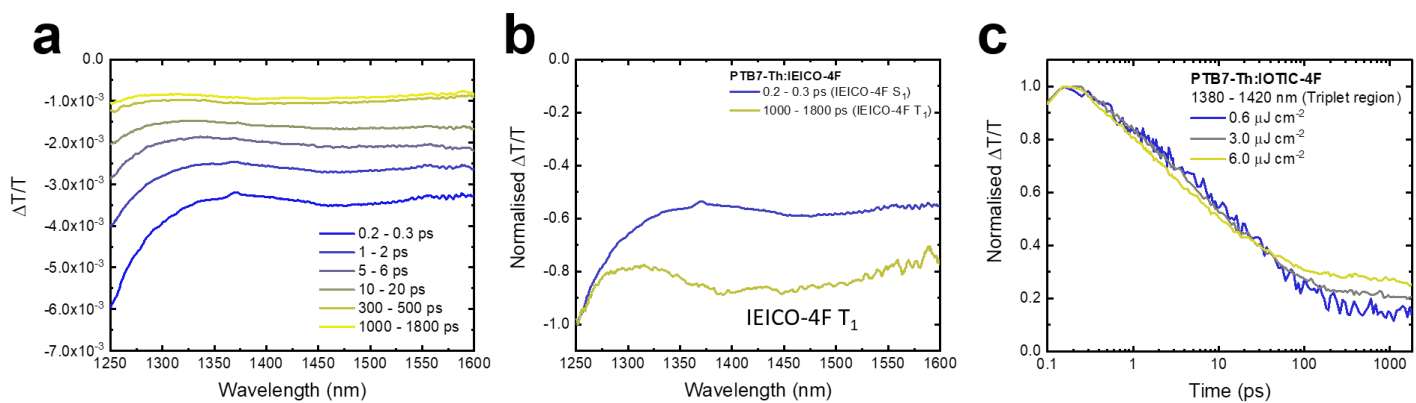

**Figure S26: (a)** The TA spectra in the NIR region of a PTB7-Th:IEICO-4F film, pumped at 800 nm for selective IEICO-4F excitation with a high fluence of 6.0 μJ cm$^{-2}$. The PIA associated with the IEICO-4F $S_1$ at 1250 nm decays away on ps timescales due to hole transfer to PTB7-Th. By 300 ps, a new broad new PIA band centred at 1400 nm is visible. This feature is assigned to the IEICO-4F $T_1$, due to the perfect spectral match with Fig. S39f. **(b)** The normalised spectra at 0.2 – 0.3 ps and 1000 – 1800 ps to more clearly show the IEICO-4F $T_1$ PIA band. **(c)** TA kinetics of a fluence series taken from the spectral region associated with the IEICO-4F $T_1$ PIA. More rapid decay can initially be seen at higher fluences, indicating bimolecular recombination processes are taking place (e.g. exciton-exciton annihilation or NGR). However, from ~50 ps onwards, the region associated with the IEICO-4F $T_1$ becomes increasingly more intense at earlier times with higher fluence. This is consistent with triplet formation via NGR, as the greater charge density in the film increases the probability of NGR events that form the $^3$CT feeder state[10–12]. Therefore, the fluence dependence of $T_1$ formation confirms PTB7-Th:IEICO-4F forms triplets via NGR.

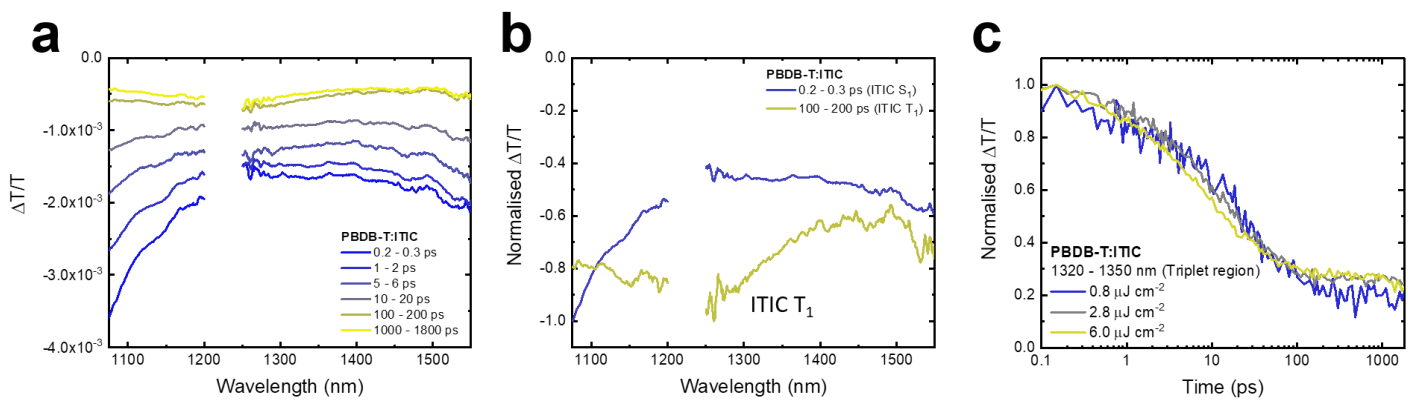

**Figure S27: (a)** The TA spectra in the NIR region of a PBDB-T:ITIC film, pumped at 700 nm for selective ITIC excitation with a high fluence of 6.0 μJ cm$^{-2}$. Features associated with the ITIC $S_1$ at 1100 nm and 1550 nm decay away on ps timescales due to hole transfer to PBDB-T. By 100 ps, a new PIA band centred at 1250 nm is visible. This feature is assigned to the ITIC $T_1$, due to the perfect spectral match with previous reports[1]. **(b)** The normalised spectra at 0.2 – 0.3 ps and 1000 – 1800 ps to more clearly show the ITIC $T_1$ PIA band. **(c)** TA kinetics of a fluence series taken from the spectral region associated with the ITIC $T_1$ PIA. More rapid decay can initially be seen at higher fluences, indicating bimolecular recombination processes are taking place (e.g. exciton-exciton annihilation or NGR). However, from ~100 ps onwards, the region associated with the ITIC $T_1$ becomes increasingly more intense at earlier times with higher fluence. This is consistent with triplet formation via NGR, as the greater charge density in the film increases the probability of NGR events that form the $^3$CT feeder state[10–12]. Therefore, the fluence dependence of $T_1$ formation confirms PBDB-T:ITIC forms triplets via NGR.

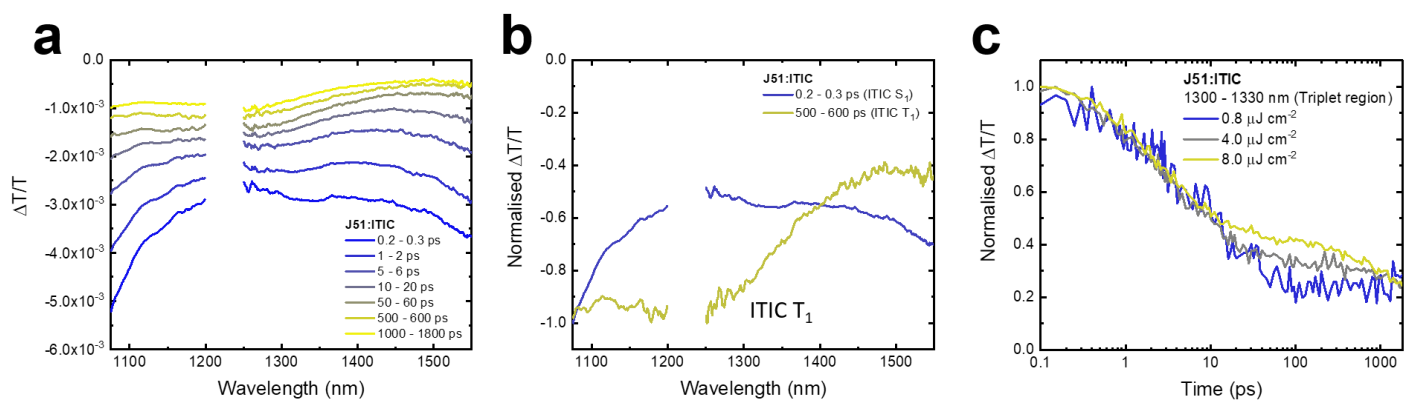

**Figure S28: (a)** The TA spectra in the NIR region of a J51:ITIC film, pumped at 680 nm for selective ITIC excitation with a high fluence of 8.0 µJ cm$^{-2}$. Features associated with the ITIC S$_1$ at 1100 nm and 1550 nm decay away on ps timescales due to hole transfer to J51. By 100 ps, a new PIA band centred at 1250 nm is visible. This feature is assigned to the ITIC T$_1$, due to the perfect spectral match with previous reports[1]. **(b)** The normalised spectra at 0.2 – 0.3 ps and 1000 – 1800 ps to more clearly show the ITIC T$_1$ PIA band. **(c)** TA kinetics of a fluence series taken from the spectral region associated with the ITIC T$_1$ PIA. From ~10 ps onwards, the region associated with the ITIC T$_1$ becomes increasingly more intense at earlier times with higher fluence. This is consistent with triplet formation via NGR, as the greater charge density in the film increases the probability of NGR events that form the $^3$CT feeder state[10–12]. Therefore, the fluence dependence of T$_1$ formation confirms J51:ITIC forms triplets via NGR.

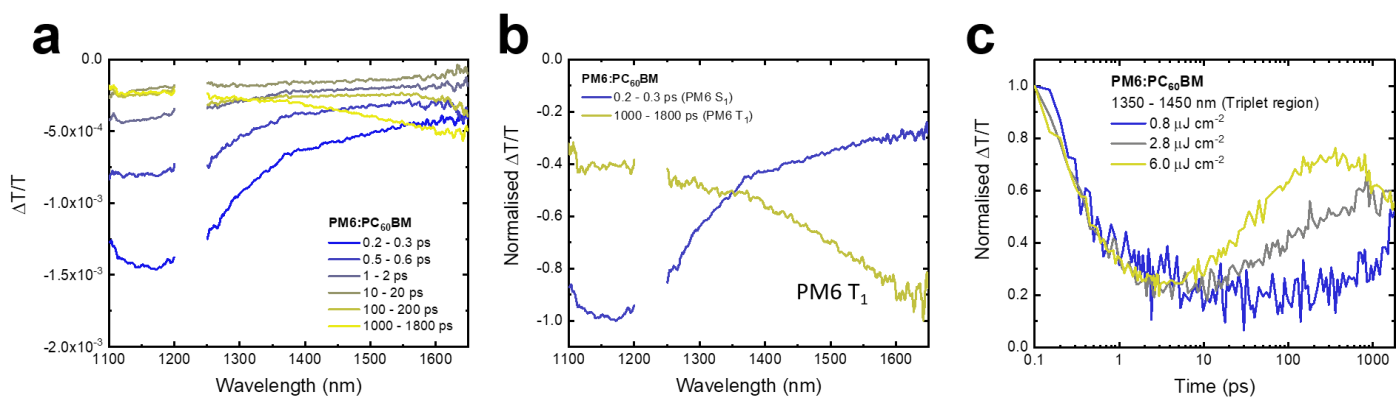

**Figure S29: (a)** The TA spectra in the NIR region of a PM6:$PC_{60}$BM film, pumped at 532 nm for preferential PM6 excitation with a moderate fluence of 2.8 μJ cm$^{-2}$. The PIA band at 1150 nm belongs to the $S_1$ of PM6 and is rapidly quenched within 1 ps, confirming that ultrafast electron transfer to $PC_{60}$BM takes place. Over a timescale of 100's ps, a new PIA band peaked towards the edge of our probe range at 1650 nm grows in. Due to the fluence dependence of its formation, we assign it to the PM6 $T_1$. **(b)** The normalised spectra at 0.2 – 0.3 ps and 1000 – 1800 ps to more clearly show the PM6 $T_1$ PIA band. **(c)** TA kinetics of a fluence series taken from the spectral region associated with the PM6 $T_1$ PIA. From ~20 ps onwards, the region associated with the PM6 $T_1$ becomes increasingly more intense at earlier times with higher fluence. This is consistent with triplet formation via NGR, as the greater charge density in the film increases the probability of NGR events that form the $^3$CT feeder state[10–12]. Therefore, the fluence dependence of $T_1$ formation confirms PM6:$PC_{60}$BM forms triplets via NGR.

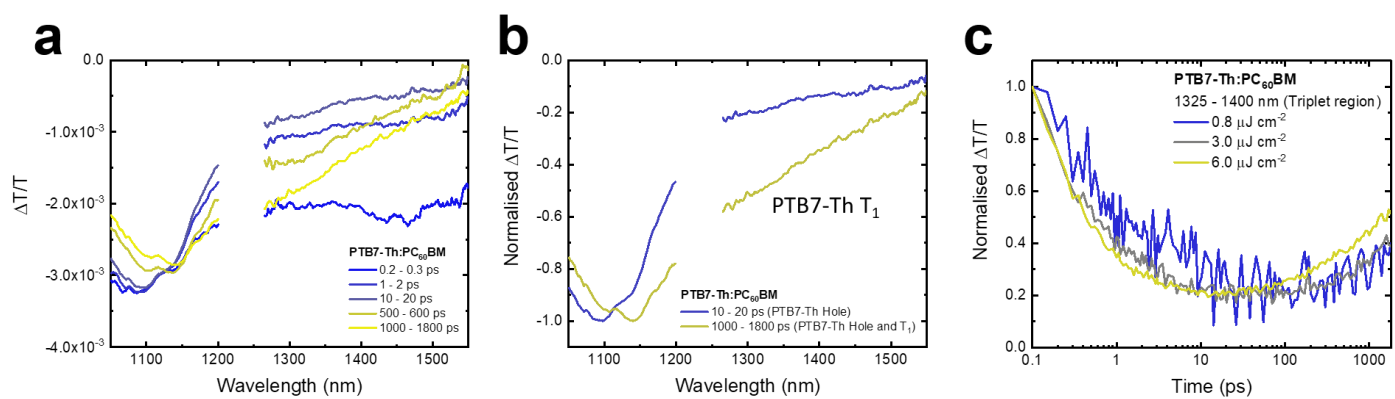

**Figure S30: (a)** The TA spectra in the NIR region of a PTB7-Th:PC$_{60}$BM film, pumped at 700 nm for preferential PTB7-Th excitation with a high fluence of 6.0 µJ cm$^{-2}$. The PIA band at 1450 nm belongs to the S$_1$ of PTB7-Th and is rapidly quenched within 1 ps, confirming that ultrafast electron transfer to PC$_{60}$BM takes place. A new PIA at 1100 nm is assigned to the charges on PTB7-Th, in-line with previous reports[7–9]. Over a timescale of 100's ps, the PTB7-Th charge PIA decreases in intensity and a new PIA that strongly overlaps with the charge begins to form around 1300 nm. Due to the fluence dependence of its formation and a previous report on non-geminate triplet formation in PTB7-Th blends with fullerenes[8], we assign it to the PTB7-Th T$_1$. **(b)** The normalised spectra at 0.2 – 0.3 ps and 1000 – 1800 ps to more clearly show the PTB7-Th T$_1$ PIA band. **(c)** TA kinetics of a fluence series taken from the spectral region associated with the PTB7-Th T$_1$ PIA. From ~100 ps onwards, the region associated with the PTB7-Th T$_1$ becomes increasingly more intense at earlier times with higher fluence. This is consistent with triplet formation via non-geminate recombination (NGR), as the greater charge density in the film increases the probability of NGR events that form the $^3$CT feeder state[10–12]. Therefore, the fluence dependence of T$_1$ formation confirms PTB7-Th:PC$_{60}$BM forms triplets via NGR.

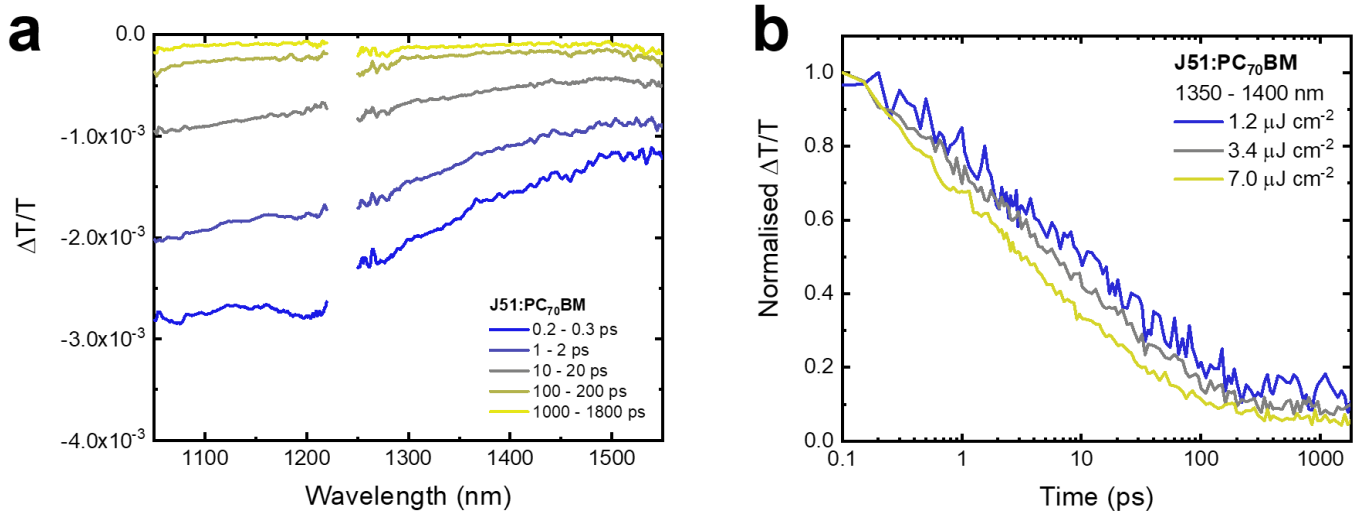

**Figure S31: (a)** The TA spectra in the NIR region of a J51:PC$_{70}$BM film, pumped at 600 nm for preferential J51 excitation with a high fluence of 7.0 µJ cm$^{-2}$. Due to its presence immediately after excitation, the PIA band at 1100 nm belongs to the S$_1$ of J51 and is quenched somewhat more slowly than in the other blends, confirming that electron transfer to PC$_{70}$BM takes place over longer timescales. After electron transfer is completed, there is no obvious formation of new PIAs, strongly suggesting that T$_1$ formation does not noticeably occur in this blend. **(b)** TA kinetics of a fluence series. At higher fluences, there is a faster decay at early times, likely due to exciton-exciton annihilation. However, there is no further fluence dependence at later timescales, confirming that J51:PC$_{70}$BM does not form triplets via NGR.

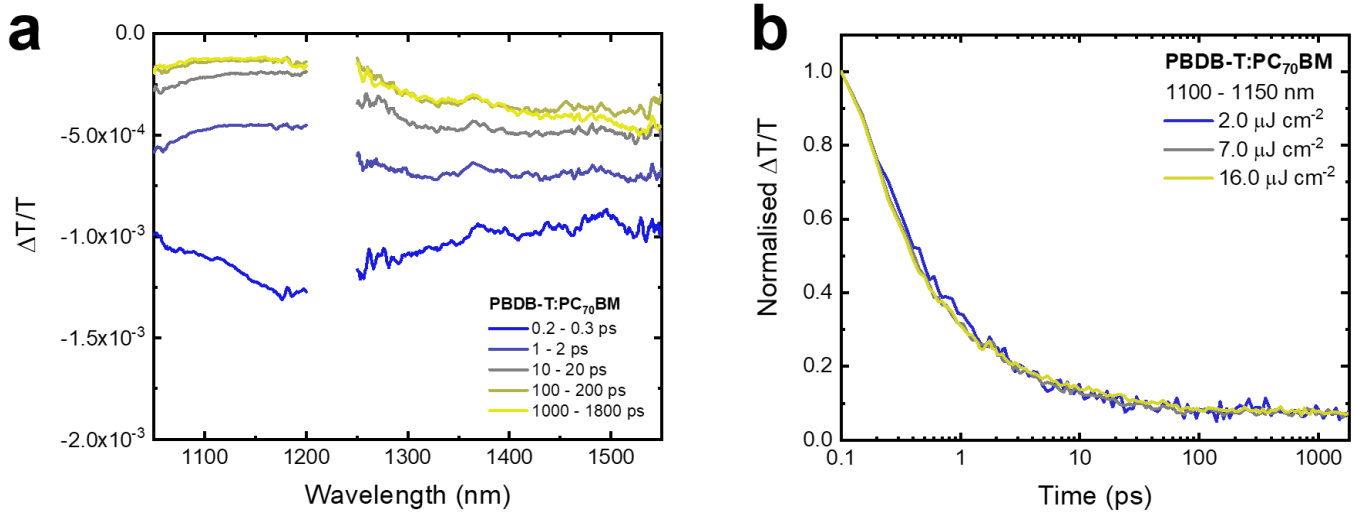

**Figure S32: (a)** The TA spectra in the NIR region of a PBDB-T:PC$_{70}$BM film, pumped at 600 nm for preferential PBDB-T excitation with a high fluence of 7.0 μJ cm$^{-2}$. Due to its presence immediately after excitation and the similarity to the S$_1$ in the related material PM6, the PIA band at 1200 nm is assigned to the S$_1$ of PBDB-T. We note that the S1 is largely quenched by 1 ps, indicating ultrafast electron transfer to PC$_{70}$BM occurs. After electron transfer is completed, there is no obvious formation of new PIAs, strongly suggesting that T$_1$ formation does not noticeably occur in this blend. **(b)** TA kinetics of a fluence series. There is no obvious fluence dependence at any times up to 1.8 ns in this blend, confirming that PBDB-T:PC$_{70}$BM does not form triplets via NGR.

# Electro-absorption analysis of NFA blends to determine $k_{dissociation}$

In OSC blends, the timescales of the CT state dissociation can be readily determined by evaluating the electro-absorption (EA) features in the TA[2]. The EA represents the Stark-shift of the material absorption spectrum by the electric field of the separating charges, with the maximum intensity reached when the CT states have dissociated into FC[2–5]. Thus, tracking the EA response provides an insight into the kinetics of charge separation. We note that the EA typically manifests as a sharp, negative signal in the TA at the steady-state absorption edge[2–5]. Here, we have investigated the kinetics of the EA formation in all NFA blends where the EA response of the D polymer is not obscured by the GSB of the NFA (i.e. those with sufficiently large offsets of the peak D and A absorption peaks). In all NFA blends analysed, we find the CT dissociation takes place with a time constant of tens of ps (reaching a maximum after ~100 – 200 ps), corresponding to a $k_{dissociation}$ of between $10^{10} – 10^{11}$ s$^{-1}$. Importantly, as the vibrational relaxation of CT states is much faster (<100 fs) than these observed dissociation timescales[14,15], the initial separation following charge transfer must occur from the same thermalized CT states as formed by NGR. Thus the dissociation timescales extracted here will be relevant when considering whether $^3$CT state re-dissociation can out-compete BCT to $T_1$, the key factor controlling whether $T_1$ formation occurs. In contrast to the NFA systems, the EA response is already near-maximum in the PM6:PC$_{60}$BM blend by 1 ps (Fig. S10), consistent with the ultrafast long-range charge separation typically observed in fullerene systems with a net driving energy[2–4].

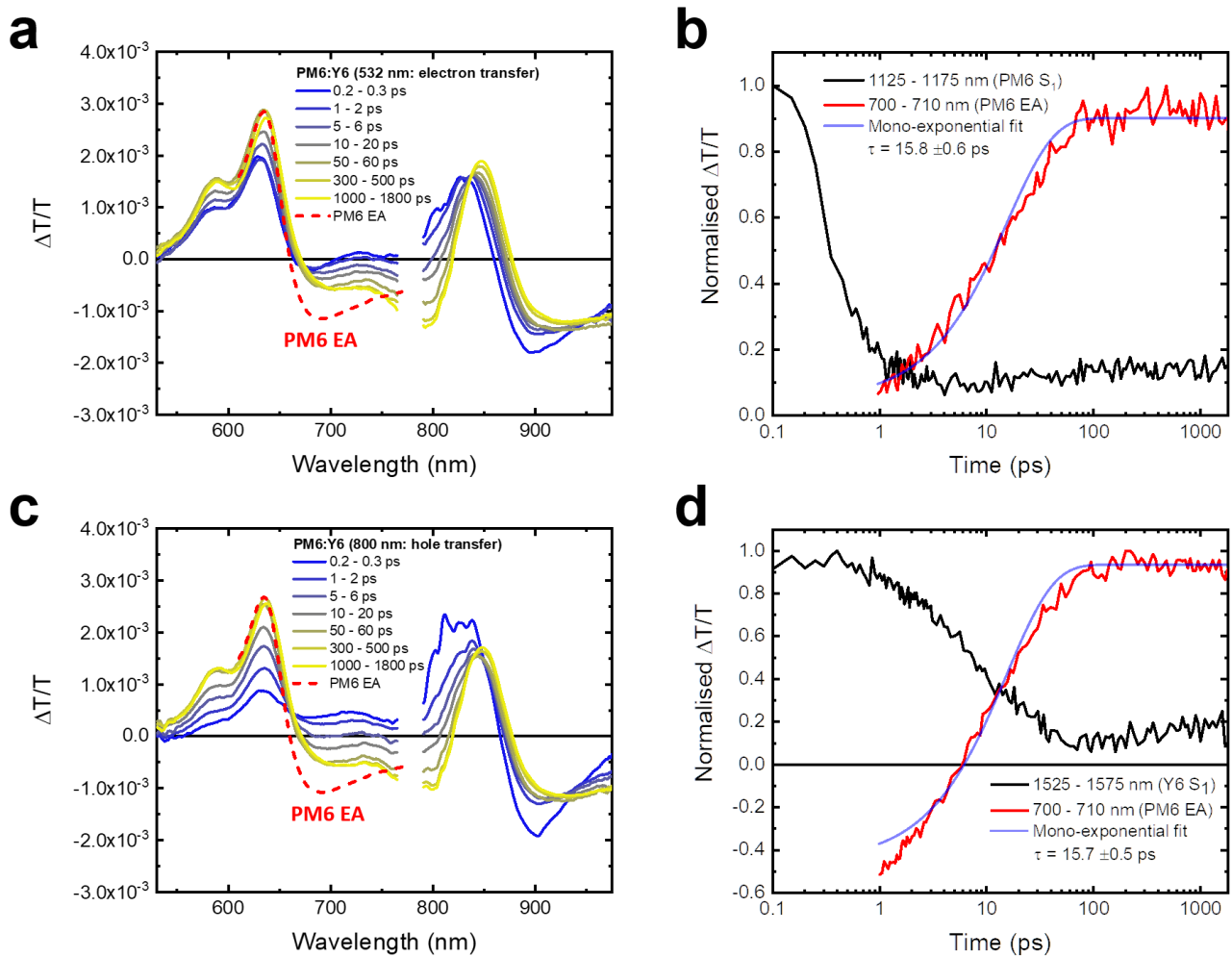

**Figure S33: (a)** The TA spectra of the PM6:Y6 blend, excited with a low fluence of 1.8 μJ cm⁻² at 532 nm for preferential PM6 excitation. The primary process occurring in this blend is electron transfer from PM6 to Y6. A new negative band between 700 – 750 nm grows in over timescales up to 100 ps. Through comparison to the EA obtained from a PM6:PC$_{60}$BM blend (dashed red line), this new feature is assigned to the EA of PM6 by the separating charges. The differences in spectral shape can be explained by the overlap with the vibronic shoulder of the Y6 GSB between 650 – 750 nm. **(b)** The TA kinetics of the PM6:Y6 blend following preferential excitation of PM6 at 532 nm. The decay of the PM6 S$_1$ shows the timescales of electron transfer in the blend, largely taking place <1 ps. The kinetic of the EA is also displayed, where it clearly grows in more slowly than the time taken for electron transfer. This represents CTE dissociation into FC in the blend, which takes up to 100 ps. The EA growth can be fitted with a mono-exponential function with a time constant of 15.8 ±0.6 ps, yielding $k_{dissociation}$ = 6.3 x 10$^{10}$ s⁻¹. Critically, as the time taken for CTE dissociation is much slower than the vibrational relaxation of CTEs (<100 fs), it will take place from the same thermalized CTEs that are formed by NGR. **(c)** The TA spectra of the PM6:Y6 blend, excited with a low fluence

of 1.0 μJ cm⁻² at 800 nm for selective Y6 excitation. The only process occurring in this blend is hole transfer from Y6 to PM6. Again, the PM6 EA feature between 700 – 750 nm can be seen to grow in over similar timescales as before, up to 100 ps. **(d)** The TA kinetics of the PM6:Y6 blend following preferential excitation of Y6 at 800 nm. The decay of the Y6 $S_1$ shows the timescales of hole transfer in the blend, largely taking place <100 ps. The kinetic of the EA is also displayed, where it grows in over identical timescales to the blend after electron transfer, with a time constant of 15.7 ±0.5 ps obtained from a mono-exponential fit. The consistency in timescales between electron and hole transfer confirms that charge separation proceeds in the same manner for both, despite the latter taking place more slowly and with a much smaller frontier molecular orbital energetic offset.

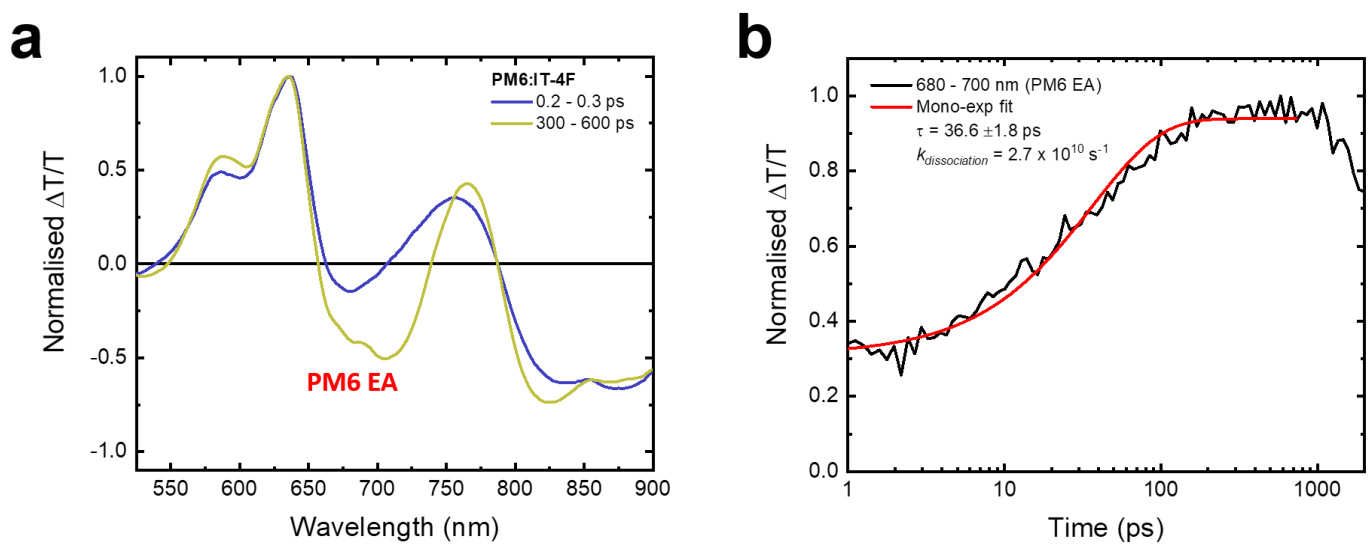

**Figure S34: (a)** The normalised TA spectra of a PM6:IT-4F film, pumped at 532 nm for preferential PM6 excitation with a fluence of 2.9 μJ cm⁻². At 0.2 ps, the PM6 and IT-4F GSBs are visible between 550 – 650 and 700 – 770 nm, respectively. By 300 – 600 ps, a new negative band has formed at the edge of the PM6 GSB at 680 nm. This feature is assigned to EA of PM6 (Fig. S10). **(b)** By tracking the kinetic from the EA region, we can visualise the separation of charge carriers, with the maximum intensity reached when they have fully separated. The EA peaks at 200 ps and can be well-described by a mono-exponential function with a time constant of 36.6 ±1.8 ps: this corresponds to a $k_{dissociation}$ =2.7x10¹⁰ s⁻¹.

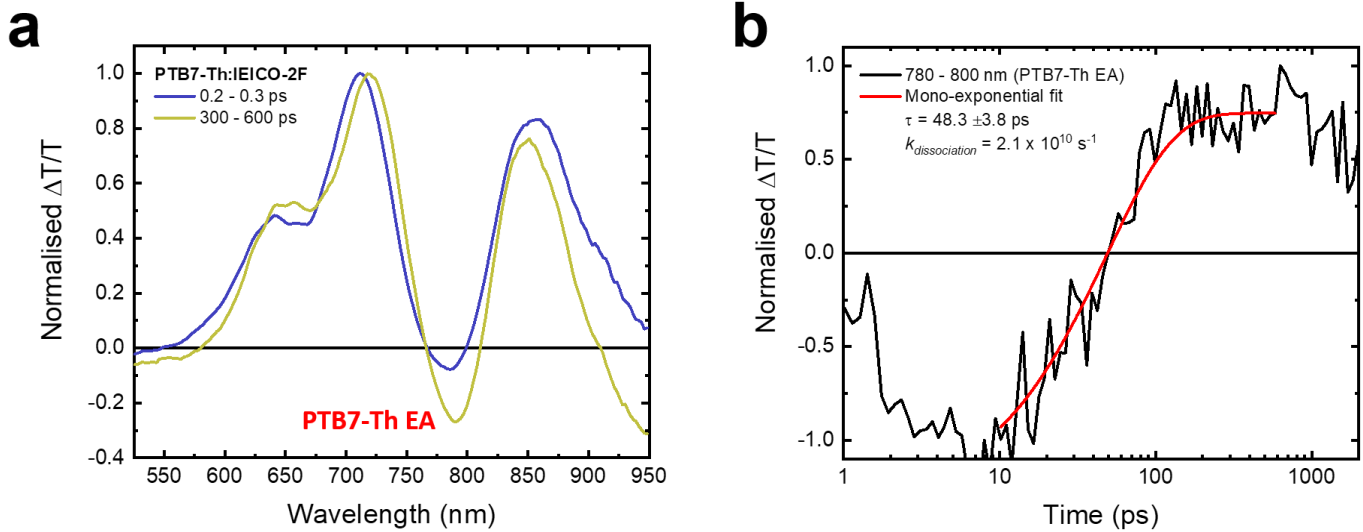

**Figure S35: (a)** The normalised TA spectra of a PTB7-Th:IEICO-2F film, pumped at 580 nm for preferential PTB7-Th excitation with a fluence of 2.1 μJ cm$^{-2}$. At 0.2 ps, the PTB7-Th and IEICO-2F GSBs are visible between 600 – 750 and 800 – 950 nm, respectively. By 300 – 600 ps, a new negative band has formed at the edge of the PTB7-Th GSB at 790 nm. This feature is assigned to EA of PTB7-Th. **(b)** By tracking the kinetic from the EA region, we can visualise the separation of charge carriers, with the maximum intensity reached when they have fully separated. The EA peaks at 150 ps and can be well-described by a mono-exponential function with a time constant of 48.3 ±3.8 ps: this corresponds to a $k_{dissociation}$ =2.1x10$^{10}$ s$^{-1}$.

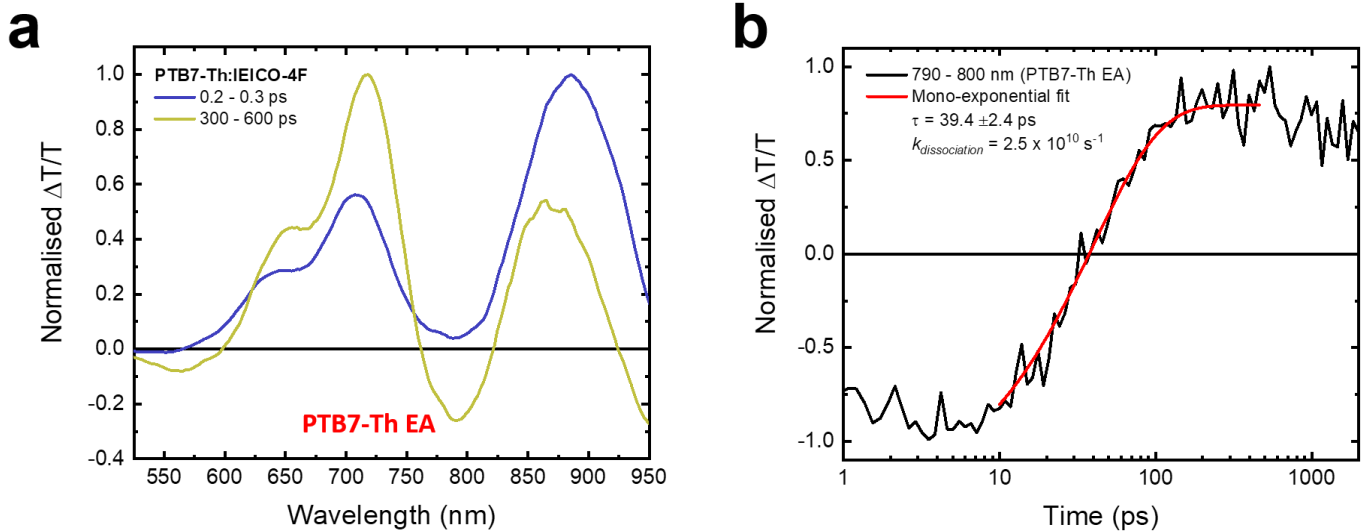

**Figure S36: (a)** The normalised TA spectra of a PTB7-Th:IEICO-4F film, pumped at 600 nm for preferential PTB7-Th excitation with a fluence of 0.5 μJ cm$^{-2}$. At 0.2 ps, the PTB7-Th and IEICO-4F GSBs are visible between 600 – 750 and 800 – 950 nm, respectively. By 300 – 600 ps, a new negative band has formed at the edge of the PTB7-Th GSB at 790 nm. This feature is assigned to EA of PTB7-Th. **(b)** By tracking the kinetic from the EA region, we can visualise the separation of charge carriers, with the maximum intensity reached when they have fully separated. The EA peaks at 150 ps and can be well-described by a mono-exponential function with a time constant of 39.4 ±2.4 ps: this corresponds to a $k_{dissociation}$ =2.5x10$^{10}$ s$^{-1}$.

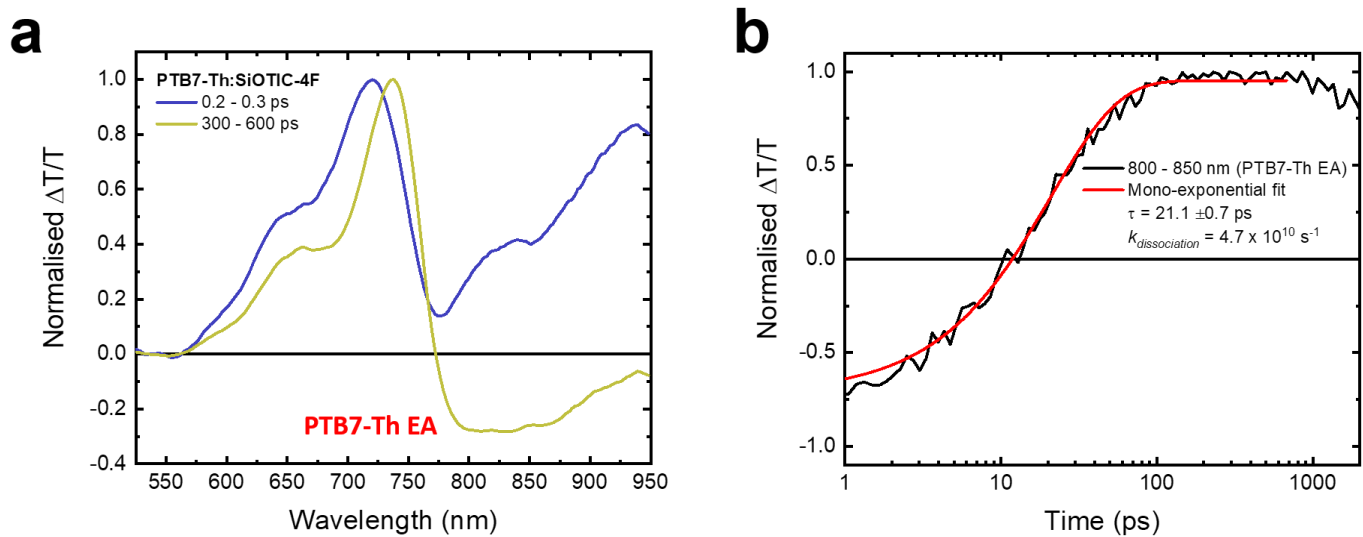

**Figure S37: (a)** The normalised TA spectra of a PTB7-Th:SiOTIC-4F film, pumped at 580 nm for preferential PTB7-Th excitation with a fluence of 2.1 µJ cm⁻². At 0.2 ps, the PTB7-Th and SiOTIC-4F GSBs are visible between 600 – 750 and 800 – 950 nm, respectively. By 300 – 600 ps, a new negative band has formed at the edge of the PTB7-Th GSB at 800 nm. This feature is assigned to EA of PTB7-Th. **(b)** By tracking the kinetic from the EA region, we can visualise the separation of charge carriers, with the maximum intensity reached when they have fully separated. The EA peaks at 100 ps and can be well-described by a mono-exponential function with a time constant of 21.1 ±0.7 ps: this corresponds to a $k_{dissociation}$ =4.7x10¹⁰ s⁻¹.

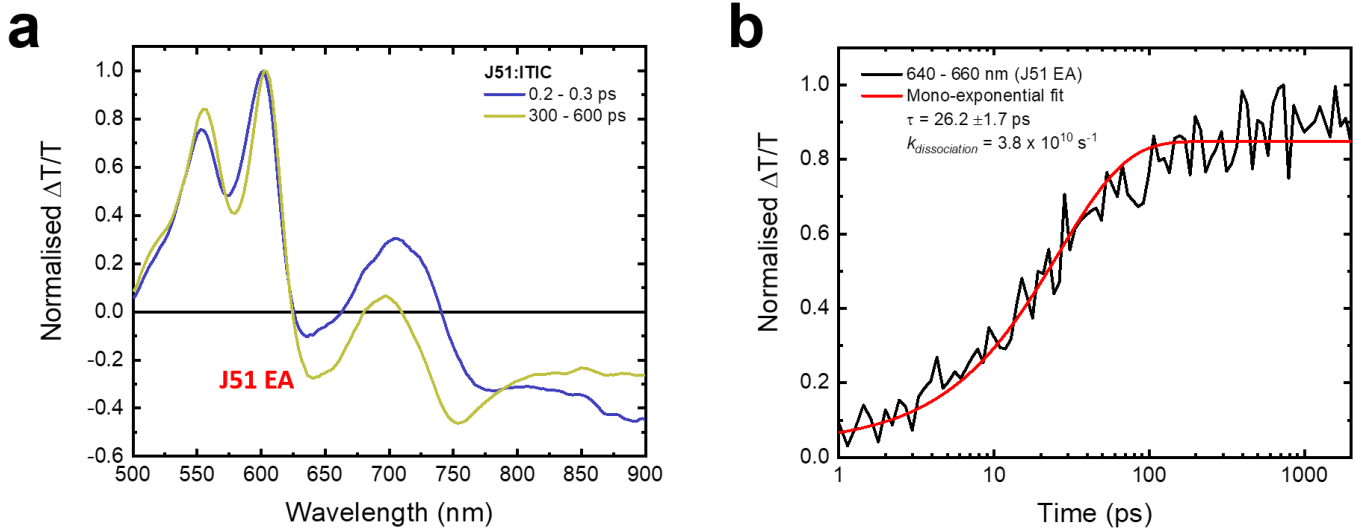

**Figure S38: (a)** The normalised TA spectra of a J51:ITIC film, pumped at 532 nm for preferential PM6 excitation with a fluence of 2.5 μJ cm⁻². At 0.2 ps, the J51 and ITIC GSBs are visible between 500 – 625 and 670 – 750 nm, respectively. By 300 – 600 ps, a new negative band has formed at the edge of the J51 GSB at 640 nm. This feature is assigned to EA of J51. **(b)** By tracking the kinetic from the EA region, we can visualise the separation of charge carriers, with the maximum intensity reached when they have fully separated. The EA peaks at 150 ps and can be well-described by a mono-exponential function with a time constant of 26.2 ±1.7 ps: this corresponds to a $k_{dissociation}$ =3.8x10¹⁰ s⁻¹.

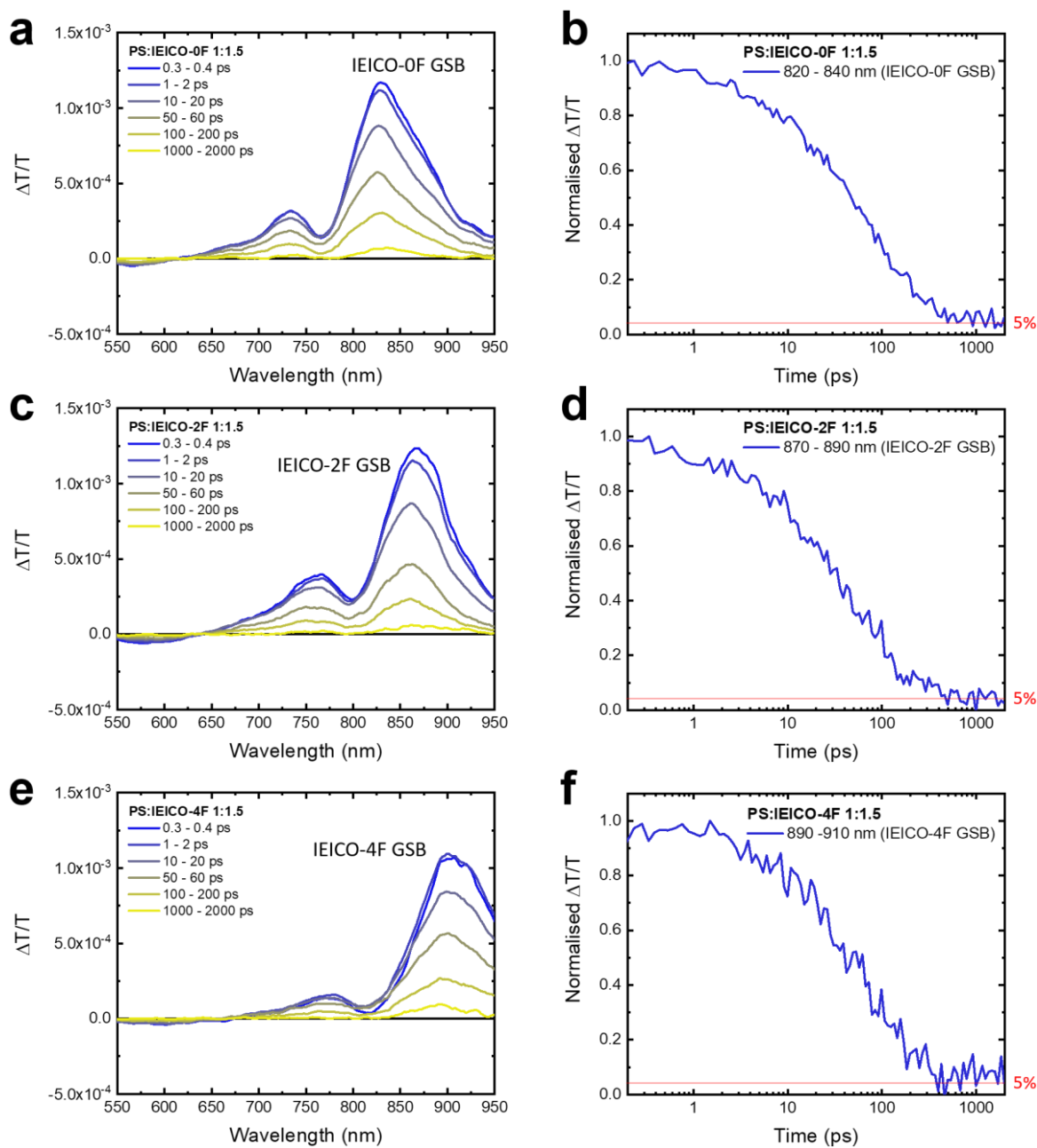

**Figure S39: (a, c, e)** The TA spectra of PS:IEICO-0F, -2F and -4F 1:1.5 films, pumped at 860, 890 and 925 nm with fluences of 0.74, 0.60 and 0.96 μJ cm$^{-2}$,respectively. Note how there is a small amount of the NFA GSB remaining at 2 ns, significantly longer than the S$_1$ lifetime of the materials. **(b, d, f)** TA kinetics of the GSB region of the IEICO derivatives. The remaining GSB intensity at 2 ns is ~5% of the peak in all materials. NIR region TA (Fig. S40) confirms that the only species present at this time are triplet excitons, formed via rapid ISC of the IEICO derivatives. Assuming singlet and triplet excitons are localised on one NFA molecule, the quantum efficiency of ISC is ~5% in all IEICO derivatives studied.

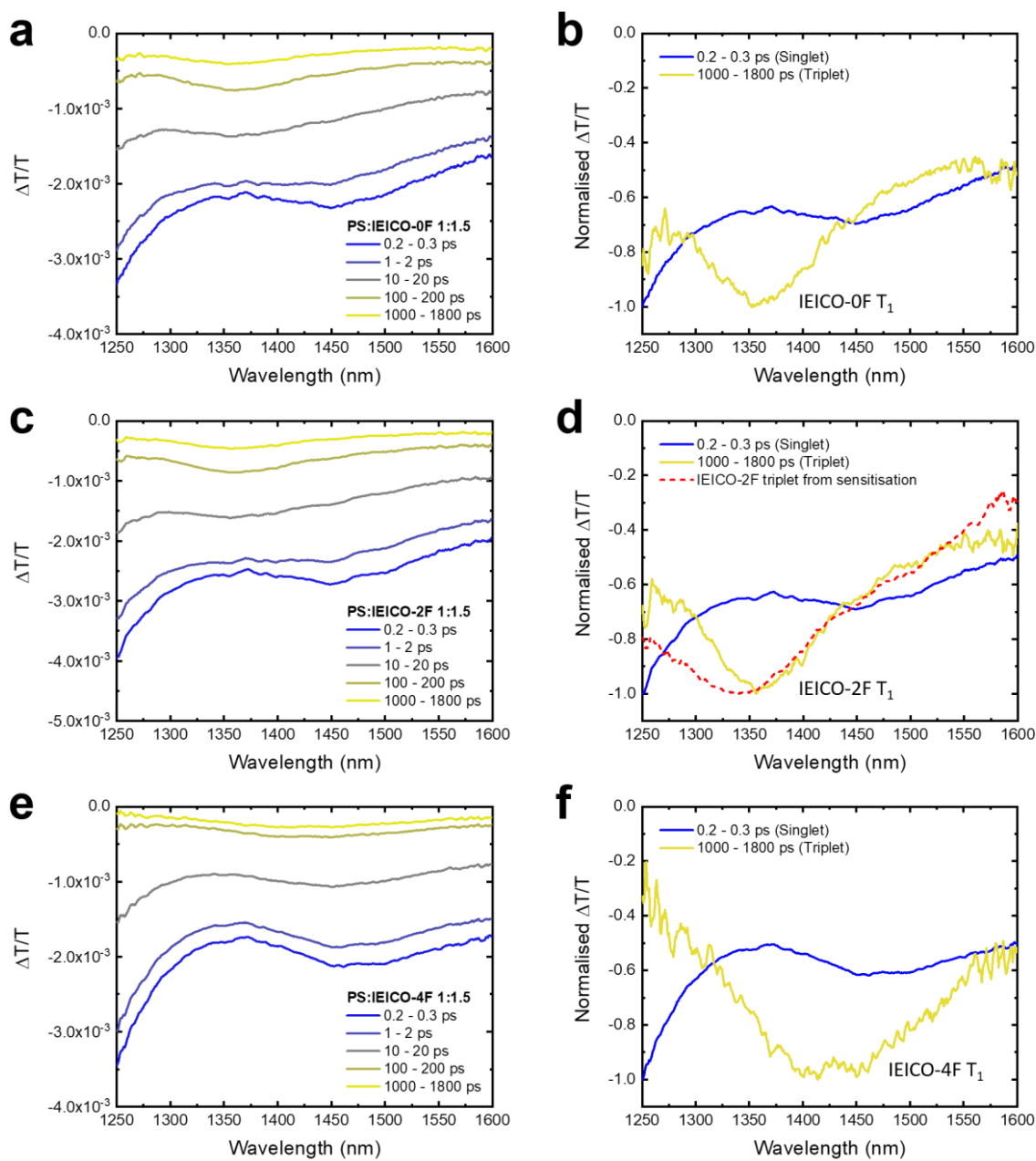

**Figure S40: (a, c, e)** The TA spectra of PS:IEICO-0F, -2F and -4F 1:1.5 films, all pumped at 800 nm with a fluence of 4.41 μJ cm⁻². After the $S_1$ PIA decays away, a new and long-lived PIA band remains that is assigned to the $T_1$ of the respective IEICO derivative. **(b, d, f)** Normalised TA spectra of the IEICO derivatives to more clearly show the $T_1$ PIA. A good match is found between the $T_1$ PIA of IEICO-2F from the sensitisation experiments and the PIA observed in the PS:IEICO-2F 1:1.5 blend, providing further evidence that this new PIA is the $T_1$ formed via rapid ISC. The discrepancy around 1250 – 1300 nm is likely due to a small amount of remaining IEICO-2F $S_1$ states in the sensitised blend; the $S_1$ lifetime will be enhanced in the PS:PtOEP:NFA 0.94:0.03:0.03 film as the high dilution of the NFA will reduce the non-radiative decay associated with aggregated molecules (concentration quenching).

# Quantifying $T_1$ formation from TA

When quantifying $T_1$ formation, previous studies have suggested that a kinetic model using triplet-charge annihilation as the $T_1$-quenching pathway provides the best description of the $T_1$ dynamics in OSC blends[10–12]. Though triplet-triplet annihilation has also been observed as a $T_1$-quenching route in fullerene OSCs[16,17], we see no increase in the charge population after $T_1$ loss in the PM6:Y6 blend (Fig. S42), which would be expected from the separation of $S_1$ states reformed by triplet-triplet annihilation[17]. Thus, we were able to successfully model the PM6:Y6 data here using just triplet-charge annihilation:

$$\frac{dN_T}{dt} = -\alpha \frac{dN_C}{dt} - \beta N_T N_C \qquad (1)$$

where $N_T$ and $N_C$ are the experimental $T_1$ and charge population densities, $\alpha$ is the fraction of recombination events that lead to $T_1$ formation, and $\beta$ is the triplet-charge annihilation rate constant. In order to quantify the number of $T_1$ and charges present in the blend, we must first determine the absorption cross section ($\sigma$) of the $T_1$ and charge species. The measured change in transmission $\frac{\Delta T}{T}$ in the TA experiment is related to the total population ($N$) by the following relation:

$$\frac{\Delta T}{T} = \sigma N \qquad (2)$$

To begin, we will calculate $\sigma$ of the charges ($\sigma_C$). In order to accurately track the population, we need to calculate $\sigma_C$ at a wavelength where there is significant absorption by the charges, but no overlap with the Y6 GSB (Fig. S6a), or any other species. Therefore, we choose to use the signal at 930 nm, as it is free from other overlapping signals that could affect the accuracy of our modelling. In the very low-fluence TA measurement of the PM6:Y6 blend (Fig. S12a), we have determined that hole transfer following selective excitation of Y6 is completed by 100 ps: at this time in the PM6:Y6 blend, $\frac{\Delta T}{T}$ = -6.40 x $10^{-4}$ at 930 nm. As we know the absorbance ($A$) of the film (Fig. S43), we can evaluate the initial number of singlet excitons created on Y6 following excitation. Using $A$ = 0.49 at 800 nm, we determine that 1.31 x $10^{12}$ singlet excitons have been created on Y6 by the 800 nm, 0.5 µJ cm$^{-2}$ pump pulse (2.4 nJ per pulse). As we know the number of excited states created, we can now estimate the number of charges generated. If we were to assume the quantum efficiency of charge transfer from Y6 $S_1$ ($\eta_{CT}$) = 100%, the number of $S_1$ initially created (1.31 x $10^{12}$) is equal to the charge population at 100 ps. We would then obtain $\sigma_C$ = 4.90 x $10^{-16}$ cm$^{-2}$ from equation 2 at 930 nm.

However, it is almost certain that $\eta_{CT} \neq$ 100%. To better evaluate the true value of $\eta_{CT}$, we begin by acknowledging that the photovoltaic internal quantum efficiency (IQE$_{PV}$) of optimised PM6:Y6 devices is ~90%[18,19]: this means that under operating conditions, 10% of the initially generated Y6 $S_1$ do not create charge carriers that are successfully extracted from the device. In order to estimate the potential loss pathways, we note that PM6:Y6 devices show exceptionally efficient carrier extraction[18]. Therefore, we expect the main loss pathway to be in the creation of charges, not the extraction. This can be rationalised by noticing that the slow hole transfer rate from Y6 to PM6 (Fig. S12) will have to compete against the relatively rapid $S_1$ decay of Y6 (Fig. S6), creating a plausible route for the decay of $S_1$. Therefore, as a conservative estimate, we consider that $\eta_{CT}$ ~90% for Y6 excitons. To confirm the validity of this assumption, we also compare the timescales for hole transfer and Y6 $S_1$ decay. The time taken for the population of Y6 $S_1$ to fall to 1/e of its initial value in the PS:Y6 film is ~100 ps. In contrast, the time for the Y6 $S_1$ to be quenched to 1/e of their initial population in the blend is ~10 ps (Fig. S44). The ratio of these two lifetimes also gives $\eta_{CT}$ ~90%, consistent with the value estimated from the IQE$_{PV}$. Because of this, our original value of $\sigma_C$ = 4.90 x $10^{-16}$ cm$^{-2}$ will be an underestimate as not every $S_1$ is dissociated; less charges than expected are leading to the observed signal. Therefore, to account for potential losses during charge generation, we divide this value by 0.9 to obtain our final $\sigma_C$ = 5.44 x $10^{-16}$ cm$^{-2}$.

Next, we must calculate $\sigma$ of the Y6 $T_1$ ($\sigma_T$). We note that a very small fraction of Y6 $S_1$ undergo intersystem crossing (ISC) to $T_1$ prior to decay. This provides us a convenient means to calculate $\sigma_T$ without having to rely on sensitisation experiments, where the fraction of $T_1$ that successfully energy transfer from a sensitizer to the target molecule can be difficult to quantify. The absorption spectrum of the PS:Y6 1:1.2 investigated is given in Fig. S45: this film was excited at 800 nm with a high fluence of 16.1 µJ cm$^{-2}$ (32.2 nJ per pulse) to create a significant population of $S_1$ from which ISC can potentially occur. As $A$ = 0.57 at 800 nm, the number of $S_1$ generated is 5.57 x $10^{13}$. By 1.8 ns, all $S_1$ states will have decayed, therefore any remaining population at this time will be solely $T_1$. This is confirmed by the absence of the $S_1$ PIA at 1550 nm and the presence of only the Y6 $T_1$ PIA at 1450 nm. Through comparing the relative intensity of the remaining Y6 GSB at 1.8 ns to the initial value, we determine that ~3% of the Y6 $S_1$ have undergone ISC to $T_1$. Therefore, the $T_1$ population at 1.8 ns is 1.67 x $10^{12}$. As $\frac{\Delta T}{T}$ = -5.70 x $10^{-4}$ at the peak of the $T_1$ PIA at 1450 nm at 1.8 ns, we obtain $\sigma_T$ = 3.42 x $10^{-16}$ cm$^{-2}$ at 1450 nm from equation 2.

The values of $\sigma_C$ and $\sigma_T$ calculated were then used to determine $N_T$ and $N_C$, using a film thickness of 90 nm for the optimised PM6:Y6 blend. We then input these values into equation 1 and fitted the TA kinetics globally over four different fluences (Fig S41). We note that kinetics following selective excitation of Y6 at 800 nm were used to simplify the charge transfer dynamics; as hole transfer from Y6 to PM6 is completed by 100 ps (Fig. S12), the data is analysed from this time onwards. From this, values of $\alpha$ = 0.91±0.03 and $\beta$ = 8.03±0.20 x $10^{-9}$ cm$^3$ s$^{-1}$ were obtained, confirming that ~90% of charges recombine via the Y6 $T_1$.

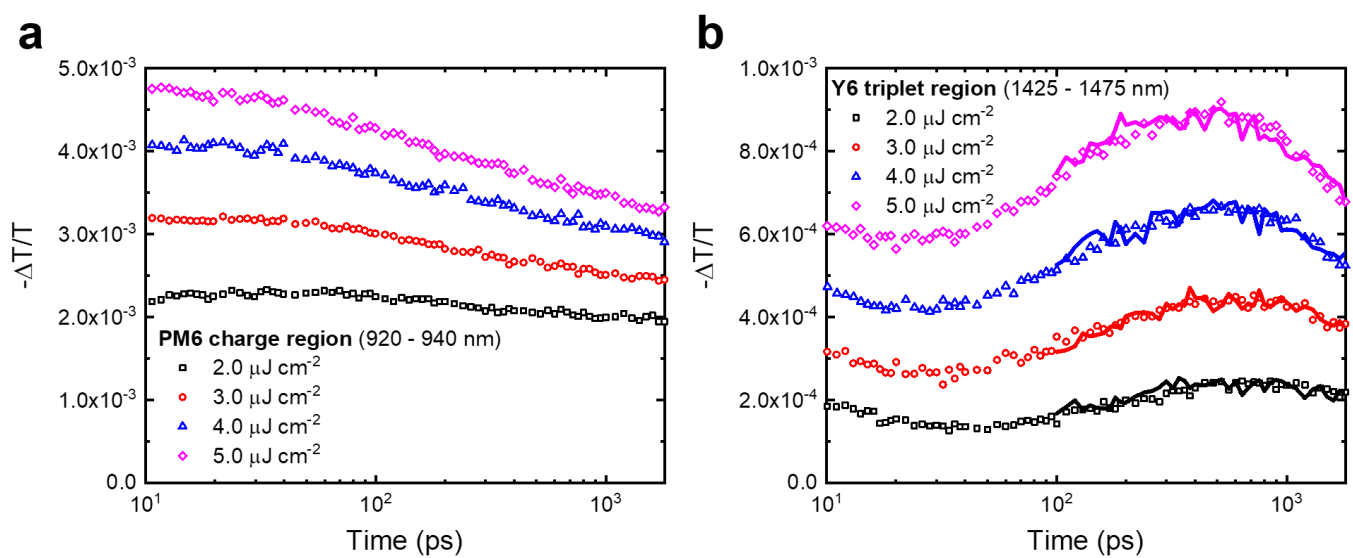

**Figure 41: (a)** The TA kinetics of the PM6:Y6 blend following selective excitation of Y6 at 800 nm at 293 K, taken around the maximum of the PM6 charge PIA between 920 – 940 nm. The PIA intensity, and therefore population, of the PM6 charges can be seen to decrease between 10 – 1800 ps. The loss of charges is significantly faster at higher excitation fluences, indicating non-geminate processes are responsible. This data was used as an input to the model described in equation 2. **(b)** The TA kinetics of the PM6:Y6 blend following selective excitation of Y6 at 800 nm at 293 K, taken around the maximum of the Y6 $T_1$ PIA between 1425 – 1475 nm. The rate and timescales over which charges are lost is clearly correlated with the increase of the Y6 $T_1$ population, indicating that the processes are related. The loss of Y6 $T_1$ population on timescales of 100's ps is due to the rapid triplet-charge annihilation occurring. The solid lines are global fits to the data using the model described in equation 2. As hole transfer from Y6 to PM6 is not completed until 100 ps (Fig. S12), as determined by an extremely low fluence measurement free from non-geminate recombination during the experimental time window, the data is only fitted for times >100 ps. Excellent agreement between the model and experimental data is obtained, revealing that ~90% of charges decay into $T_1$ on Y6.

For PTB7-Th:IEICO-2F, we have shown that $T_1$ formation is not a measurable loss pathway. However, it is important to be able to put an upper bound on the fraction of excited states that could recombine via the IEICO-2F $T_1$ without being detected. With this in mind, we note that the smallest signal we can reliably detect in our TA measurements is $\frac{\Delta T}{T}$ = 1 x 10$^{-5}$ at 1350 nm (Fig. S47c). Therefore, in order to not be observed, the absorption by $T_1$ states on IEICO-2F must result in a signal lower than this baseline value. We begin by determining $\sigma_T$ for IEICO-2F. Using $A$ = 0.32 at 800 nm for our PS:IEICO-2F film (Fig. S47b) and an excitation fluence of 4.41 µJ cm$^{-2}$ (17.0 nJ per pulse), we calculate the initial $S_1$ population after excitation to be 9.26 x 10$^{12}$. Assuming the $T_1$ yield is 5% (Fig. S39d), the number of $T_1$ states present at 1.8 ns after all $S_1$ have decayed is 4.63 x 10$^{11}$. From this population and a signal intensity of 4.1 x 10$^{-4}$ at 1.8 ns in the NIR region TA of PS:IEICO-2F (Fig. S47d), we calculate $\sigma_T$ = 8.86 x 10$^{16}$ cm$^{-2}$ for IEICO-2F (equation 2). Using this $\sigma_T$, the number of IEICO-2F $T_1$ states that would give a signal of 1 x 10$^{-5}$ is 1.13 x 10$^{10}$. Turning now to the PTB7-Th:IEICO-2F blend; after excitation at 620 nm with a fluence of 3.80 µJ cm$^{-2}$ (7.6 nJ per pulse), we calculate that 5.52 x 10$^{12}$ $S_1$ states are formed, using $A$ = 0.46 at 620 nm (Fig. S47a). From the ratio of the number of $S_1$ states formed (7.79 x 10$^{12}$) and the smallest number of IEICO-2F $T_1$ detectable (1.13 x 10$^{10}$), we determine that for $T_1$ formation not to be observed, the fraction of recombination proceeding via the IEICO-2F $T_1$ must be less than 0.15%.

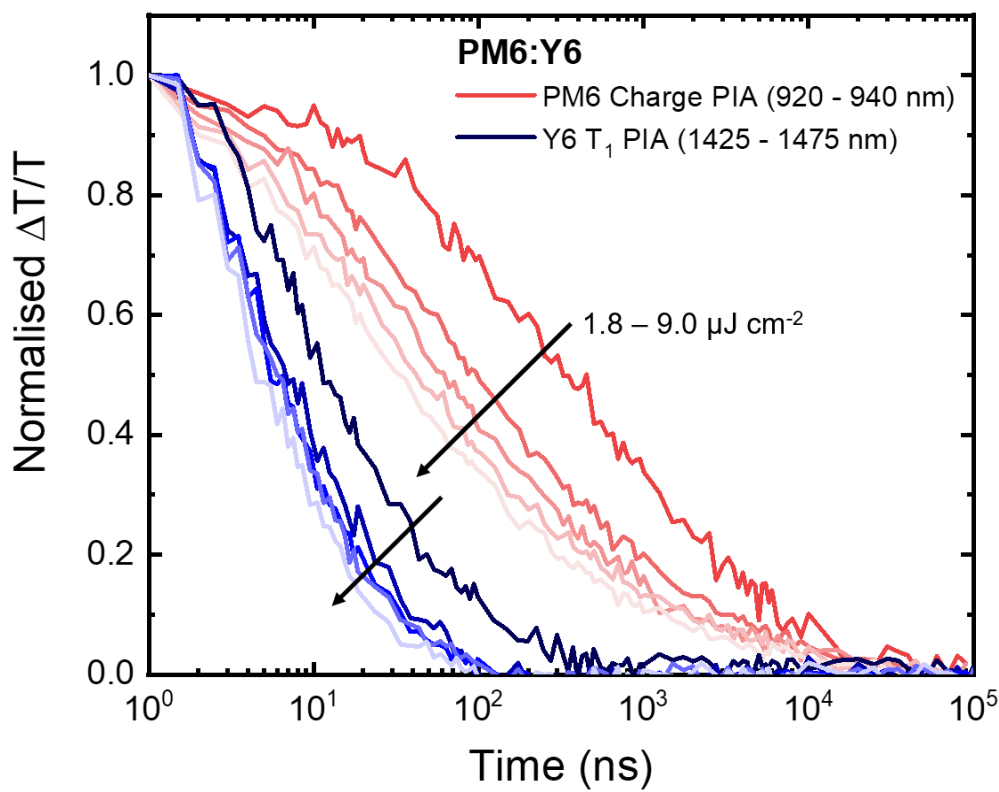

**Figure S42:** The ns-TA kinetics taken from the PM6 charge (red lines, 920 – 940 nm) and Y6 $T_1$ (blue lines, 1425 – 1475 nm) PIAs of a PM6:Y6 film, pumped at 532 nm for preferential PM6 excitation. A fluence series was performed, with fluences of 1.8, 3.6, 5.4, 7.2 and 9.0 μJ cm⁻² used. We note that if a significant amount of triplet-triplet annihilation (TTA) was occurring, we would expect an increase in the number of charges over the timescales of $T_1$ quenching[17]. This is because TTA forms one $S_1$ state from two $T_1$ states, with the $S_1$ able to undergo charge transfer again, increasing the charge population. However, we clearly notice no increase in the PM6 charge PIA intensity over the timescales of 1 – 100 ns when $T_1$ quenching is taking place. From this, we conclude that the primary $T_1$ quenching route in our PM6:Y6 blend is via triplet-charge annihilation (TCA). Therefore, in order to avoid over-parameterisation, we introduce TCA as the only $T_1$ quenching pathway in our modelling of the $T_1$-charge dynamics. The validity of only including TCA is confirmed by the excellent agreement between the modelling and experimental data.

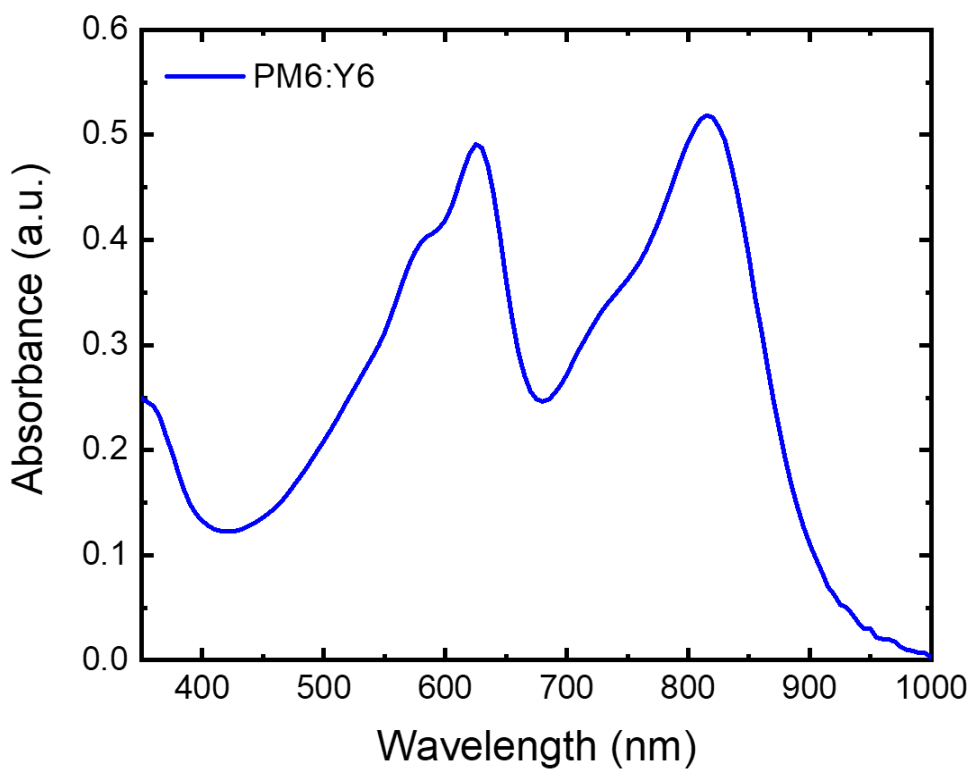

**Figure S43:** The absorbance spectrum of the PM6:Y6 film used in the TA measurements. This film was fabricated in an identical fashion to the optimised devices. The absorbance of the film at 800 nm is 0.49 and the thickness is 90 nm.

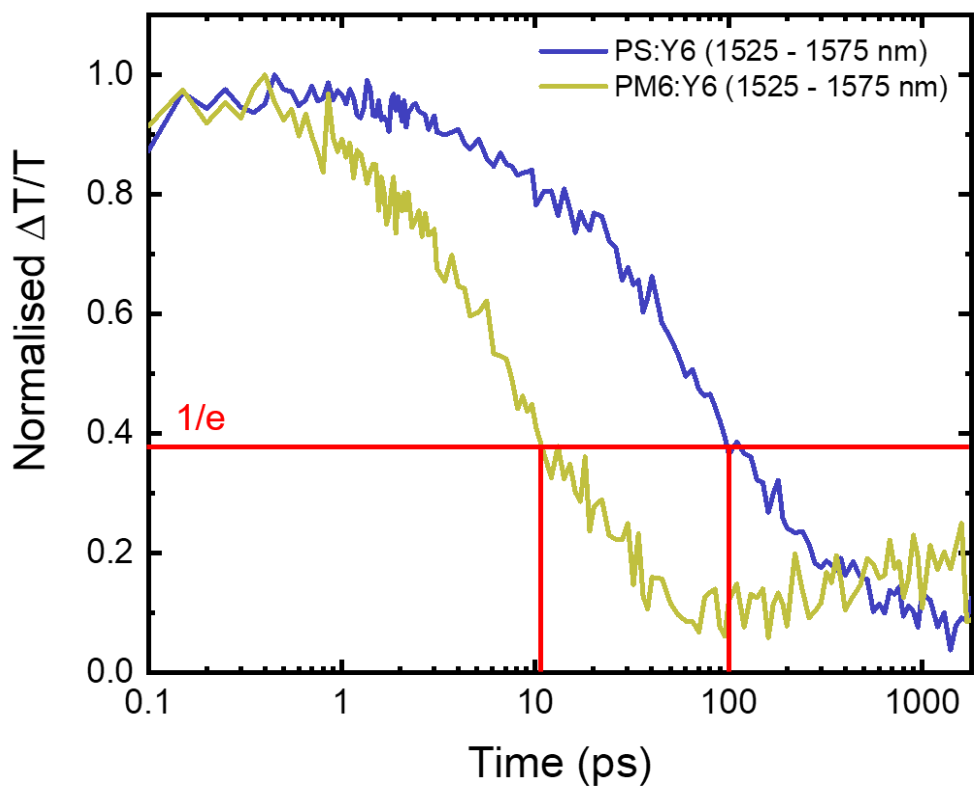

**Figure S44:** The TA kinetics of the Y6 $S_1$ PIA region (1525 – 1575 nm) in a PS:Y6 1:1.2 and PM6:Y6 film, excited at 800 nm with fluences of 1.8 and 1.0 µJ cm$^{-2}$, respectively. The time taken for the magnitude of the Y6 $S_1$ PIA to fall to 1/e in the PS:Y6 film is ~100 ps, whilst the time taken in the PM6:Y6 films is ~10 ps. From the ratio of these two lifetimes, $\eta_{CT}$ is estimated to be ~90%. The slight rise in the blend signal after 100 ps is due to the kinetic also capturing the edge of the Y6 $T_1$ PIA at 1450 nm.

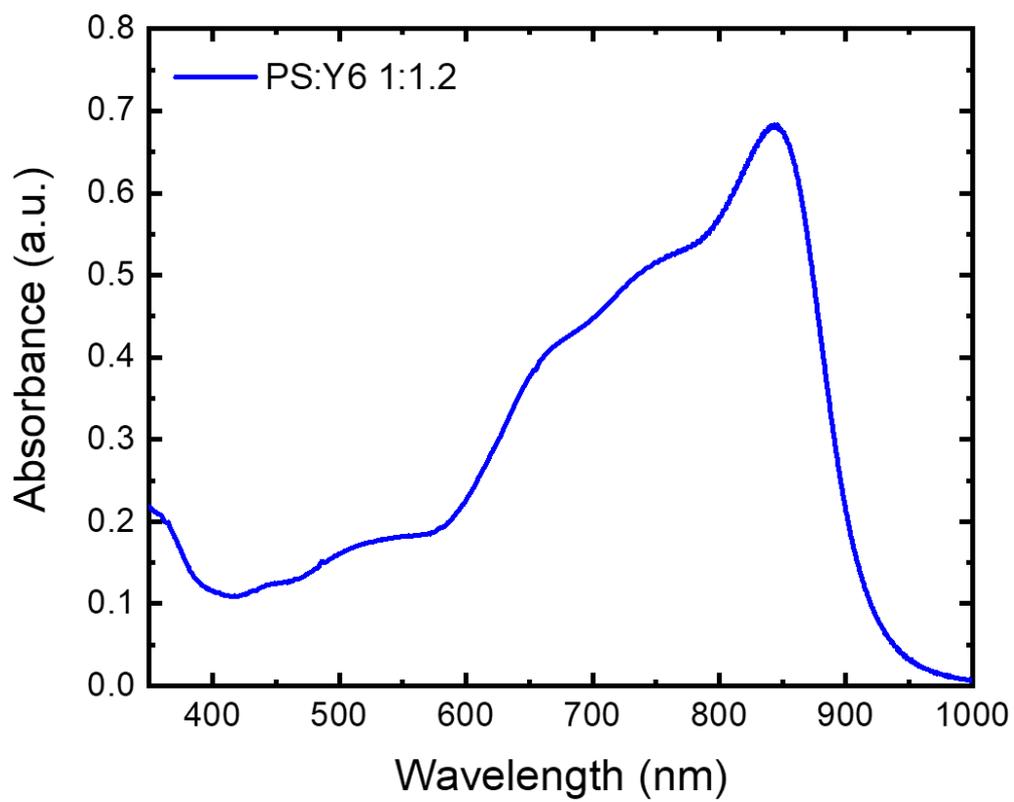

**Figure S45:** The absorbance spectrum of the PS:Y6 1:1.2 film used in the TA measurements. The absorbance of the film at 800 nm is 0.57.

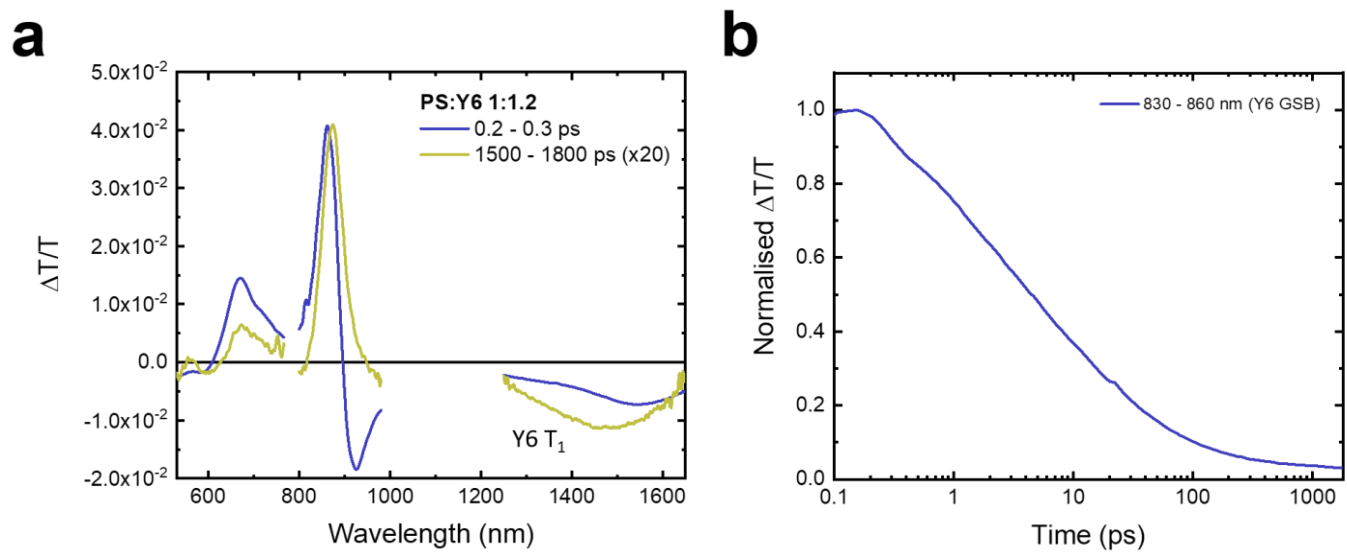

**Figure S46: (a)** The TA spectra of a PS:Y6 1:1.2 film, excited at 800 nm with a high fluence of 16.1 μJ cm$^{-2}$ (pulse energy = 80 nJ). At 0.2 ps, the Y6 GSB is visible between 600 – 900 nm, with two distinct vibronic peaks. There are two Y6 $S_1$ PIAs in the NIR region: one sharp peak adjacent to the Y6 GSB at 910 nm and a weaker, broad feature peaked at 1550 nm. By 1.8 ns (scaled by a factor of 20 for clarity), the Y6 $S_1$ PIA has fully decayed, leaving behind only the Y6 $T_1$ PIA at 1450 nm. Therefore, we assume that any remaining GSB can be solely attributed to Y6 molecules bleached by triplet excitons. This allows us to determine the population of Y6 excited states remaining in the film and therefore $\sigma_T$. The intensity of the Y6 $T_1$ at 1.8 ns is -5.70 x 10$^{-4}$ at 1450 nm. **(b)** The kinetic of the Y6 GSB region. As the sharp Y6 $S_1$ PIA band at 920 nm overlaps with the peak of the GSB at 870 nm, we have analysed the GSB between 830 – 860 nm to provide a better estimate of the fraction of Y6 excited states remaining at 1.8 ns. We find this value to be ~3% from the ratio of the maximum GSB signal intensity just after excitation.

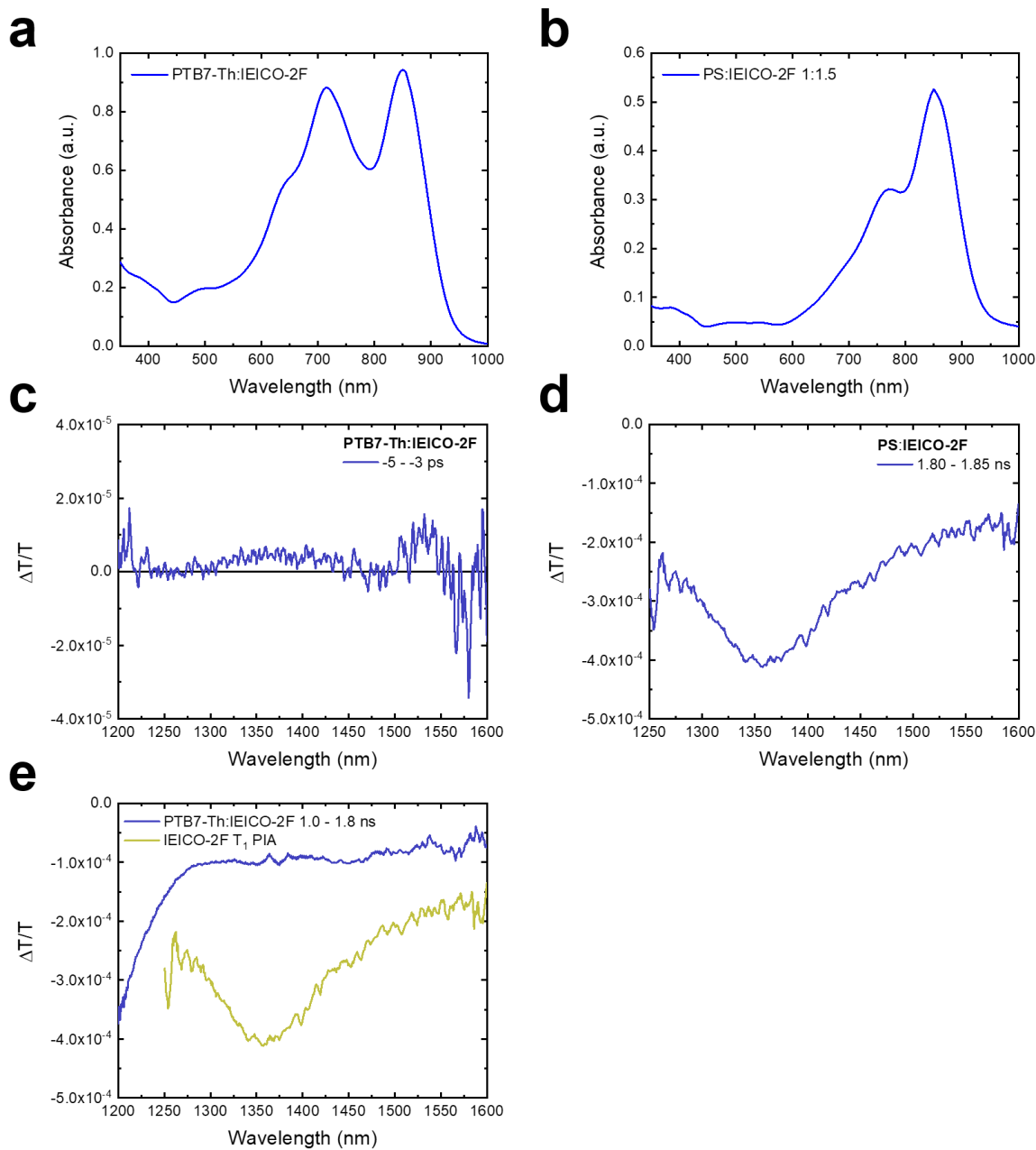

**Figure S47: (a)** The absorbance spectrum of the PTB7-Th:IEICO-2F film used in the TA measurements. This film was fabricated in an identical fashion to the optimised devices. The absorbance of the film at 620 nm is 0.27 and the thickness is 90 nm. **(b)** The absorbance spectrum of the PS-Th:IEICO-2F 1:1.5 film used in the TA measurements. The absorbance of the film at 800 nm is 0.32. **(c)** The baseline averaged between -5 and -3 ps for the measurement of the PTB7-Th:IEICO-2F blend that is displayed in the main text (Figure 2d). From this, we assess that the smallest signal reliably resolvable around 1350 nm is 1 x $10^{-5}$. **(d)** The TA spectra at 1.80 – 1.85 ns of a PS:IEICO-2F 1:1.5 film, pumped at 800 nm with a fluence of 4.41 μJ $cm^{-2}$ (pulse energy = 17 nJ). The remaining signal is attributed to the IEICO-

2F $T_1$ formed after ISC. From this, we calculate $\sigma_T$ = 8.86 x $10^{16}$ cm$^{-2}$ for IEICO-2F. **(e)** The TA spectra of PTB7-Th:IEICO-2F after excitation at 620 nm with a fluence of 3.80 μJ cm$^{-2}$. At 1.0 – 1.8 ns, it is clear that there is no observable PIA band at 1350 nm that would correspond to the IEICO-2F $T_1$. The IEICO-2F $T_1$ PIA, taken from Fig. S47d, is overlaid for clarity.

# Theory of triplet excited states studied by trEPR spectroscopy

Triplet states consist of two strongly coupled unpaired electrons that can be described by a spin Hamiltonian comprising the exchange term and the dipole-dipole term[20–22]. The exchange term depends exponentially on the distance of the two unpaired electrons, while the dipole-dipole term follows an inverse cubed dependence[20]. Since molecular triplet states are characterized by a short distance between the two unpaired electrons, the exchange term usually overwhelms the dipolar term and completely separates the energy levels of the triplet and singlet states[20]. As a result, the triplet sublevels of molecular triplet states are not "mixed" with the singlet level and therefore the exchange term is usually neglected in the simulation of molecular triplet states. The dipole-dipole term energetically splits the three sublevels of a triplet state even in the absence of an externally-applied magnetic field: it is also referred to as the Zero-Field Splitting (ZFS) interaction[20,21]. The eigenvalues of the ZFS Hamiltonian (X, Y and Z) are commonly expressed in terms of the ZFS parameters D and E that are defined as D = -3/2Z and E = 1/2(Y-X)[21]. The D parameter defines the strength of the dipolar coupling and is directly related to the delocalization of the triplet state, whilst the E parameter represents the deviation of the triplet delocalization from axial symmetry. It is important to note that the EPR line position depends only on the relative sign of D and E and therefore the absolute sign is often unknown.

Standard trEPR spectroscopy is carried out under the presence of an external magnetic field (about 340 mT). Therefore, to simulate the trEPR spectra of triplet states, both the electronic Zeeman and the dipole-dipole terms should be taken into account:

$$H = \mu_B \boldsymbol{B_0} g \boldsymbol{S} + \boldsymbol{SDS} \tag{3}$$

where $\boldsymbol{g}$ is the Zeeman g-tensor and $\boldsymbol{D}$ is the ZFS tensor, $\boldsymbol{B_0}$ is the external magnetic field vector, $\boldsymbol{S}$ is the spin operator, and $\mu_B$ is the Bohr magneton. In the high-field approximation, the spin sub-levels of the triplet state are commonly referred to as $T_+$, $T_0$ and $T_-$ and their eigenvalues depend on the relative strength between the Zeeman and ZFS interactions and the direction of the magnetic field. For every molecular orientation, there are two allowed transitions between the three triplet sublevels ($\Delta m_s = \pm 1$) that correspond to two peaks in the trEPR spectrum. Their magnetic field position is determined by the eigenvalues of the spin Hamiltonian, while their intensity is determined by the spin-polarization mechanism, as discussed below. In a disordered material, such as the organic layers studied in this work, the full trEPR spectrum can be calculated as the convolution of the contributions from all the randomly oriented molecules in the film. This is commonly referred to as a *powder spectrum*.

An interesting aspect of trEPR spectroscopy of triplet states is that although the time resolution is typically a few hundred ns, trEPR spectra allow us to obtain information about the photophysical processes that led to the creation of the triplet states observed, even if they occurred faster than the intrinsic experimental time resolution; this is due to the phenomenon of spin-polarization[23]. Spin-polarization occurs because the triplet states that are generated after a short laser pulse are far from thermal equilibrium and therefore the corresponding trEPR spectra show signals of enhanced absorption (*a*) and emission (*e*). This spin polarization pattern allows us to determine whether the triplet has been generated by an intersystem crossing (ISC) mechanism or a geminate back-charge transfer (BCT) process[23].

ISC from $S_1$ to $T_1$ is promoted by the spin-orbit interaction and is characterized by a strong anisotropy of the populating rates of the three triplet sublevels ($m_s$ = -1,0,+1). ISC triplets can have several different spin polarization patterns, namely *aaaeee*, *eeeaaa*, *eeaeaa* and *aaeaee*[23]. Geminate BCT can be understood in the framework of the "spin correlated radical pair (SCRP) mechanism"[24–27]. The standard geminate recombination pathway starts from the $^1$CT state, which in EPR spectroscopy is termed a SCRP due to the strong magnetic interactions between the two unpaired spins of the CT state. Distinct from localized molecular excited states, the $^1$CT$_0$ and $^3$CT$_0$ spin sublevels of a SCRP are "mixed" together because of hyperfine and electron Zeeman interactions. Thus, two distinct pathways can lead to the presence of BCT triplet polarisation patterns:

(i)  The SCRP $^3$CT$_0$ sublevel formed by mixing with the $^1$CT$_0$ undergoes a spin-allowed BCT to an energetically low-lying molecular triplet $T_0$ state, generating an excess spin population in the $T_0$. This results in a spin polarisation pattern of *aeeaae* (D<0) or *eaaeea* (D>0) for the triplet exciton[23].

(ii)  The mixing of the $^1$CT$_0$ and $^3$CT$_0$ SCRP sublevels opens a spin-allowed recombination pathway for $^3$CT$_0$ to the $S_0$ ground state via $^1$CT$_0$. This process results in an excess population remaining in the $^3$CT$_+$ and $^3$CT$_-$, which undergo a spin-allowed BCT to the molecular triplet $T_+$ and $T_-$ sublevels, generating an excess spin population in the $T_+$ and $T_-$. This results in a spin polarisation pattern of *aeeaae* (D>0) or *eaaeea* (D<0) for the triplet exciton[23].

As a matter of clarity, molecular triplet states generated by non-geminate BCT cannot be detected by trEPR spectroscopy due to the lack of spin-polarization; the spin-statistical recombination of uncorrelated FCs results in an equal population of the $^1CT_0$, $^3CT_+$, $^3CT_0$ and $^3CT_-$. As a result, trEPR spectroscopy is well-suited as a complementary technique to the TA experiments.

## The impact of temperature on the trEPR results

It is worth making clear that standard trEPR experiments of organic semiconducting films are commonly performed at low temperatures (~80 K) to slow down the spin relaxation processes, which results in improved signal-to-noise. Therefore, it is important to consider whether the low temperature may affect the photo-generation dynamics. In the systems studied, ISC triplets are formed from singlet excitons that don't reach the D/A interface for charge transfer. In relation to this, we note that the low temperatures used will reduce the non-radiative vibrational relaxation of singlet excitons[28], allowing more time for ISC to occur before exciton decay; this provides a route for enhanced ISC triplet yields at the low temperatures in the trEPR measurements. However, at room temperature where vibrational decay can occur rapidly, we believe that the presence of a significant amount of ISC triplets is unlikely[29], as most un-dissociated singlets will simply decay directly to the ground state. In contrast, the formation of geminate BCT triplets at room temperature in fullerene acceptor (FA) OSCs is not expected to be significantly different to 80 K. This is because it has been demonstrated that the initial rapid (<1 ps) charge separation process of the geminate CT state takes place coherently via a wave-like propagation of the electron through the band-like states of fullerene aggregates[2]. As this process is largely independent of temperature[2,30], the charge separation at 80 K will resemble that at room temperature. For NFA OSCs where the charge separation can take up to 100 ps[1,31,32], it is reasonable to assume this takes place via thermalized CT states[14,15]. Therefore, charge separation likely occurs via a thermally-assisted hopping mechanism[33], though we note little dependence on the efficiency of charge separation with temperature was reported in the PM6:Y6 blend[19]. In light of this we consider that, if anything, BCT triplet formation is more likely to take place at low temperature due to the slower charge separation timescales providing more opportunity for the mixing of $^1CT_0$ and $^3CT_0$; this typically takes place on ~ns timescales[34,35]. Therefore, if BCT triplets are not present at 80 K in the NFA blends, they are exceedingly unlikely to then be present at room temperature.

## trEPR results

We carried out trEPR spectroscopy to investigate in detail the structure and the dynamics of the excited triplet states in the FA and NFA blends, as well as the neat materials. The trEPR spectra acquired at two or three different delay times after the laser pulse are reported in Figures S48 – S67, together with the best-fit spectral simulations. In Table S2, we summarize the main results obtained from the spectral simulations. In the following, we divide the analysis and the discussion of the trEPR spectra in three sections: (1) neat donor and acceptor films, (2) NFA blends and (3) FA blends.

### Neat donor and acceptor films

The trEPR spectra of the neat donor and acceptor films are characterized by the presence of up to two different signals (one spectrally-narrow and the other much broader) with a different time evolution. The narrow bandwidth EPR signal extends for a few Gauss (~10 G) and possesses a g-value and a polarization pattern (either in enhanced absorption or emission) which are typical of free charges generated upon photon absorption[36]. The detection of free charges highlights the capability of the neat materials to generate FC despite the absence of an electron-donor/acceptor counterpart[37]. This can be rationalized by the presence of electron-withdrawing and -acceptor units within the polymer chain which favour the photo-induced charge transfer process (either intra- or inter-molecular)[37] and the subsequent charge separation of the CT state. In line with literature results[38], the EPR signal of these photo-generated charges decays very rapidly (a few µs) due to the rapid charge recombination process occurring, even at low temperatures (80 K).

The broad bandwidth EPR signal can be attributed to a localized excited triplet excitons[23,39–41]. To confirm our hypothesis, we carried out the spectral simulations for all the studied samples that exhibited substantial triplet formation. Unfortunately, the triplet signal in ITIC and IT-4F was too weak for successful simulation. The obtained spectroscopic parameters are summarised in Table S2. From the simulations, we obtained important information on the zero-field splitting (ZFS) parameters and the non-equilibrium populations of triplet sublevels (spin polarization). These ZFS parameters are directly related to the delocalization and the symmetry of the triplet states[42]. The obtained ZFS parameters suggest that the observed triplet states are delocalized over few monomeric units (~1-2), in-line with other photovoltaic polymers in literature[20]. The spin polarization is related to the triplet populating mechanism, which in our neat films is always ISC from $S_1$ to $T_1$[42]. The good ISC yields in the neat polymer films is due to the longer lifetimes of the singlet excitons at low temperature, resulting from decreases rates of non-radiative vibrational decay[28].

It is worth mentioning that two polymers (PM6 and PBDB-T) show a slightly different behaviour. In both polymers, a third signal with different spectral features and time evolution is observed. In PM6, from the spectral simulations, we conclude that this signal can be attributed to a second triplet state generated via ISC. Furthermore, the ZFS parameters of triplet 2 (Table S2) suggest it is more delocalized than triplet 1, which can be rationalized with the presence of a hybrid locally excited charge-transfer (HLCT) triplet state[43]. In PBDB-T, the third signal presents a polarization pattern typical of geminate BCT triplet states. The presence of a BCT triplet is in line with the good photo-generation efficiency of the film which is confirmed by the strong EPR signal of photo-generated charges in the EPR spectrum.

Finally, we noticed that many of the SOC-ISC triplets detected via trEPR show an inversion in the spin polarization pattern with time. This effect can be understood taking a closer look at the populating and decay rates of the triplet sublevels. In both cases, the populating and decay kinetic constants possess a similar mathematical description: $k_\mu^{pop} \propto \left| \left\langle \psi_{T_\mu} \middle| H_{ISC} \middle| \psi_{S_1} \right\rangle \right|^2$ and $k_\mu^{dec} \propto \left| \left\langle \psi_{T_\mu} \middle| H_{ISC} \middle| \psi_{S_0} \right\rangle \right|^2$. Since $\psi_{S_1}$ and $\psi_{S_0}$ possess similar symmetry, it is probable that the spin-population of the triplet levels that are populated more rapidly will also decay faster. As a result, the triplet polarization evolves with time, showing an inversion of the spin polarization pattern whilst retaining the same spectral shape[41].

## NFA blends

Once again, all trEPR spectra of NFA blends, apart from J51:ITIC, are characterized by the presence of two main signals: one narrow and one broad. The narrow signal in the centre of the EPR spectrum can be attributed to charges photo-generated following photon absorption. This signal is very intense as a consequence of the high charge photo-generation efficiency in the studied donor-acceptor blends. In PTB7-Th:IEICO-2F, PBDBT:ITIC and J51:ITIC, the signal shows an interesting time evolution: at shorter times (1 μs), the an *ea* polarization pattern typical of SCRP is observed. The presence of SCRPs highlights that in these blends, the charges are still magnetically interacting at shorter times and are therefore not fully separated. At longer times (5 μs), the signal evolves into a single peak, typical of free photo-generated charges. This observation suggests that in these three blends, the charge generation process at 80 K is slower compared to the other NFA blends, which also show only a single peak at shorter times. In contrast, the PM6:ITIC blend exhibits a SCRPs signal at both 1 and 5 µs, potentially indicating even slower charge generation. This is consistent with the relatively lower performance observed in this blend.

The broader signal is attributed to the triplet excitons generated via ISC from $S_1$ to $T_1$. In the studied NFA blends, this signal is much weaker when compared to the pristine polymer films, due to the faster singlet exciton quenching rates resulting from an efficient charge transfer process. The triplet excitons appear particularly weak in PTB7-Th:IEICO-4F and J51:ITIC, where no spectral simulation were able to be performed. For the other blends, we carried out the spectral simulations: importantly, by comparing the obtained ZFS values to those measured in the neat films, we could elucidate whether the detected triplets are localized on the donor or the acceptor (Table S2). In all PTB7-Th NFA blends, the observed triplet excitons are localized on the acceptor, whilst in contrast, the triplet is localised on the donor in the PM6 and PBDB-T blends. Furthermore, in PM6:ITIC and PM6:IT-4F blends, we detect two ISC triplets: this is consistent with the observations from the neat PM6 film. As the absence of ISC triplets in TA measurements demonstrate (with the exception of PTB7-Th:IEICO-0F), ISC triplets are unlikely to be observed at room temperature under normal device operating conditions[29]. Finally, the trEPR signal of the ISC triplet exciton in PM6:Y6 blend is very weak, rendering simulation difficult. However, due to the large D value, we believe the triplet exciton is localized on PM6, similar to the situation in PM6:ITIC and PM6:IT-4F.

## FA blends

In the FA blends, we again largely observe the same two signals (one with a narrow bandwidth and the other much broader) that are attributed to photo-generated CT states and charges, and triplet excitons formed via ISC. The exception is PBDB-T:$PC_{70}$BM, where the triplet signal is too weak to simulate. In addition to these two signals, PTB7-Th:$PC_{60}$BM, PM6:PCBM and J51:$PC_{70}$BM also show a second triplet signal, which has not been observed before. From the ZFS parameters (Table S2), we conclude that this triplet is localized on the fullerene. In addition, its populating rates clearly demonstrate that the triplet is generated by a geminate BCT mechanism, with a characteristic *eaaeea* polarisation pattern. The presence of triplet excitons generated by geminate BCT is in stark contrast to the results obtained for the NFA blends and can be rationalized by the high miscibility of FA with the side chains of the polymer[44–47], resulting in the formation of mixed polymer/fullerene regions[44–47]. If the acceptor concentration in these mixed regions is below the percolation threshold for efficient electron transport, charge separation will be impeded, resulting in geminate recombination[48,49]. Thus, we propose that poorly-connected fullerene regions provide the opportunity for geminate $^3$CT formation via $S_0$-$T_0$ spin-mixing from $^1$CT states on ns timescales[34,35], followed by BCT to $T_1$, increasing losses. In contrast, many efficient NFA OSCs have been shown to possess good phase purity[18,50–52], which has previously been shown to facilitate CTE dissociation and reduce $T_1$ formation[53].

| Material class | Material | Triplet 1 | [D E] (G) | Triplet 2 | [D E] (G) | D or A? | Charges |
|---|---|---|---|---|---|---|---|
| **Neat donor film** | PM6 | ISC | [500 30] | ISC | [50 0] | D | FC |
| | PTB7-Th | ISC | [360 60] | | | D | FC |
| | PBDB-T | ISC | [500 55] | BCT | [500 55] | D | FC |
| | J51 | ISC | [460 70] | | | D | FC |
| **Neat acceptor film** | Y6 | ISC | [300 50] | | | A | |
| | ITIC | ISC (very weak) | no sim | | | A | FC |
| | IT-4F | ISC (very weak) | no sim | | | A | FC |
| | IEICO-2F | ISC (very weak) | [290 60] | | | A | |
| | SiOTIC-4F | ISC | [300 60] | | | A | |
| **NFA blends** | PM6:Y6 | ISC (very weak) | [450 80] | | | D* | FC |
| | PM6:ITIC | ISC | [450 30] | ISC | [50 0] | D | CT |
| | PM6:IT-4F | ISC | [440 30] | ISC | [50 0] | D | FC |
| | PTB7-Th:IEICO-0F | ISC | [290 60] | | | A | FC |
| | PTB7-Th:IEICO-2F | ISC | [290 60] | | | A | CT -> FC |
| | PTB7-Th:IEICO-4F | ISC (very weak) | no sim | | | | FC |
| | PTB7-Th:SiOTIC-4F | ISC | [300 60] | | | A | FC |
| | PBDB-T:ITIC | ISC | [480 50] | | | D | CT -> FC |
| | J51:ITIC | | | | | | CT -> FC |
| **FA blends** | PM6:PC$_{60}$BM | ISC | [470 10] | BCT | [420 30] | D | CT -> FC |
| | PTB7-Th:PC$_{60}$BM | ISC | [350 50] | BCT | [350 50] | D | CT |
| | PBDB-T:PC$_{70}$BM | very weak | no sim | | | | FC |
| | J51:PC$_{70}$BM | ISC | [140 30]* | BCT | [55 -7] | ?* + A | FC |

*very weak signal: the simulation and the attribution is not clear.

**Table S2:** A summary of the best-fit spectral simulations of the trEPR measurements reported in Figures S48 – S67. The samples are split into four categories: neat donor films, neat

acceptor films, non-fullerene acceptor (NFA) blends and fullerene acceptor (FA) blends. For each blend, the ZFS parameters and the populating mechanism of the triplet states are reported. The ZFS parameters are given in units of Gauss. From the ZFS parameters, we assigned the triplet either to the donor (D) or the acceptor (A). In addition, the presence of charges (either a CT state or FC) is reported.

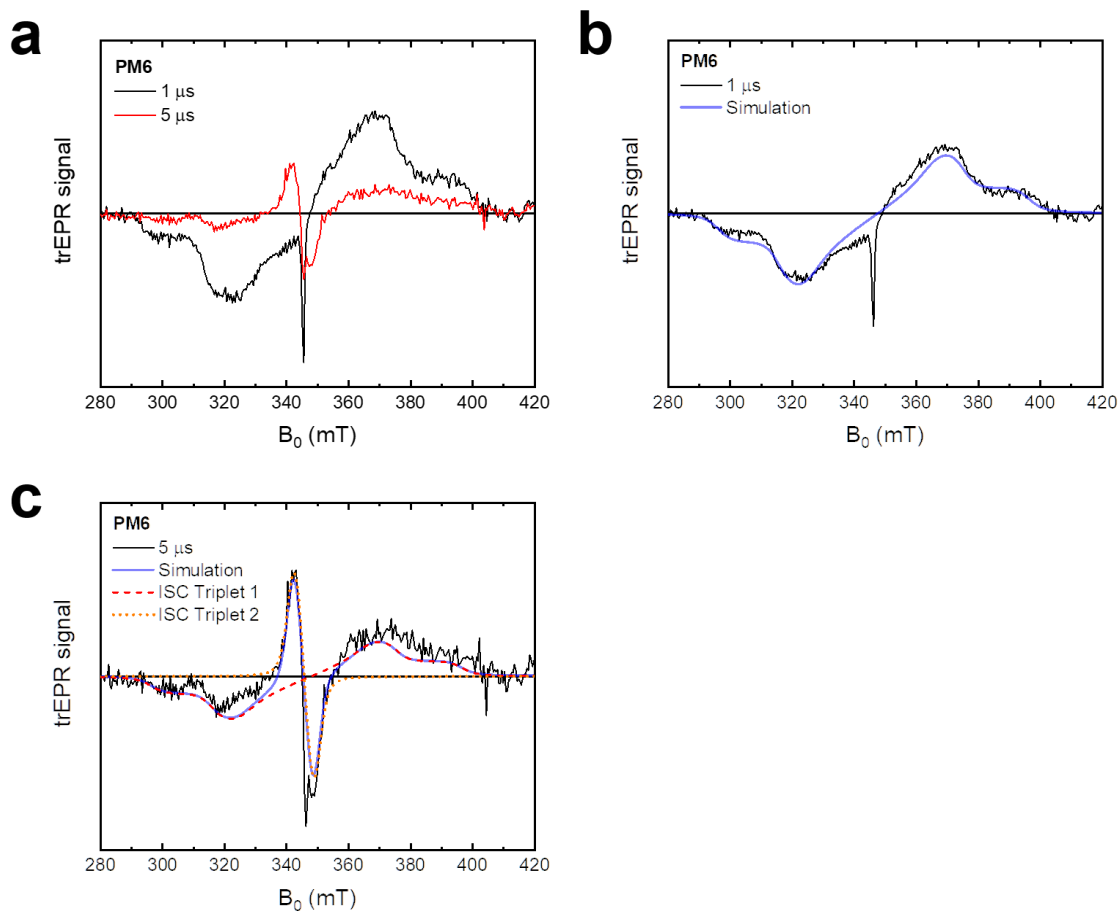

**Figure S48: (a)** The trEPR spectra of a neat PM6 film, taken at representative time points of 1 and 5 μs after excitation at 532 nm. Absorption (*a*) is up, emission (*e*) is down. The sharp peak at ~346 mT is a signature of free polarons, whilst the broader signal between 290 – 410 mT is assigned to triplet excitons. As time progresses, the broader triplet signature largely disappears and a new, narrower triplet feature between 335 – 355 mT forms. This indicates the triplet is less localised, suggestive of a HLCT triplet state. **(b)** The trEPR spectra at 1 μs is shown, with the simulation overlaid. An *eeeaaa* polarisation pattern, indicative of a triplet exciton formed via ISC, is obtained. The [D E] parameters of the triplet state are [500 30]. **(c)** The trEPR spectra at 5 μs is shown, with the simulation overlaid. The combination of two triplets with an *eeeaaa* polarisation pattern are required to fit the spectra, indicative of triplet excitons formed via ISC. The first triplet has [D E] parameters of [500 30], indicating it is the same triplet state as previously observed at 1 μs. The second, narrower triplet has [D E] parameters of [50 0]. The smaller [D] parameter confirms that this triplet state is more delocalised than triplet 1.

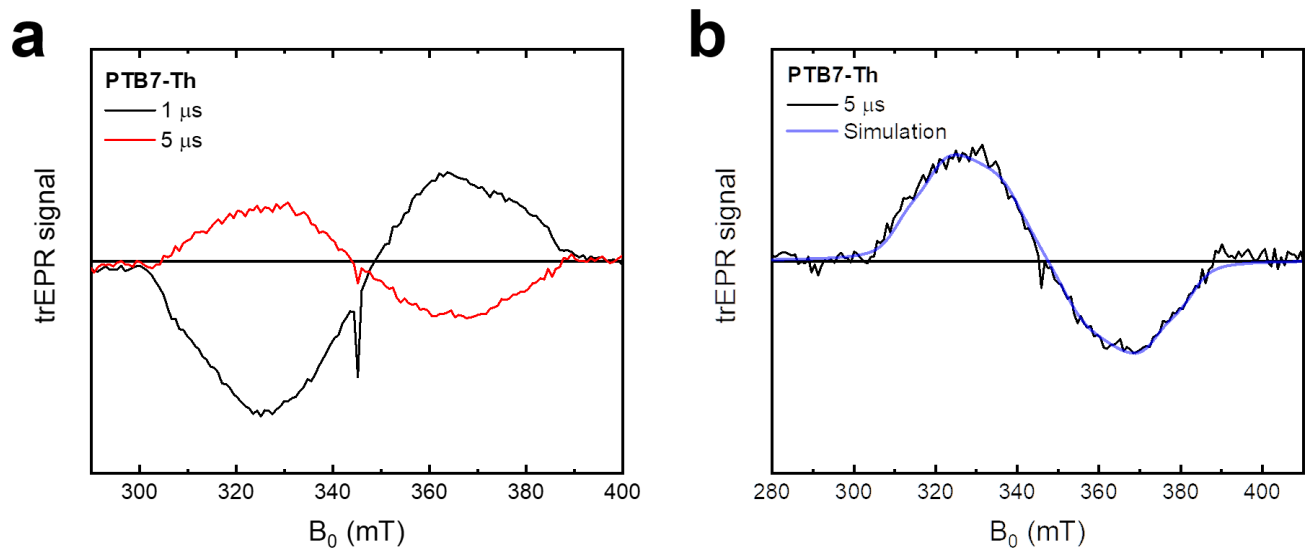

**Figure S49: (a)** The trEPR spectra of a neat PTB7-Th film, taken at representative time points of 1 and 5 µs after excitation at 532 nm. Absorption (*a*) is up, emission (*e*) is down. The sharp peak at ~346 mT is a signature of free polarons, whilst the broader signal between 310 – 390 mT is assigned to triplet excitons. As time progresses, the triplet signal inverts. This is likely due to unequal decay rates from the three high-field triplet states[54]. **(b)** The trEPR spectra at 5 µs is shown, with the simulation overlaid. An *aaaeee* polarisation pattern, indicative of a triplet exciton formed via ISC, is obtained. The [D E] parameters of the triplet state are [360 60].

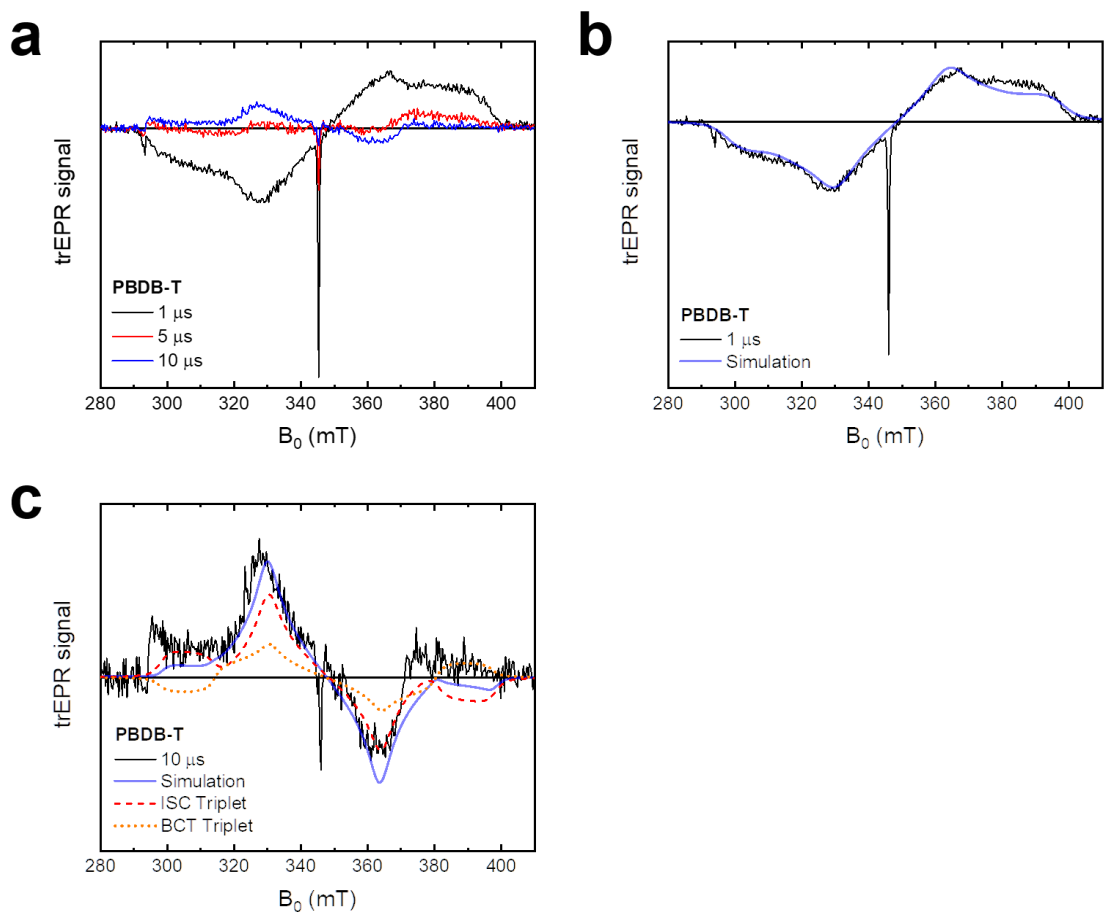

**Figure S50: (a)** The trEPR spectra of a neat PBDB-T film, taken at representative time points of 1, 5 and 10 μs after excitation at 532 nm. Absorption (*a*) is up, emission (*e*) is down. The sharp polaron peak at ~346 mT is particularly pronounced in PBDB-T, suggesting the pathway to free charge generation is quite efficient. Also present is a broad triplet signal between 290 – 400 mT, which evolves to display a new, complex polarisation pattern by 10 μs. **(b)** The trEPR spectra at 1 μs is shown, with the simulation overlaid. An *eeeaaa* polarisation pattern, indicative of a triplet exciton formed via ISC, is obtained. The [D E] parameters of the triplet state are [500 55]. **(c)** The trEPR spectra at 10 μs is shown, with the simulation overlaid. Two triplet species are required to fit the spectra, both with [D E] parameters of [500 55]. The first has an *aaaeee* polarisation pattern, confirming it is formed via ISC. This is likely the same triplet as visible at 1 μs, but inverted. Again, we assign this to unequal decay rates from the three high-field triplet states[54]. The second triplet species has an *eaaeea* polarisation, suggesting it is caused by a BCT process. This can be rationalised by the intense polaron feature in PBDB-T, suggesting charge generation is particularly efficient.

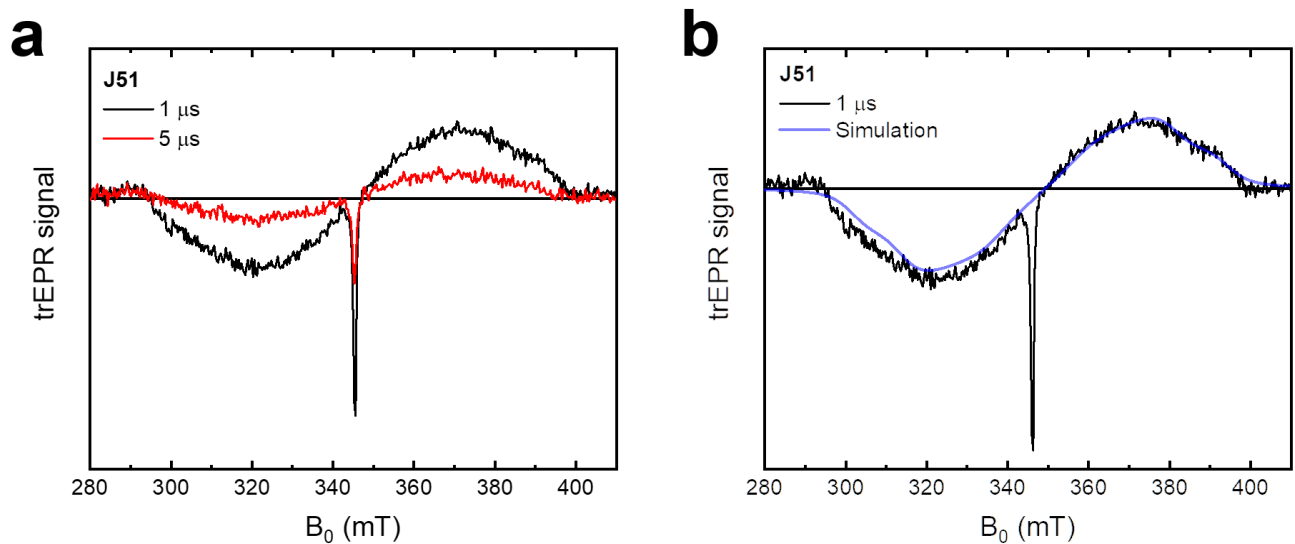

**Figure S51: (a)** The trEPR spectra of a neat J51 film, taken at representative time points of 1 and 5 µs after excitation at 532 nm. Absorption (*a*) is up, emission (*e*) is down. The sharp peak at ~346 mT is a signature of free polarons, whilst the broader signal between 290 – 400 mT is assigned to triplet excitons. **(b)** The trEPR spectra at 1 µs is shown, with the simulation overlaid. An *eeeaaa* polarisation pattern, indicative of a triplet exciton formed via ISC, is obtained. The [D E] parameters of the triplet state are [460 70].

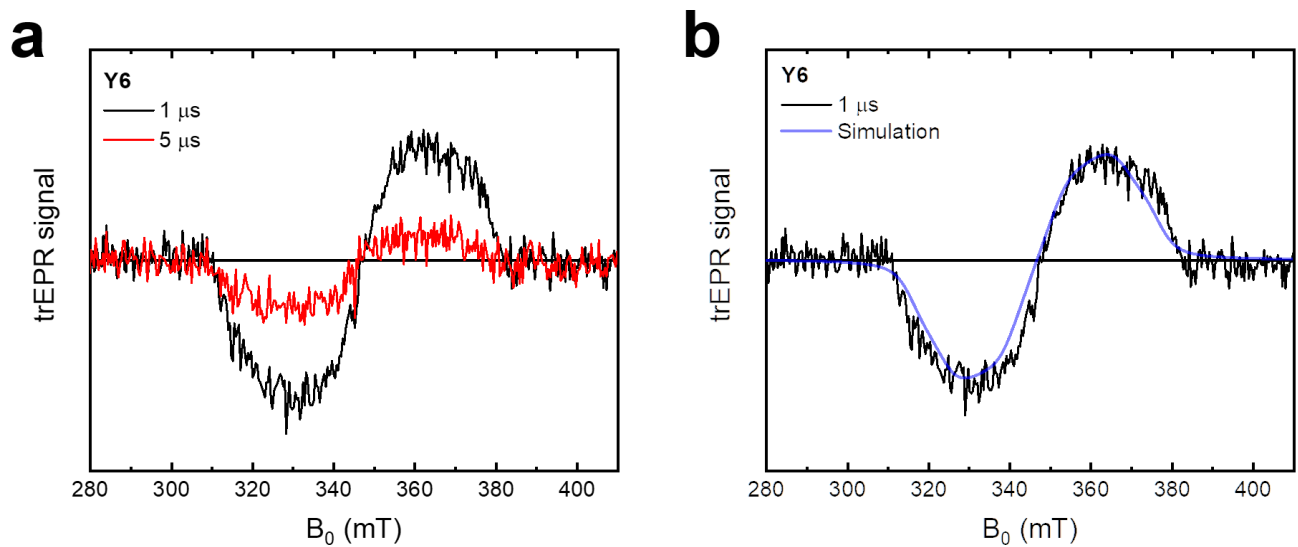

**Figure S52: (a)** The trEPR spectra of a neat Y6 film, taken at representative time points of 1 and 5 µs after excitation at 532 nm. Absorption (*a*) is up, emission (*e*) is down. The broad signal between 310 – 380 mT is assigned to triplet excitons. **(b)** The trEPR spectra at 1 µs is shown, with the simulation overlaid. An *eeeaaa* polarisation pattern, indicative of a triplet exciton formed via ISC, is obtained. The [D E] parameters of the triplet state are [300 50].

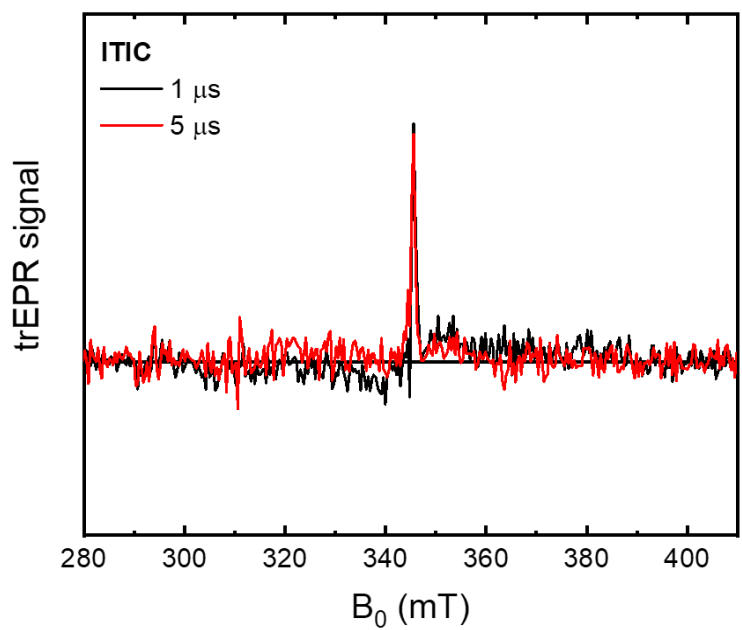

**Figure S53: (a)** The trEPR spectra of a neat ITIC film, taken at representative time points of 1 and 5 µs after excitation at 532 nm. Absorption (*a*) is up, emission (*e*) is down. The sharp peak at ~346 mT is a signature of free polarons. A broad feature between 320 – 360 mT is weakly visible at 1 µs, which appears to have an *eeeaaa* polarisation pattern indicative of a triplet exciton formed via ISC. However, due to the extremely low intensity of the signal, it is not possible to perform a simulation to confirm this.

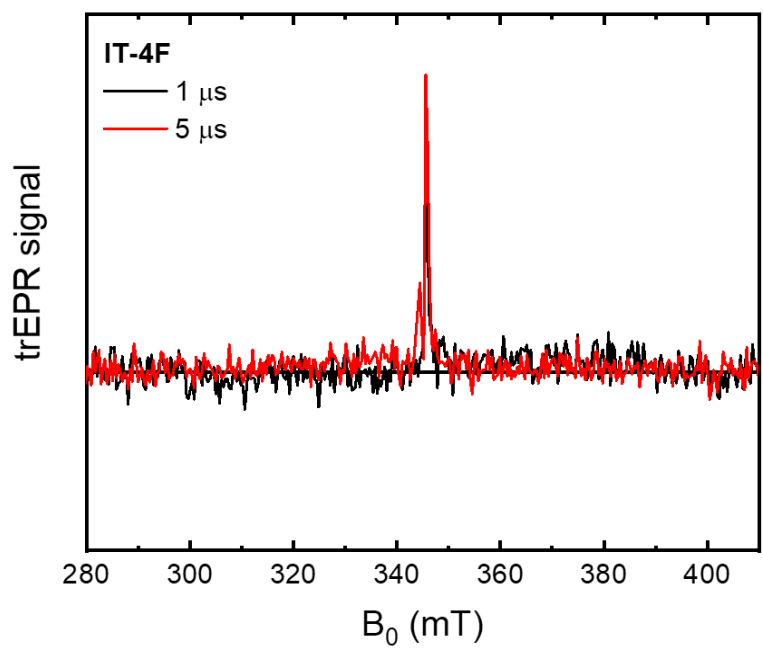

**Figure S54: (a)** The trEPR spectra of a neat IT-4F film, taken at representative time points of 1 and 5 µs after excitation at 532 nm. Absorption (*a*) is up, emission (*e*) is down. The sharp peak at ~346 mT is a signature of free polarons. No obvious features that could be associated with triplet excitons are visible in this sample.

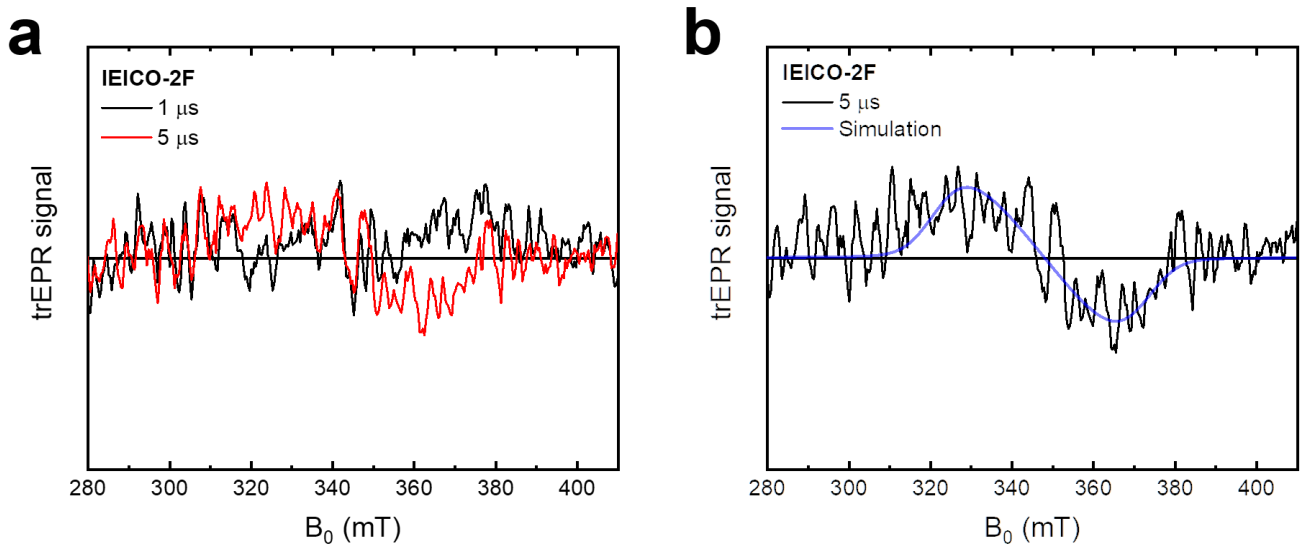

**Figure S55: (a)** The trEPR spectra of a neat IEICO-2F film, taken at representative time points of 1 and 5 μs after excitation at 532 nm. Absorption (*a*) is up, emission (*e*) is down. A broad feature between 300 – 380 mT is weakly visible at 5 μs, which appears to have an *aaaeee* polarisation pattern indicative of a triplet exciton formed via ISC. **(b)** The simulation of the signal at 5 μs, confirming an *aaaeee* polarisation pattern with [D E] parameters of [290 60].

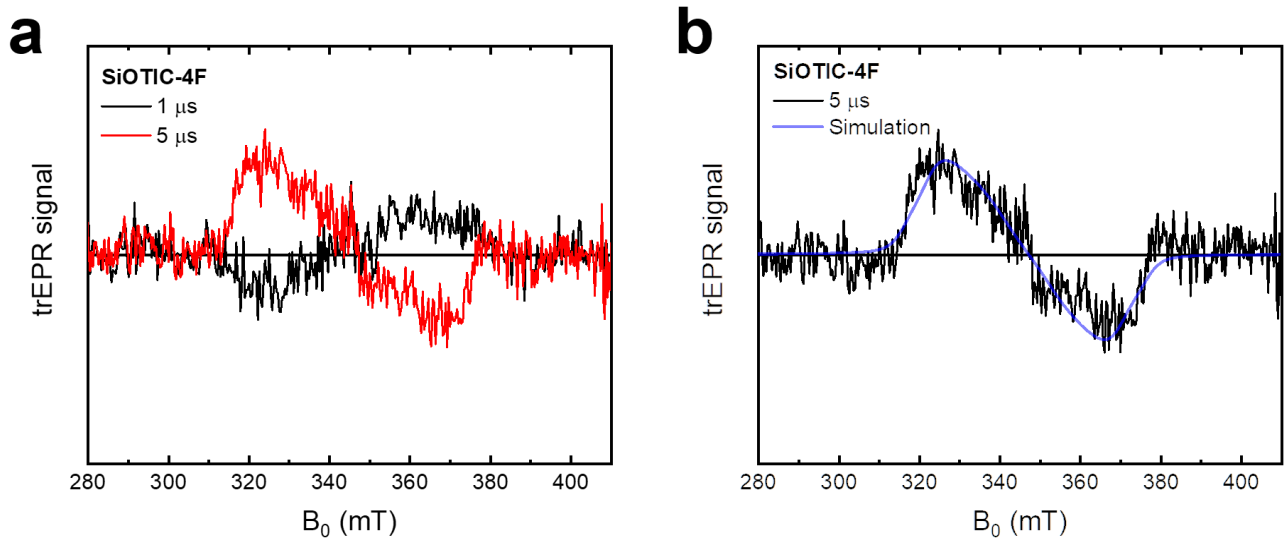

**Figure S56: (a)** The trEPR spectra of a neat SiOTIC-4F film, taken at representative time points of 1 and 5 µs after excitation at 532 nm. Absorption (*a*) is up, emission (*e*) is down. The broad signal between 310 – 380 mT is assigned to triplet excitons. As time progresses, the triplet signal inverts. This is likely due to unequal decay rates from the three high-field triplet states[54]. **(b)** The trEPR spectra at 5 µs is shown, with the simulation overlaid. An *aaaeee* polarisation pattern, indicative of a triplet exciton formed via ISC, is obtained. The [D E] parameters of the triplet state are [300 60].

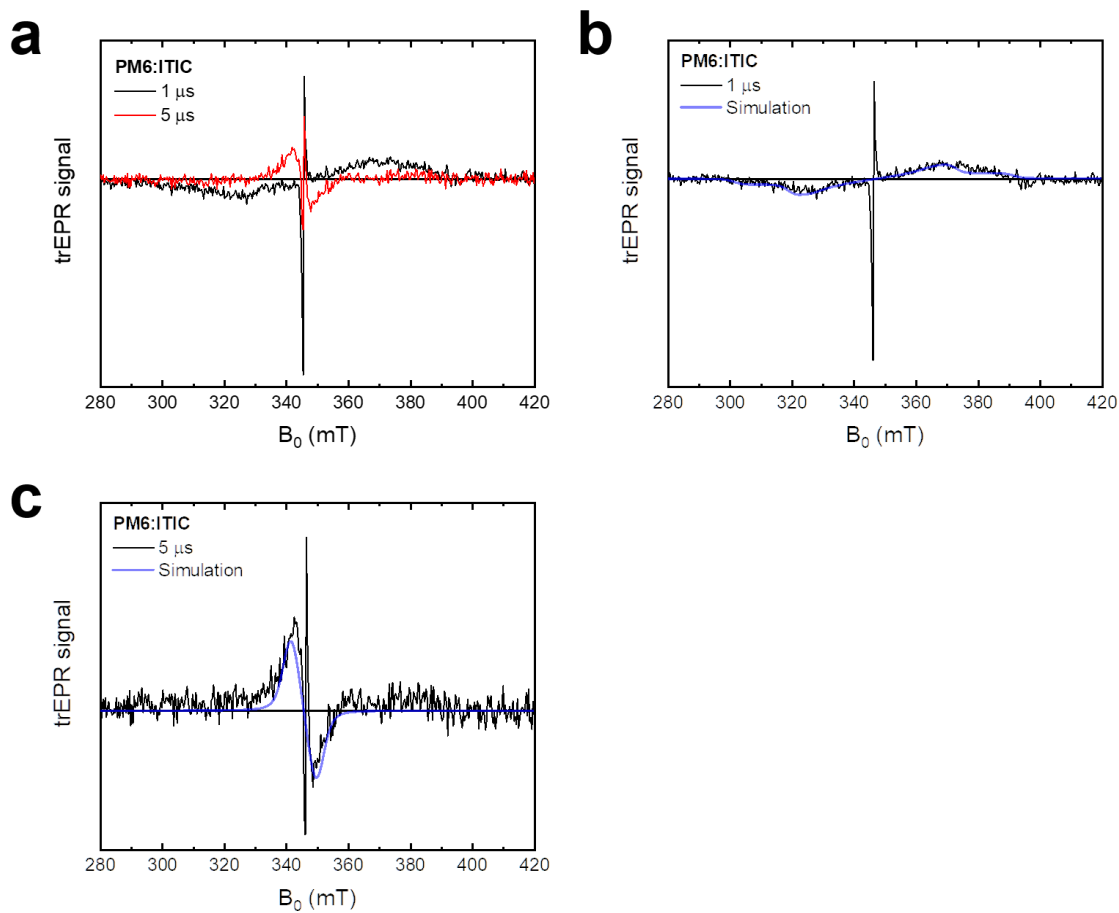

**Figure S57: (a)** The trEPR spectra of a PM6:ITIC blend film, taken at representative time points of 1 and 5 μs after excitation at 532 nm. Absorption (*a*) is up, emission (*e*) is down. The sharp *ea* feature at ~346 mT is a signature of a CT state, whilst the broader signal between 290 – 410 mT is assigned to triplet excitons. As time progresses, the broader triplet signal disappears and a new, narrower triplet feature between 335 – 355 mT forms. **(b)** The trEPR spectra at 1 μs is shown, with the simulation overlaid. An *eeeaaa* polarisation pattern, indicative of a triplet exciton formed via ISC, is obtained. The [D E] parameters of the triplet state are [450 30], which are comparable to the spectra of triplet 1 observed in neat PM6 (Fig. S43). Therefore, it is highly likely that the triplet is localised on PM6. **(c)** The trEPR spectra at 5 μs is shown, with the simulation overlaid. This narrower triplet has [D E] parameters of [50 0] and is an excellent match to the spectrum of triplet 2 in PM6 (Fig. S48). Given the similarities in spectra and time evolution of the triplet signals, it is likely they originate from the ISC of un-dissociated singlets on PM6. Importantly, no triplets with a polarisation pattern characteristic of BCT are observed in this blend.

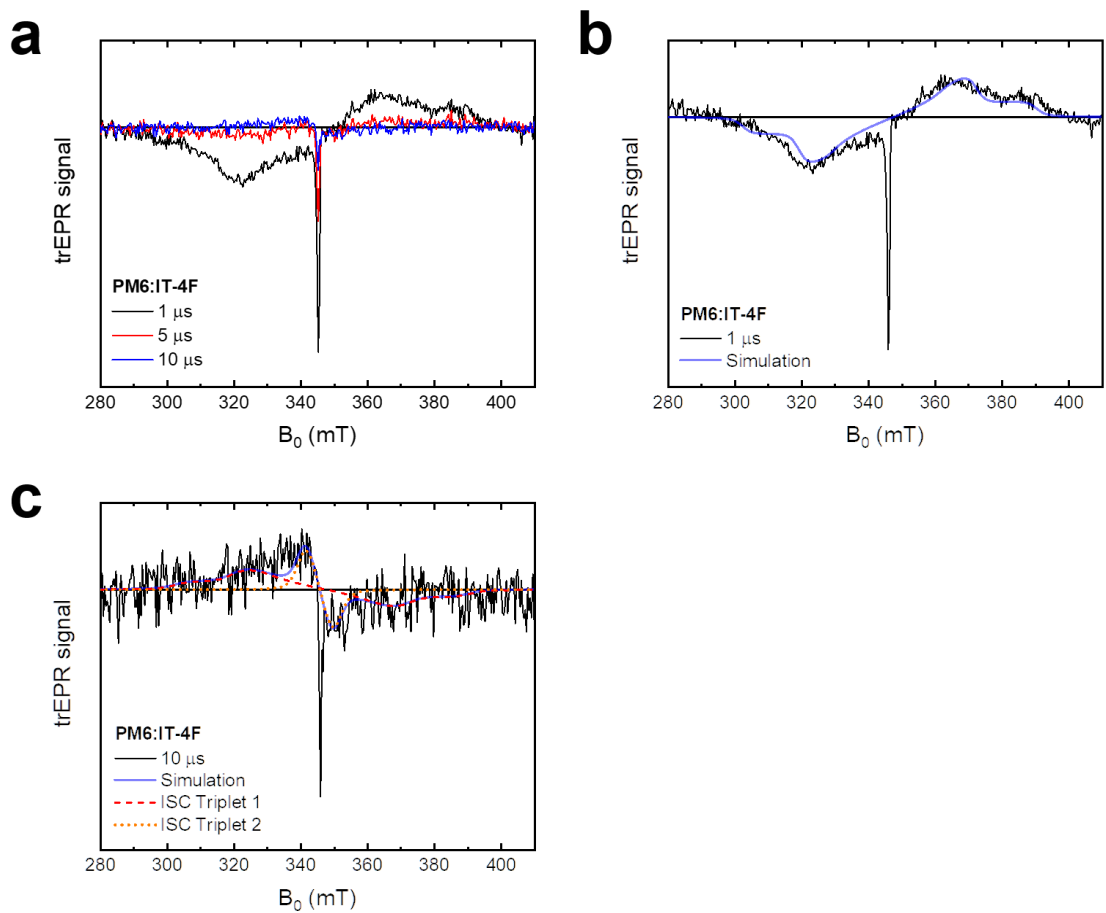

**Figure S58: (a)** The trEPR spectra of a PM6:IT-4F blend film, taken at representative time points of 1, 5 and 10 µs after excitation at 532 nm. Absorption (*a*) is up, emission (*e*) is down. The sharp feature at ~346 mT is a signature of free polarons, whilst the broader signal between 290 – 410 mT is assigned to triplet excitons. As time progresses, the broader triplet signal largely disappears and a new, narrower triplet feature between 335 – 355 mT forms. **(b)** The trEPR spectra at 1 µs is shown, with the simulation overlaid. An *eeeaaa* polarisation pattern, indicative of a triplet exciton formed via ISC, is obtained. The [D E] parameters of the triplet state are [440 30], which are comparable to the spectra of triplet 1 observed in neat PM6 (Fig. S43). Therefore, it is highly likely that the triplet is localised on PM6. **(c)** The trEPR spectra at 10 µs is shown, with the simulation overlaid. Two triplet species are required to achieve a good fit: the first triplet is the same as observed at 1 µs, except inverted. The second triplet is narrower, with [D E] parameters of [50 0] and is an excellent match to the spectrum of triplet 2 in PM6 (Fig. S48). Given the similarities in spectra and time evolution of the triplet signals, it is likely they originate from the ISC of un-dissociated singlets on PM6. Importantly, no triplets with a polarisation pattern characteristic of BCT are observed in this blend.

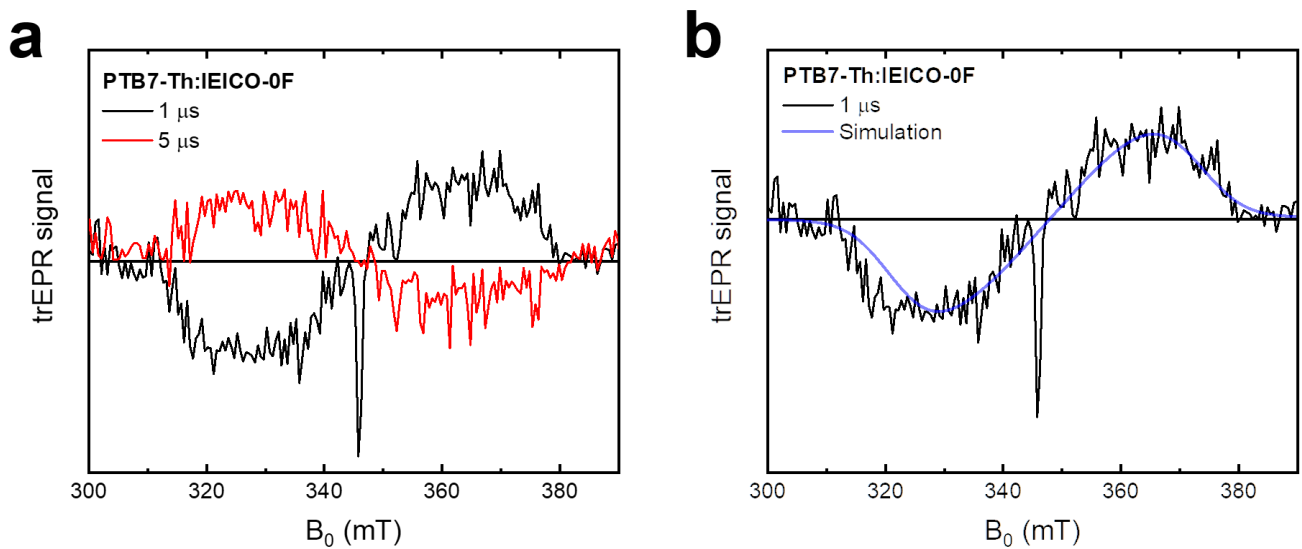

**Figure S59: (a)** The trEPR spectra of a PTB7-Th:IEICO-0F blend film, taken at representative time points of 1 and 5 µs after excitation at 532 nm. Absorption (*a*) is up, emission (*e*) is down. The sharp feature at ~346 mT is a signature of free polarons, whilst the broad signal between 310 – 380 mT is assigned to triplet excitons. As time progresses, the triplet signal inverts. This is likely due to unequal decay rates from the three high-field triplet states[54]. We note that the ISC triplet signal in this blend is significantly more intense relative to the polaron signal than in the PTB7-Th:IEICO-4F blend (Fig. S60), suggesting that the ISC of undissociated singlet excitons is enhanced; this is consistent with the observations in the TA (Fig. S24). **(b)** The trEPR spectra at 1 µs is shown, with the simulation overlaid. An *eeeaaa* polarisation pattern, indicative of a triplet exciton formed via ISC, is obtained. The [D E] parameters of the triplet state are [290 60], which exactly match those obtained from the similar IEICO-2F (Fig. S55). Therefore, the triplet is assigned to the direct ISC of un-dissociated excitons on IEICO-0F.

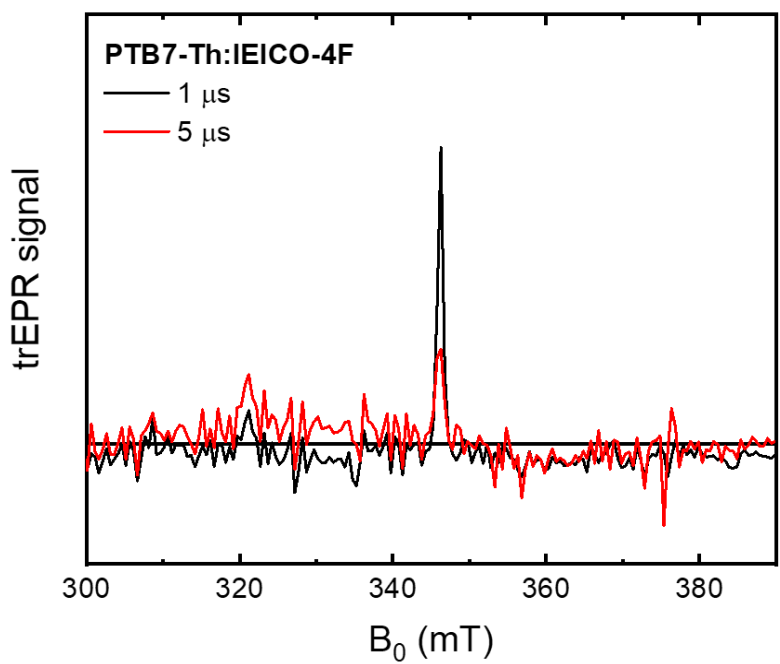

**Figure S60:** The trEPR spectra of a PTB7-Th:IEICO-4F blend film, taken at representative time points of 1 and 5 µs after excitation at 532 nm. Absorption (*a*) is up, emission (*e*) is down. The sharp feature at ~346 mT is a signature of free polarons, with no obvious triplet signals present. The absence of triplets, especially in comparison to the prominent triplet in the PTB7-Th:IEICO-0F blend, can be assigned to the more rapid hole transfer in this blend. This quenches singlet excited states on the NFA faster, leaving less opportunity for ISC.

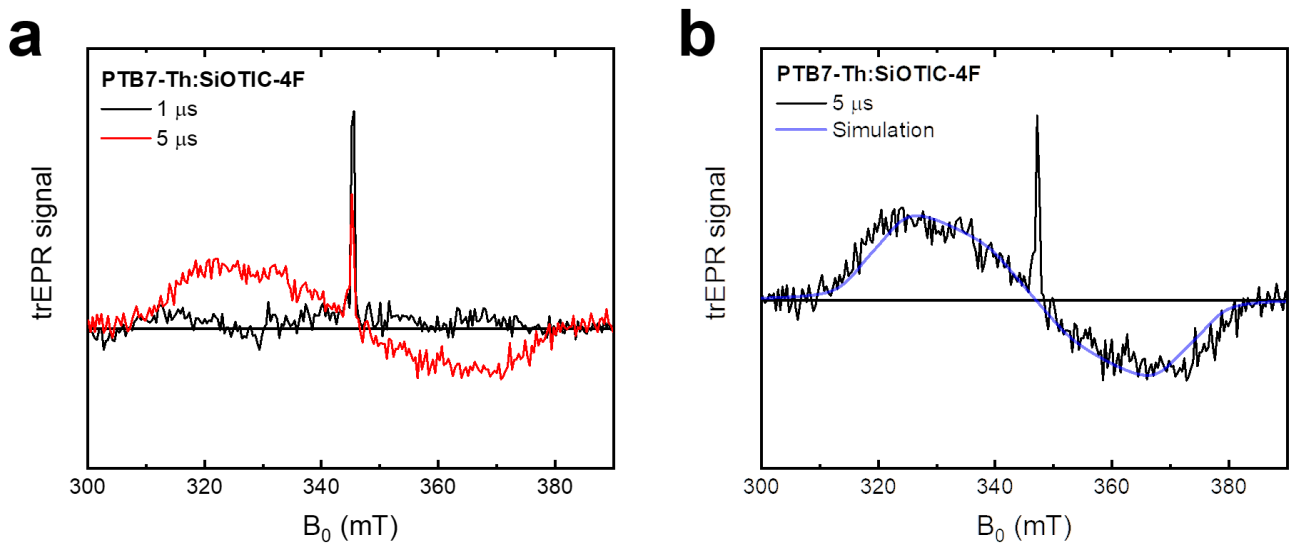

**Figure S61: (a)** The trEPR spectra of a PTB7-Th:SiOTIC-4F blend film, taken at representative time points of 1 and 5 μs after excitation at 532 nm. Absorption (*a*) is up, emission (*e*) is down. The sharp feature at ~346 mT is a signature of free polarons, whilst the broad signal between 310 – 380 mT is assigned to triplet excitons. **(b)** The trEPR spectra at 5 μs is shown, with the simulation overlaid. An *aaaeee* polarisation pattern, indicative of a triplet exciton formed via ISC, is obtained. The [D E] parameters of the triplet state are [300 60], which exactly match those obtained from the neat SiOTIC-4F film (Fig. S56). Therefore, the triplet is assigned to the direct ISC of un-dissociated excitons on SiOTIC-4F.

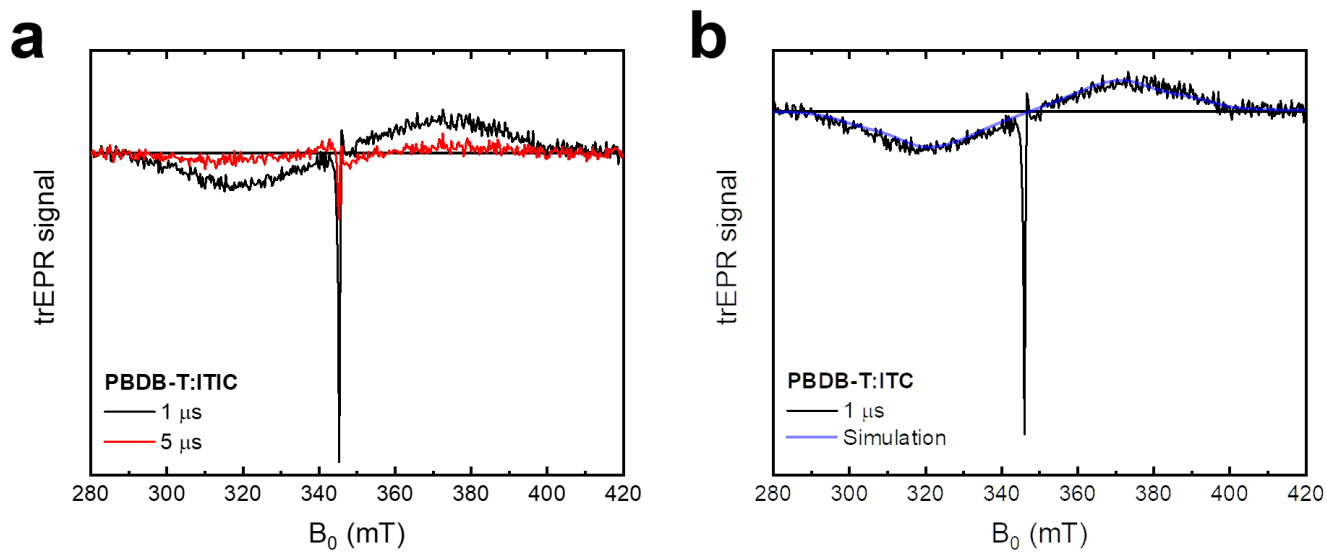

**Figure S62: (a)** The trEPR spectra of a PBDB-T:ITIC blend film, taken at representative time points of 1 and 5 μs after excitation at 532 nm. Absorption (*a*) is up, emission (*e*) is down. The sharp *ea* feature at ~346 mT is a signature of a CT state, which separates into free polarons by 5 μs. The broad signal between 290 – 400 mT is assigned to triplet excitons. **(b)** The trEPR spectra at 1 μs is shown, with the simulation overlaid. An *eeeaaa* polarisation pattern, indicative of a triplet exciton formed via ISC, is obtained. The [D E] parameters of the triplet state are [480 50], which are comparable to the values obtained from the neat PBDB-T film (Fig. S50). Therefore, the triplet is assigned to the direct ISC of un-dissociated excitons on PBDB-T.

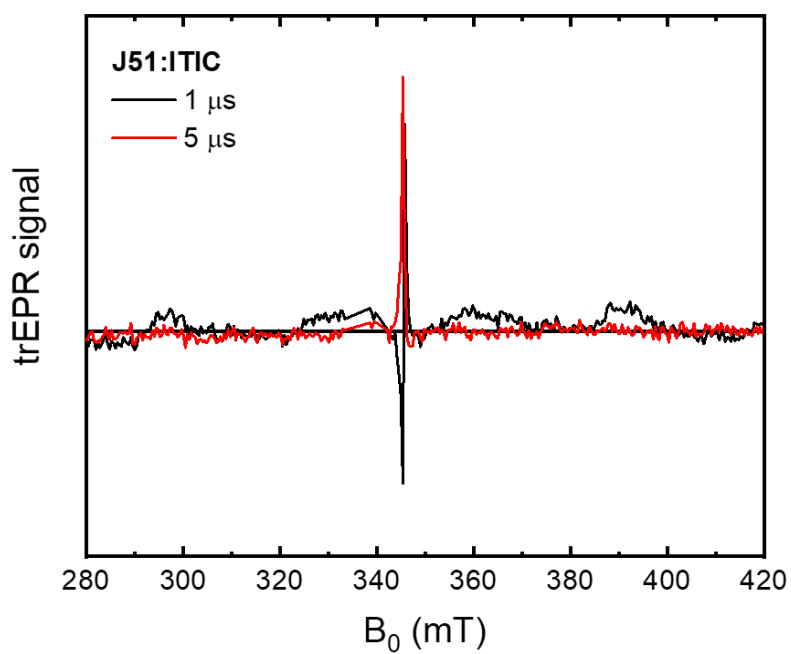

**Figure S63:** The trEPR spectra of a PTB7-Th:IEICO-4F blend film, taken at representative time points of 1 and 5 µs after excitation at 532 nm. Absorption (*a*) is up, emission (*e*) is down. The sharp *ea* feature at ~346 mT is a signature of a CT state, which separates into free polarons by 5 µs. No obvious triplet signals are present.

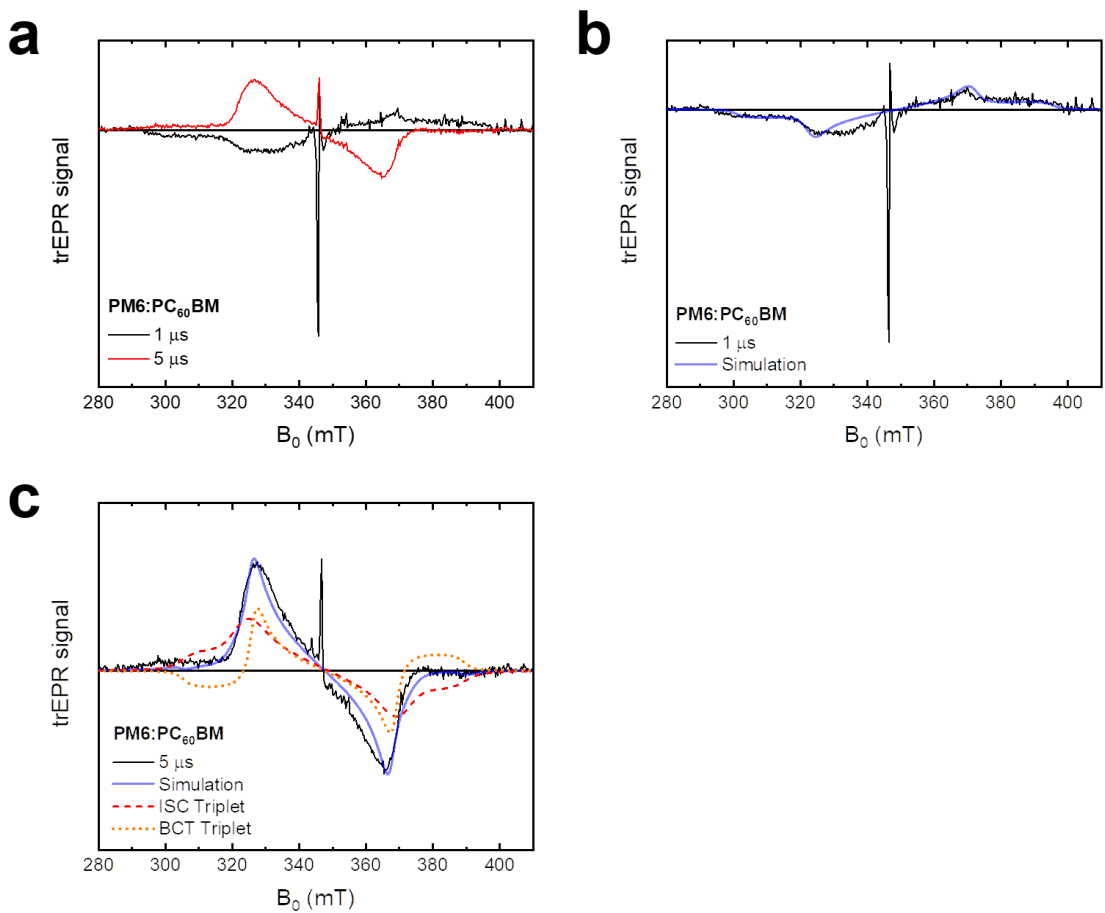

**Figure S64: (a)** The trEPR spectra of a PM6:PC$_{60}$BM blend film, taken at representative time points of 1 and 5 μs after excitation at 532 nm. Absorption (*a*) is up, emission (*e*) is down. The sharp *ea* feature at ~346 mT is a signature of a CT state, which separates into free polarons by 5 μs. The broader signal between 290 – 410 mT is assigned to triplet excitons. As time progresses, the initial triplet signal inverts and the polarisation pattern becomes more complex. **(b)** The trEPR spectra at 1 μs is shown, with the simulation overlaid. An *eeeaaa* polarisation pattern, indicative of a triplet exciton formed via ISC, is obtained. The [D E] parameters of the triplet state are [470 10], which are comparable to the spectra of triplet 1 observed in neat PM6 (Fig. S8). Therefore, it is highly likely that the triplet is localised on PM6. **(c)** The trEPR spectra at 5 μs is shown, with the simulation overlaid. Two triplet species are required to achieve a good fit: the first triplet is the same as observed at 1 μs, except inverted. The second triplet has similar [D E] parameters of [420 30], confirming that it is located on PM6 (Fig. S48), but an *eaaeea* polarisation pattern that is characteristic of a triplet formed via BCT. Therefore, it is clear that geminate BCT triplet formation occurs in this blend.

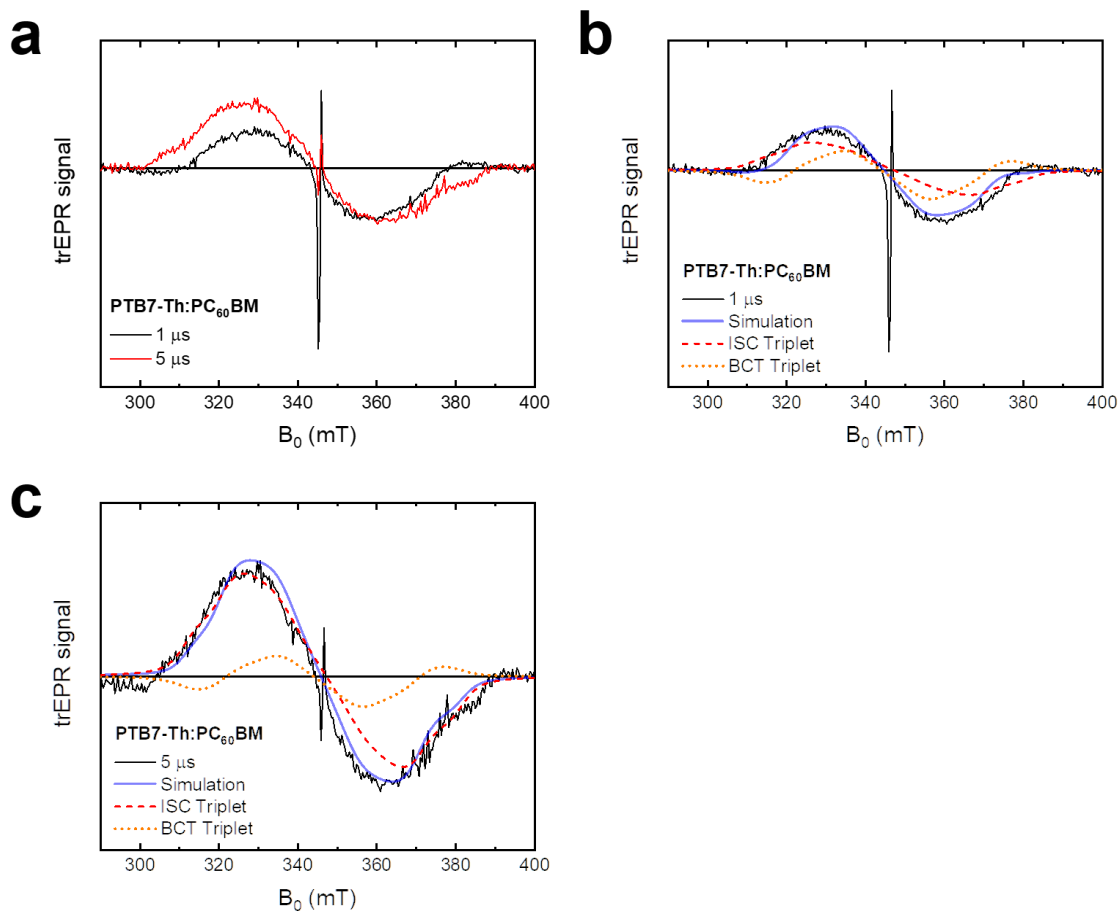

**Figure S65: (a)** The trEPR spectra of a PTB7-Th:PC$_{60}$BM blend film, taken at representative time points of 1 and 5 μs after excitation at 532 nm. Absorption (*a*) is up, emission (*e*) is down. The sharp *ea* feature at ~346 mT is a signature of a CT state. The broader signal between 300 – 390 mT is assigned to triplet excitons. **(b)** The trEPR spectra at 1 μs is shown, with the simulation overlaid. Two triplet contributions are required to simulate the spectra, both with [D E] parameters of [350 50]: these values are highly comparable to those obtained from the neat PTB7-Th film (Fig. S49), confirming that the triplet is localised on PTB7-Th. The first triplet has an *aaaeee* polarisation pattern, indicative of a triplet exciton formed via ISC. The second triplet has an *eaaeea* polarisation pattern that is characteristic of a triplet formed via BCT. Therefore, it is clear that geminate BCT triplet formation occurs in this blend. **(c)** The trEPR spectra at 5 μs is shown, with the simulation overlaid. The same two ISC and BCT triplet species contributions as used before are required to simulate the spectrum.

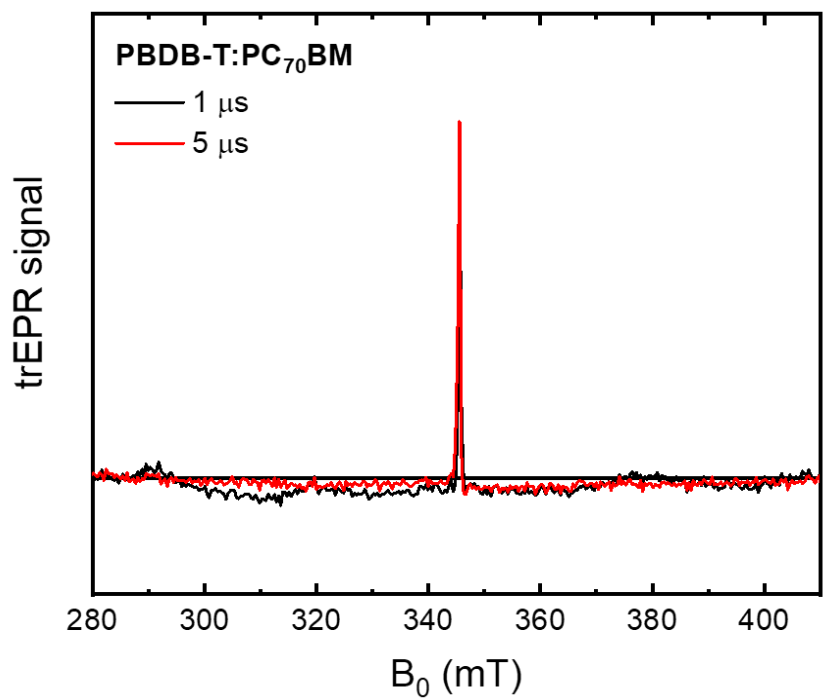

**Figure S66:** The trEPR spectra of a PBDB-T:PC$_{70}$BM blend film, taken at representative time points of 1 and 5 µs after excitation at 532 nm. Absorption (*a*) is up, emission (*e*) is down. The sharp feature at ~346 mT is a signature of free polarons, with no obvious triplet signals present.

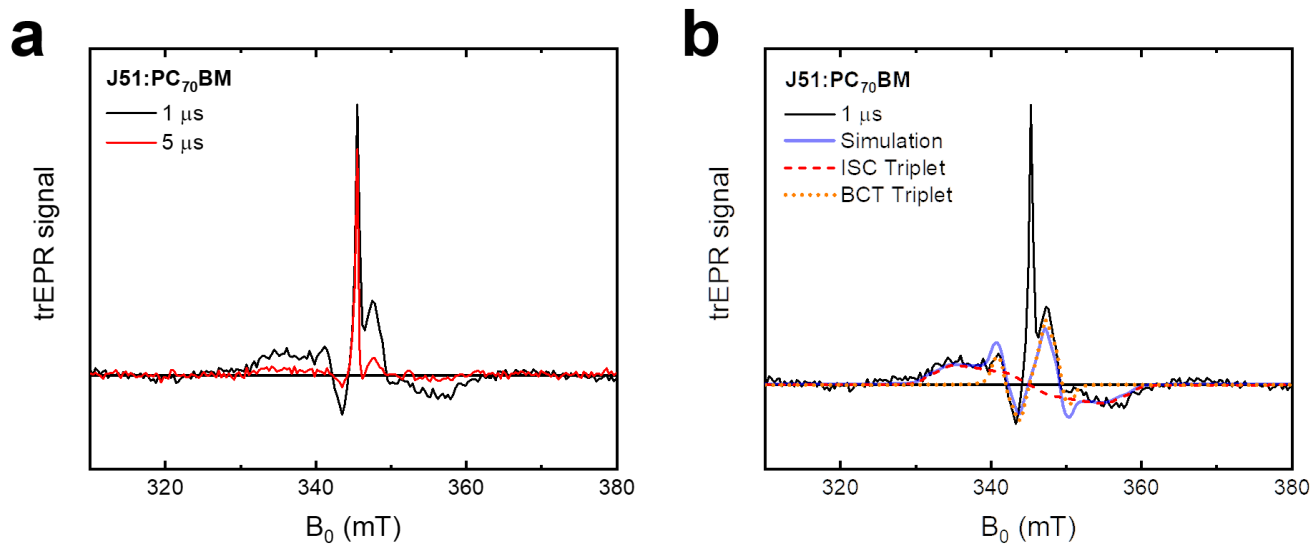

**Figure S67: (a)** The trEPR spectra of a J51:ITIC blend film, taken at representative time points of 1 and 5 µs after excitation at 532 nm. Absorption (*a*) is up, emission (*e*) is down. The sharp feature at ~346 mT is a signature of free polarons. The broader signal between 330 – 360 mT is assigned to triplet excitons, which have largely disappeared by 5 µs. **(b)** The trEPR spectra at 1 µs is shown, with the simulation overlaid. The first triplet has an *aaaeee* polarisation pattern, indicative of a triplet exciton formed via ISC. The second triplet has an *eaaeea* polarisation pattern that is characteristic of a triplet formed via BCT. The [D E] parameters of the ISC triplet state are [140 30], with [55 -7] obtained for the BCT triplet. The narrow [D] parameter confirms that the BCT triplet is located on $PC_{70}BM$, whilst the assignment of the ISC triplet is not clear due to its low intensity.

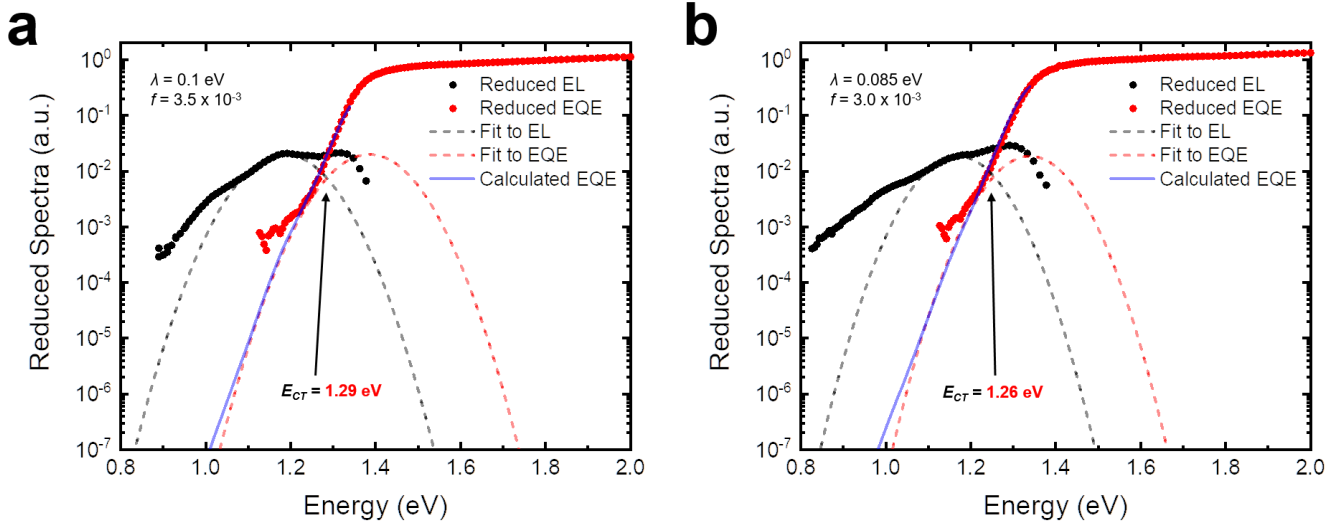

**Figure S68: (a)** The reduced emission and absorption spectra for PTB7-Th:IEICO-2F with the fits obtained from Marcus theory included. From this, a charge transfer state energy ($E_{CT}$) = 1.29 eV is obtained. **(b)** The reduced emission and absorption spectra for PTB7-Th:IEICO-4F with the fits obtained from Marcus theory included. From this, $E_{CT}$ = 1.26 eV is obtained.

The equations used to perform the Marcus theory fitting and obtain the $E_{CT}$[55]:

$$EQE_{PV,CT}(E) = \frac{f}{E\sqrt{4\pi\lambda k_B T}} \exp\left(\frac{-(E_{CT} + \lambda - E)^2}{4\lambda k_B T}\right) \qquad (4)$$

$$EQE_{EL,CT}(E) = E\frac{f}{\sqrt{4\pi\lambda k_B T}} \exp\left(\frac{-(E_{CT} - \lambda - E)^2}{4\lambda k_B T}\right) \qquad (5)$$

$$EQE_{PV}(E) \propto EL(E)E^{-2} \exp\left(\frac{E}{k_B T}\right) \qquad (6)$$

where, $k_B$ is Boltzmann's constant, $E$ is the photon energy, and $T$ is the absolute temperature. The fit parameters are $E_{CT}$, which is the energy at the point of intersection between the CT state absorption and emission, $\lambda$, which is the reorganization energy, and $f$, which is a measure of the strength of the donor-acceptor coupling.

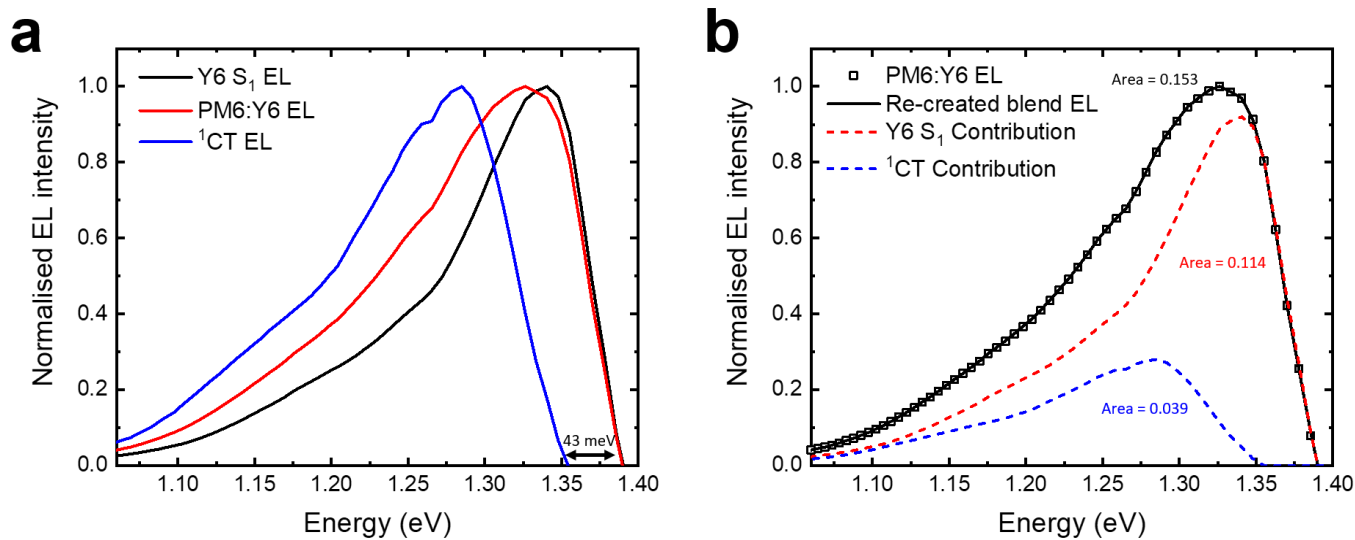

**Figure S69: (a)** The normalised electroluminescence (EL) spectra of neat Y6 and PM6:Y6 devices. The EL spectra of neat Y6 and the PM6:Y6 blend are extremely similar, with an identical emission onset in both films. If emission was occurring solely from $^{1}$CT in PM6:Y6, a small red-shift in the emission onset would be expected, corresponding to the reported ~50 meV difference between the Y6 $S_1$ and the $^{1}$CT obtained from fitting the reduced photovoltaic EQE and EL spectra[18]. The fact that the onset is the same for both neat Y6 and PM6:Y6 implies that a portion of the blend emission comes from the Y6 $S_1$. Subtracting the neat Y6 EL from the blend EL, both normalised to the EL onset region between 1.38 – 1.36 eV, gives the form of the PM6:Y6 $^{1}$CT EL spectra. The onset of the $^{1}$CT EL is 43 meV lower in energy than the neat Y6 EL: this is completely consistent with the ~50 meV $S_1$-CT energy difference previously reported for this blend[18], providing further evidence that a significant fraction of the blend EL is coming from the Y6 $S_1$. When running the device under forward bias as a light emitting diode, we note that the injected charge carriers will initially form CT states. Therefore, the Y6 $S_1$ emission must originate from $S_1$ states reformed via thermal activation from the $^{1}$CT. This suggests an equilibrium of $S_1$, CT states and FC is present in the blend. **(b)** The normalised blend EL spectra, re-created by a 0.77:0.23 ratio of the normalised Y6 $S_1$ EL to the $^{1}$CT EL. Integrating under the EL curve gives an area ratio of the $S_1$:$^{1}$CT EL contributions of roughly 3:1. This ratio suggests that the total fraction of the total blend EL originating from the Y6 $S_1$ is 75%.

# Quantum chemical calculations

## Time-Dependent Density Functional Theory calculations

To begin our computational study, we shall focus on several representative models for 1:1 D/A complexes, comprising a tetramer of PTB7-Th interacting with IEICO-2F or SiOTIC-4F. It is pertinent to note here that the PTB7-Th:IEICO-2F blend did not exhibit non-geminate $T_1$ formation, whilst the PTB7-Th:SiOTIC-4F blend did. Additionally, as both blends utilise the same donor polymer, PTB7-Th, this allows for a more consistent investigation of factors involving the NFA that affect triplet formation. In all the calculations, the alkyl chains were replaced with methyl groups to reduce the computational costs. Gas-phase ground state 1:1 complexes were optimized at the DFT level with a range-separated hybrid (RSH) ωB97X-D functional using 6-31G(d,p) basis set[56]. The D/A intermolecular equilibrium distance was found to be in range of 3.5–4.0 Å for all investigated configurations. In order to account for the solid-state environment, we tuned the range-separation parameter ω in the presence of polarizable continuum model (PCM) by setting the dielectric constant of toluene $\varepsilon = 2.37$ and utilizing optimized gas-phase geometries[57]. In this approach[58], for each system of interest an optimal value of ω was found by aligning the negative eigenenergies of HOMO orbitals for the N and the (N+1)-electron system with their respective vertical ionization potentials (IP) (barring relaxation effects). The overall error function to be minimize is given as follows:

$$J^2(\omega) = \sum_{i=0}^{1} (\varepsilon_{HOMO}(N + i, \omega) + IP(N + i, \omega))^2 \qquad (7)$$

The non-empirical "optimal" tuning ω in PCM yielded $\omega_{PCM} = 0.011 \; Bohr^{-1}$ for the PTB7-Th:IEICO-2F complex and $\omega_{PCM} = 0.014 \; Bohr^{-1}$ for PTB7-Th:SiOTIC-4F. Subsequent TD-DFT + PCM calculations were carried out with the optimally tuned $\omega_{PCM}$ parameter for each complex, targeting the [1]CT and [3]CT energies of the D/A dyads which together with the energies of local excitations are summarised in Table S3. Interestingly, as explained in the main text in Figure 3a, the energy ordering of the CT states is inverted in the PTB7-Th:IEICO-2F complex, with the [3]CT higher than the [1]CT by 70 meV as a result of hybridisation between local exciton and CT states. On the other hand, the typical energy ordering is restored in the other two complexes due to the lack of hybridisation effects. Indeed, in PTB7-Th:SiOTIC-4F, the [3]CT is lower in energy of 18 meV with respect to the [1]CT.

To check the accuracy of the optimally-tuned functionals, we benchmarked the excitation energies by employing more robust *screened* RSH (SRSH) functionals[59]. In this approach, solid-state polarization effects are introduced by adjusting two additional $\alpha$ and $\beta$ parameters within the exchange-correlation density functional along with $\omega$. For LC-ωhPBE functional[60], the exchange-correlation energy expression reads as:

$$E_{xc}^{SRSH} = (\alpha + \beta)E_{x,HF}^{LR} + (1 - \alpha - \beta)E_{x,PBE}^{LR} + \alpha E_{x,HF}^{SR} + (1 - \alpha)E_{x,PBE}^{SR} + E_{c,PBE} \qquad (8)$$

where $\alpha$ quantifies the fraction of Hartree-Fock (HF) exchange included in the short-range (SR) domain, while $\alpha + \beta$ quantifies the fraction of HF exchange included in the long-range (LR) part and the PBE correlation is used for the whole range. For any choice of $\alpha$, the condition $\alpha + \beta = 1$ ensures 100% of HF exchange in the LR part and the correct asymptotic behaviour of the Coulomb potential in gas-phase. To introduce the effect of the surrounding medium, we imposed the asymptotic convergence of Coulomb potential to $\frac{1}{\varepsilon r}$ rather than to $\frac{1}{r}$, and by fixing $\alpha = 0.2$, we deduced the $\beta$ parameter from $\alpha + \beta = \frac{1}{\varepsilon}$, so that $\beta = \frac{1}{\varepsilon} - \alpha = 0.221$, where $\varepsilon = 2.37$ (a typical value used for a variety of organic molecules). In these calculations, the optimally tuned $\omega_{vac}$ (in vacuum) value have been retained: for PTB7-Th:IEICO-2F we have found $\omega_{vac} = 0.080\ Bohr^{-1}$, while for PTB7-Th:SiOTIC-4F $\omega_{vac} = 0.079\ Bohr^{-1}$.

The SRSH TD-DFT calculations have been carried out for G0_PTB7-Th:IEICO-2F and G0_PTB7-Th:SiOTIC-4F complexes, featuring the smallest and the largest energy difference between ³CT and ¹CT, respectively. In PTB7-Th:IEICO-2F, the ³CT has been found to be higher in energy than the ¹CT by 45 meV, a result which is on par with that obtained with the optimally tuned functionals (the energy of ¹CT is 1.41 eV with an oscillator strength of 0.171 and the energy of ³CT is 1.46 eV). In contrast, the PTB7-Th:SiOTIC-4F complex shows a more stable ³CT in energy than its ¹CT by 34 meV, with the energy of ¹CT of 1.45 eV (with an oscillator strength of 0.003) and the energy of ³CT of 1.42 eV. The consistency of the results obtained by the two approaches that introduce solid-state screening effects in a different fashion indicates a weak dependence on methodology, which reinforces the robustness of the conclusions drawn from the theoretical data. DFT and TD-DFT calculations were carried out with Gaussian16 suite[61].

## Back charge-transfer rate calculations

Since the exact back-charge transfer rate between $^3$CT and T$_1$ is difficult to obtain experimentally, we computed the values theoretically for all the representative polymer/NFA complexes from figure S66. Here, we first defined auxiliary diabatic states (D$^+$)(A$^-$), (D$^*$)(A$^0$), and (D$^0$)(A$^*$) where D is for donor (polymer), and A is for acceptor (NFA), being either in ionized (+/-), ground (0) or excited T$_1$ (*) state, respectively. These states represent $^3$CT and T$_1$ in the absence of configurational mixing. With this definition, the BCT rate was calculated by employing Marcus-Levich-Jortner theory[62,63], using the following expression:

$$k_{CT \to T1\left(\frac{D}{A}\right)} = \frac{2\pi}{\hbar} H^2_{CT \to T1\left(\frac{D}{A}\right)} \sqrt{\frac{1}{4\pi\lambda_s k_B T}} \sum_{n=0}^{\infty} exp^{-S} \frac{S^n}{n!} exp^{-\frac{\left(\Delta E_{CT \to T1\left(\frac{D}{A}\right)} + \lambda_s + n\hbar\omega_i\right)^2}{4\lambda_s k_B T}} \qquad (9)$$

where besides for the fundamental Boltzmann constant, $k_B$, and temperature, $T$ of 298.15 K, three most important contributions, namely, the difference in the diabatic energies, $\Delta E_{CT \to T1(D/A)}$, coupling parameter between the triplet states, $H_{CT \to T1(D/A)}$, and Huang-Rhys factors, $S = \lambda_i / \hbar\omega$, were calculated for each D/A complex as described below. In turn, the remaining, external reorganization energy, $\lambda_s$, together with $\hbar\omega$, were taken from literature.

To compute the couplings, we utilized the Generalized Mulliken-Hush theory that targets minimizing the transition dipole moment between the adiabatic $^3$CT and T$_1$, $\mu_{CT \to T1(D/A)}$[64]. For the two-state model, the coupling is given as follows:

$$H_{CT \to T1\left(\frac{D}{A}\right)} = \frac{\mu_{CT \to T1\left(\frac{D}{A}\right)} \left(E^{ad}_{T1\left(\frac{D}{A}\right)} - E^{ad}_{CT}\right)}{\sqrt{\left(\mu_{T1\left(\frac{D}{A}\right)} - \mu_{CT}\right)^2 - 4\mu^2_{CT \to T1\left(\frac{D}{A}\right)}}} \qquad (10)$$

This scheme is particularly convenient since it deals only with the observable adiabatic energies, $E^{ad}$, and state dipole moments, $\mu$, that are readily available from TD-DFT calculations using the Gaussian software suite. The transition dipole moments between the (excited) $^3$CT and T$_1$ states were extracted by post-processing the TD-DFT wavefunctions using the Multiwfn software[65].

The $\lambda_i$ internal reorganization energy entering Eq. 9 through the Huang-Rhys factor $S$ reflects changes in the geometry of the D/A complex upon BCT. These were calculated at the DFT level based on 4-point total energy differences between reactants and products and assuming additive contributions from the polymer donor and NFA acceptor[66].

Finally, using the two-state model, the difference in diabatic energies was directly deduced from the adiabatic-to-diabatic transformation[67,68], given as:

$$\begin{pmatrix} E_{CT}^{diab} & H_{CT \to T1\left(\frac{D}{A}\right)} \\ H_{CT \to T1\left(\frac{D}{A}\right)} & E_{T1\left(\frac{D}{A}\right)}^{diab} \end{pmatrix} = U \begin{pmatrix} E_{CT}^{ad} & 0 \\ 0 & E_{T1\left(\frac{D}{A}\right)}^{ad} \end{pmatrix} U^T \tag{11}$$

where U is a unitary matrix, commonly referred to as a rotation matrix for a mixing angle between adiabatic ³CT and T₁, $\alpha$, given as:

$$U = \begin{pmatrix} \cos\alpha & \sin\alpha \\ -\sin\alpha & \cos\alpha \end{pmatrix} \tag{12}$$

Once the coupling was computed from eq. 8, the mixing angle was determined from $\sin 2\alpha = 2H_{CT \to T1(D/A)}/\Delta E^{ad}$, leading to a rapid evaluation of the diabatic energy differences as

$$\Delta E^{diab} = \Delta E^{ad} \cdot \cos 2\alpha = \Delta E^{ad} \cdot \sqrt{\frac{\Delta E_{ad}^2 - 4H_{CT \to T1\left(\frac{D}{A}\right)}^2}{\Delta E_{ad}^2}} = (sign)\sqrt{\Delta E_{ad}^2 - 4H_{CT \to T1\left(\frac{D}{A}\right)}^2} \tag{13}$$

with the sign taken in consistency with $\Delta E^{ad}$.

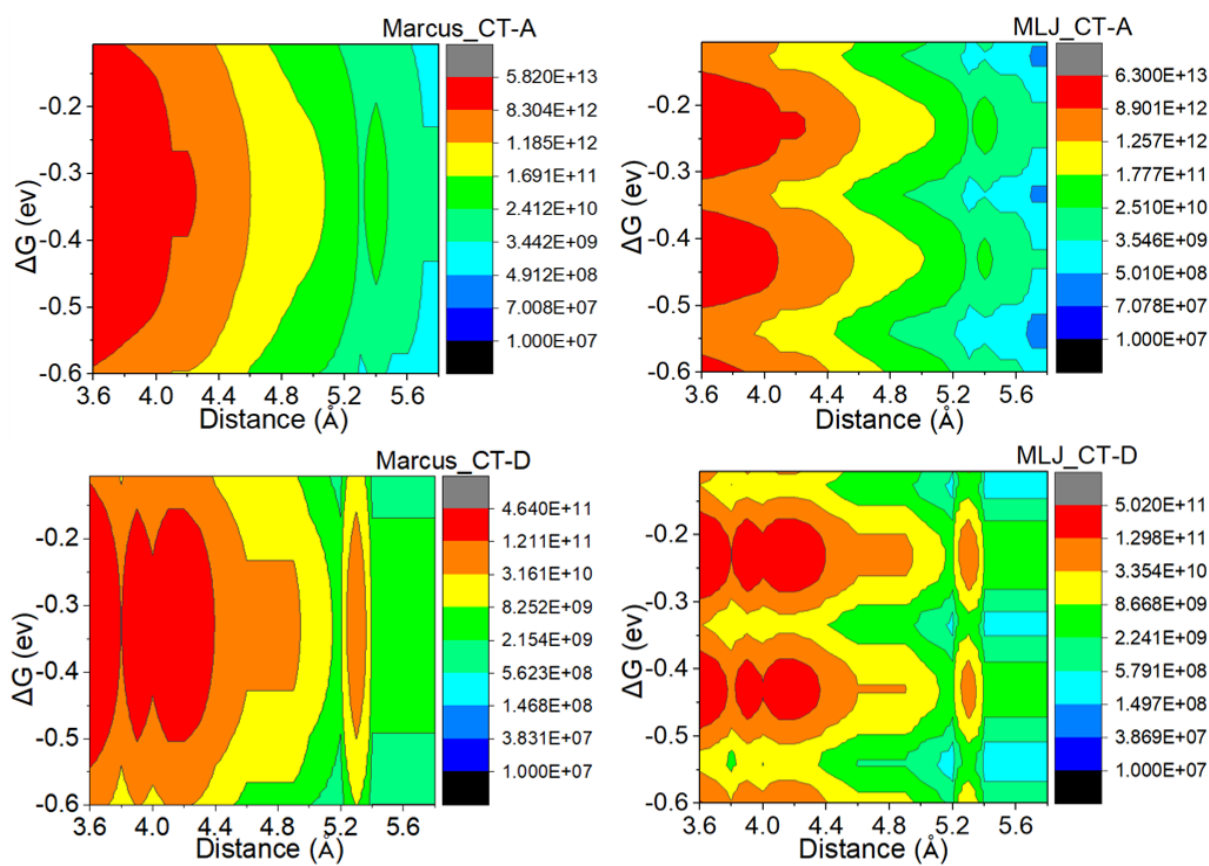

**Figure S70:** BCT recombination rates from the triplet charge-transfer state to the localized triplet excitation localized on the polymer donor and NFA acceptor, as computed for PTB7-Th:IEICO-2F(G0) using Marcus (left) and Marcus-Levich-Jortner (right) rate expressions. MLJ includes quantum tunnelling through an effective high-frequency vibration of energy 0.2 eV. The total reorganization energy is 0.33 eV in the Marcus calculations and is split into an internal part $\lambda_i$ of 0.3 eV and an external part $\lambda_s$ of 0.03 eV in the MLJ calculations. The rates are shown for a range of energy differences between the involved states spanning around the TD-DFT values (along the vertical axis) and as a function of the distance between the conjugated backbones of the interacting donor and acceptor (along the horizontal axis). Recurrences in the MLJ rates are due to tunnelling through successive quantum vibrational states. Recombination into the deeper-lying NFA acceptor is predicted to be one-to-two orders of magnitude faster than to the polymer donor, essentially because of a larger excitonic coupling. Most importantly, the rates decrease exponentially with increasing intermolecular separation as a result of the decreasing wavefunction overlap and excitonic interaction, from values in the range ps$^{-1}$ at close distances to ~ns$^{-1}$ when the molecules are further separated by less than 2 Å. $G_i$ (i=0,1,2,…) represents one possible local minimum on the ground-state potential energy surface of the complex, as probed using dispersion-corrected DFT calculations by changing the initial configuration (namely by translating one molecule with respect to another longitudinally and/or laterally). The data reported on Fig. 3a of the manuscript correspond to rates obtained using the following set of parameters: $\lambda_i = 0.3$ eV, $\lambda_s = 0.03$ eV, $\Delta G \approx$ -0.3 eV[69].

## Singlet and triplet hybridization

A first hint towards hybridization in the singlet manifold is obtained from the sharing of the oscillator strengths among the lowest adiabatic states of the complexes[9,18,70]. From Table S3, it clearly appears that configurational mixing is particularly important in the PTB7-Th:IEICO-2F(G0 and G1) case, where the CT-like singlet borrows significant intensity from the closely-lying localized excited states, while it is less effective in other geometries of the same system, as well as in the other blends. To proceed, it is informative to focus on the PTB7-Th:SiOTIC-4F(G0) vs PTB7-Th:IEICO-2F(G1) complexes, as both systems share the same polymer donor and similar face-to-face orientation, but only the latter shows an inversion of the state ordering with the $^3$CT state being higher in energy than the $^1$CT. Because the energy separation between the involved electronic states is similar in the two cases, we hypothesized that the difference in hybridization must be due to the excitonic interactions. Under the reasonable assumption that the wavefunctions for $^3$CT and $T_1$ states are captured by single electronic configurations based on the two-level models shown in Figure S72, the excitonic coupling between the many-body wavefunctions can be cast in terms of one-electron transfer integrals among molecular orbitals. For the dominant (as indicated by the BCT rate calculations above) coupling to the NFA acceptor, the relevant transfer integral is between the HOMOs of D and A. Such a matrix element scales with the spatial overlap between the orbitals. We thus plotted the (diabatic) HOMOs of the isolated D and A on a common grid and computed their overlap in Figure S75[71]. We see a substantial difference between the two systems; while for PTB7-Th:IEICO-2F(G1), the two orbitals interact in-phase giving rise to a constructive overlapping pattern (with most contributions being of the same sign), the corresponding orbitals are out-of-phase and yield destructive interactions with alternating regions of positive and negative overlap in PTB7-Th:SiOTIC-4F. Note that this is fully consistent with the 3D transition density cube plots between the adiabatic states involved in the BCT reaction (and that directly enters the GMH excitonic coupling through the corresponding transition dipole moment, $\mu_{CT \rightarrow T1(D/A)}$), see Figure S75. The transition density distributions for $CT \rightarrow T1(A)$ directly echo the symmetry patterns defined by the orbitals, with contributions along the heterojunction adding up constructively (destructively) to yield a large (smaller) transition moment dipole moment of 2.51D (0.71D) in PTB7-Th:IEICO-2F(G1) (PTB7-Th:SiOTIC-4F(G0)).

| PTB7-Th:IEICO-2F | | | | | | | |
|---|---|---|---|---|---|---|---|
| G0 | | G1 | | G2 | | G3 | |
| Singlet (ev) | $f_{osc}$ | Singlet (ev) | $f_{osc}$ | Singlet (ev) | $f_{osc}$ | Singlet (ev) | $f_{osc}$ |
| 1.41 | 0.443 | 1.42 | 0.554 | 1.58 | 0.552 | 1.56 | 0.937 |
| 1.52 | 0.318 | 1.60 | 0.569 | 1.59 | 0.046 | 1.61 | 1.563 |
| 1.59 | 1.341 | 1.65 | 1.057 | 1.67 | 1.553 | 1.67 | 0.043 |
| 1.74 | 0.032 | 1.78 | 0.006 | 1.85 | 0.066 | 1.78 | 0.001 |
| 1.78 | 0.001 | 1.87 | 0.004 | 1.89 | 0.008 | 1.83 | 0.048 |
| **PTB7-Th:SIOTIC-4F** | | | | **PTB7-Th:IEICO-4F** | | | |
| G0 | | G1 | | G0 | | | |
| Singlet (ev) | $f_{osc}$ | Singlet (ev) | $f_{osc}$ | Singlet (ev) | | $f_{osc}$ | |
| 1.40 | 0.004 | 1.42 | 0.016 | 1.38 | | 0.375 | |
| 1.52 | 0.304 | 1.47 | 0.889 | 1.49 | | 0.265 | |
| 1.55 | 1.584 | 1.61 | 0.966 | 1.57 | | 1.509 | |
| 1.76 | 0.112 | 1.75 | 0.033 | 1.72 | | 0.035 | |
| 1.77 | 0.085 | 1.80 | 0.004 | 1.75 | | 0.004 | |

**Table S3**: Vertical excitation energies and oscillator strengths for the D/A complexes from figure S72.

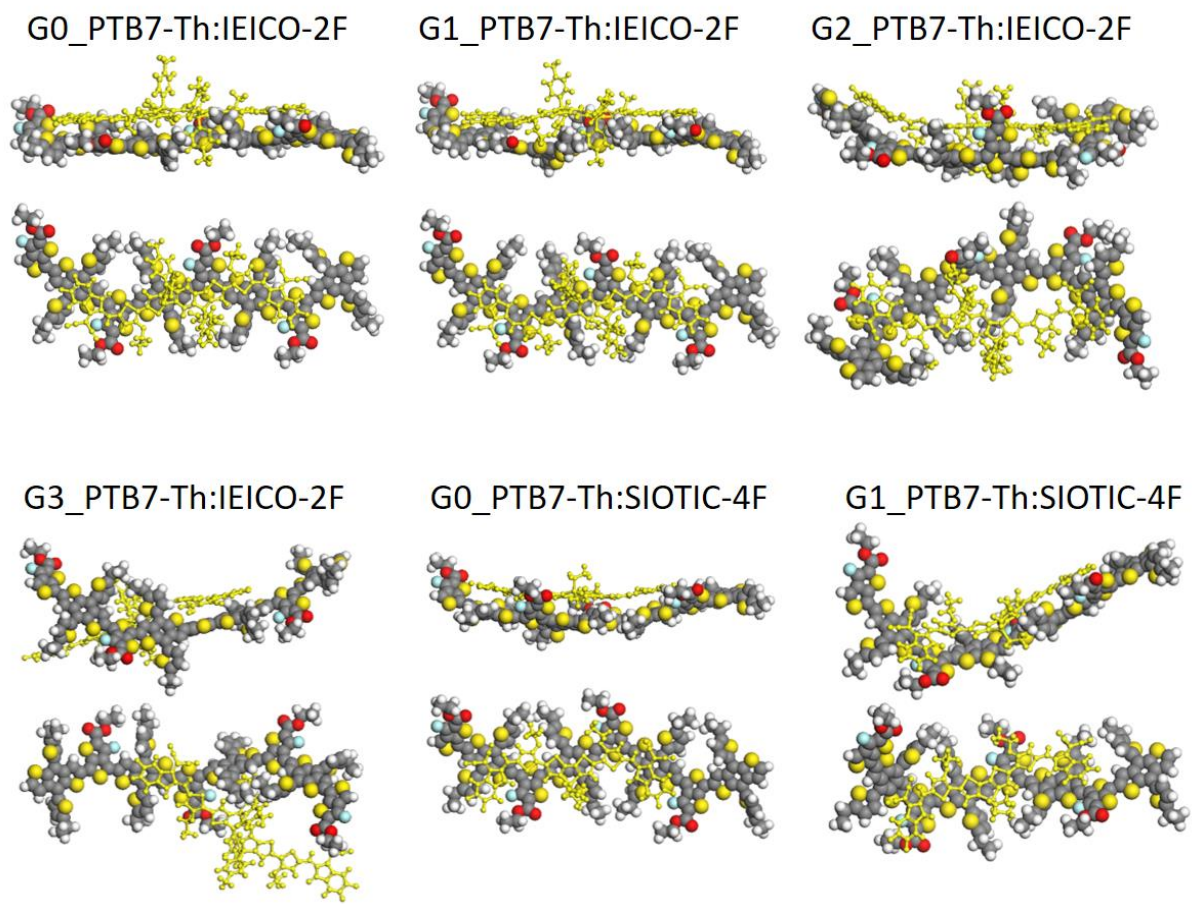

**Figure S71:** Top- and side-view of polymer/NFA complexes investigated in this work. $G_i$ (i=0,1,2,…) represents one possible local minimum on the ground-state potential energy surface of the complex, as probed by changing the initial configuration (namely by translating one molecule with respect to another longitudinally and/or laterally).

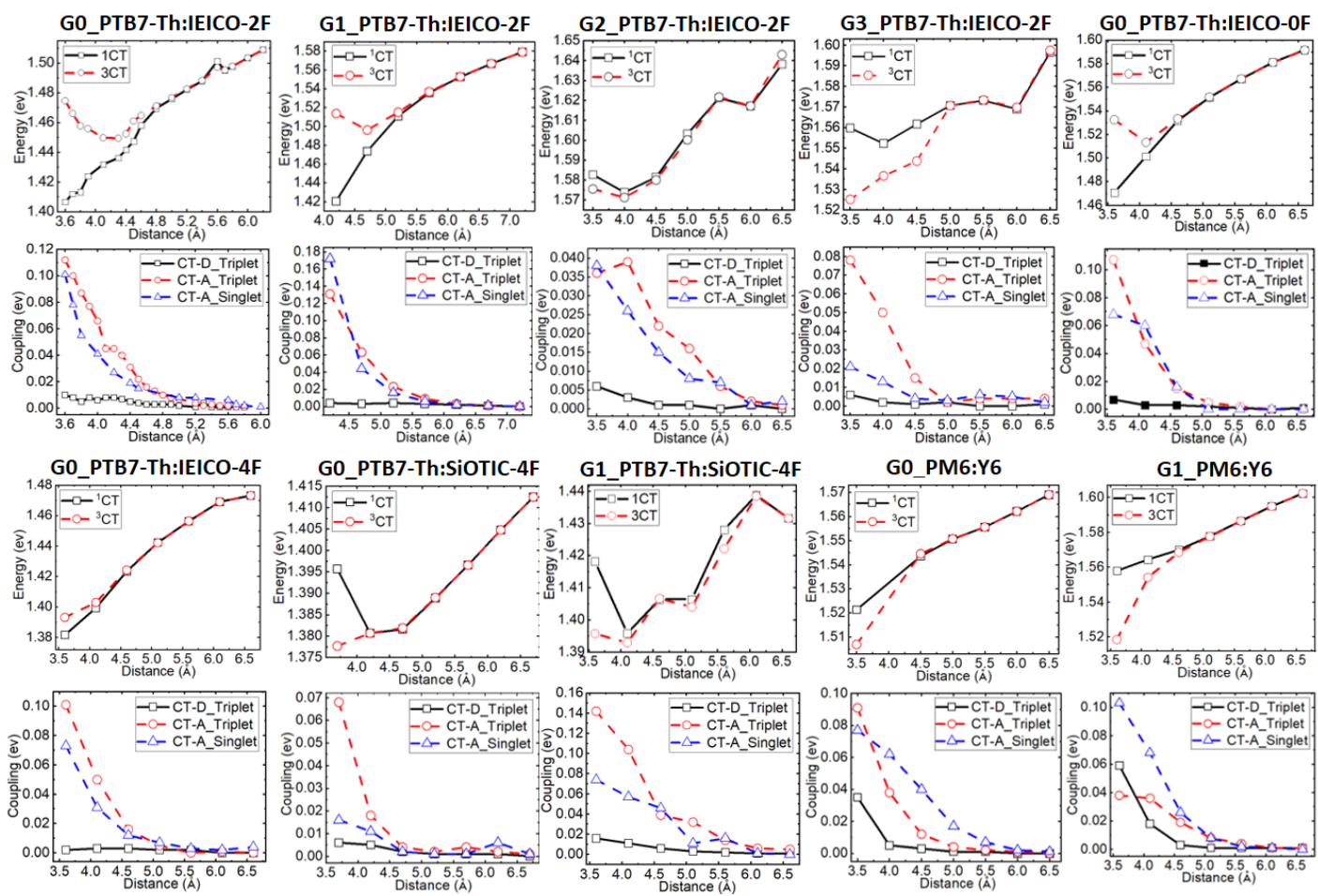

**Figure S72:** (top) [1]CT and [3]CT energies and (bottom) couplings between CT and local excitations for the D/A complexes from figure S70. Note that hybridization, as manifested with an inverted energy ordering of the triplet and singlet CT-like states, is only present for the G0 and G1 geometries of the PTB7-Th:IEICO-2F complex, while G2 and G3 lead to the usual situation with the [3]CT being stabilized over the [1]CT by exchange coupling. Thus, not surprisingly, the local microstructure has a strong impact on configurational mixing. It is also interesting to compare G0_PTB7-Th:IEICO-2F with G0_PTB7-Th:IEICO-0F and G0_PTB7-Th:IEICO-4F. While the two former blends behave similarly, the degree of hybridization is strongly reduced (the lowest singlet and triplet CT-like states are now quasi-degenerate) in G0_PTB7-Th:IEICO-4F because of the larger energy mismatch between the interacting states (associated with the pulling down of the frontier energy levels on the NFA acceptor when grafting additional fluorine atoms). Thus, the energy alignment between the local and the charge-transfer excitations is also critical. None of the geometrical structures generated for PTB7-Th:SiOTIC-4F or for PM6:Y6 (with the constraint of no D/A intermolecular F···F interactions at distances <1 nm, imposed from our previous solid-state NMR studies on the PM6:Y6 blend[18]) result in hybridisation and an inversion in the ordering of the [1]CT and [3]CT states.

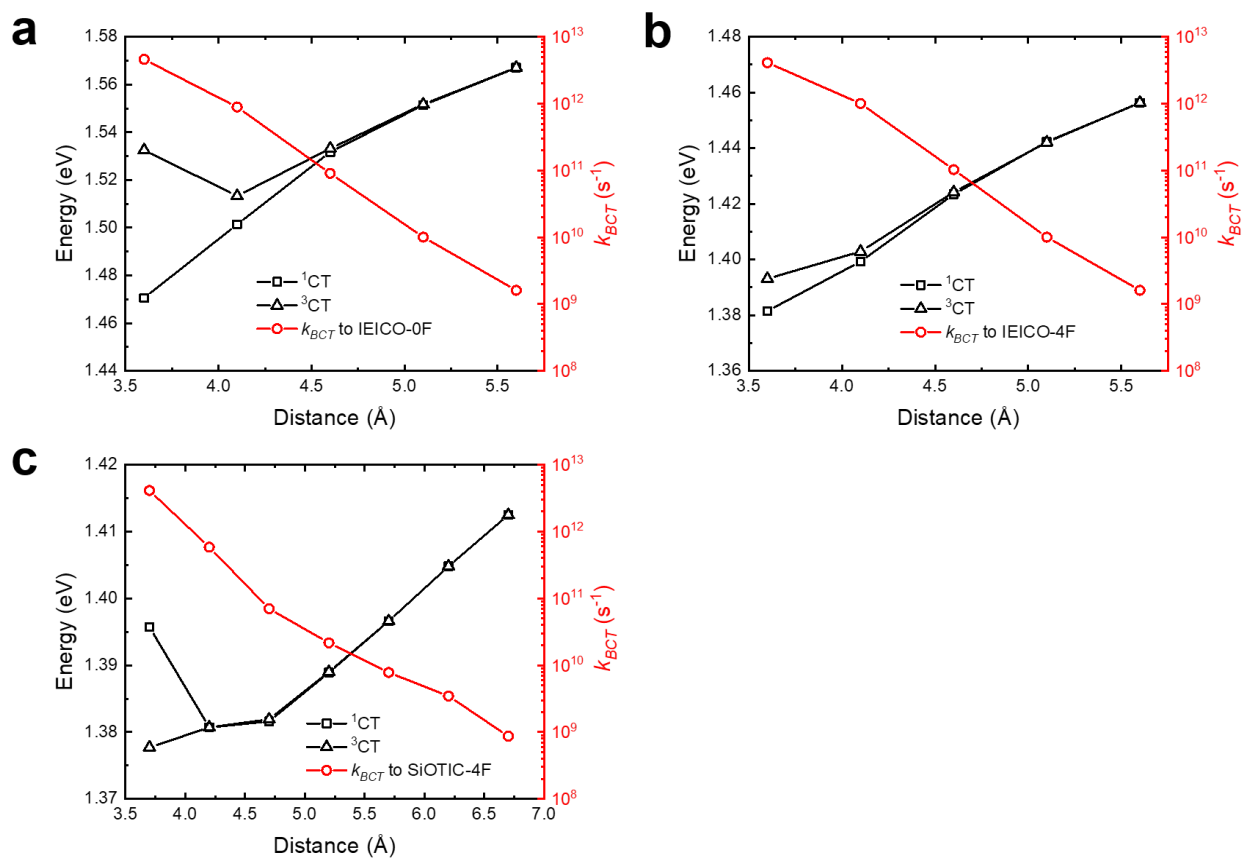

**Figure S73: (a)** The BCT rate from $^3$CT to the IEICO-0F T$_1$ for the PTB7-Th:IEICO-0F "G0" complex as a function of D/A separation, overlaid on the $^1$CT and $^3$CT state energies. As with the PTB7-Th:IEICO-2F complex, the inversion of the $^1$CT and $^3$CT as a result of hybridisation can clearly be seen. Additionally, the most stable $^3$CT configuration is no longer at the equilibrium geometry; this increases the separation of the charges in the $^3$CT state, slowing the BCT process by an order of magnitude. This result is completely consistent with the experimental observations, where there is no evidence for triplet formation via BCT in the PTB7-Th:IEICO-0F blend. **(b)** The BCT rate from $^3$CT to the IEICO-4F T$_1$ for the PTB7-Th:IEICO-4F "G0" complex as a function of D/A separation, overlaid on the $^1$CT and $^3$CT state energies. In contrast to the PTB7-Th:IEICO-0F and -2F complexes, the $^3$CT is only slightly destabilised as a result of hybridisation, as discussed in Fig. S72. As the most stable $^3$CT configuration is still at the equilibrium geometry, it is no longer energetically unfavourable for the charges to approach each other; the BCT rate is not decreased. This is consistent with experimental observations, where IEICO-4F triplets are formed via BCT. **(c)** The BCT rate from $^3$CT to the SiOTIC-4F T$_1$ for the PTB7-Th:SiOTIC-4F "G0" complex as a function of D/A separation, overlaid on the $^1$CT and $^3$CT state energies. Due to the weak electronic coupling between D and A in this blend, CT-LE hybridisation is not observed and the $^1$CT and $^3$CT energy ordering is as expected from exchange interactions. The BCT rate is therefore not decreased and triplet formation via BCT is observed in this blend.

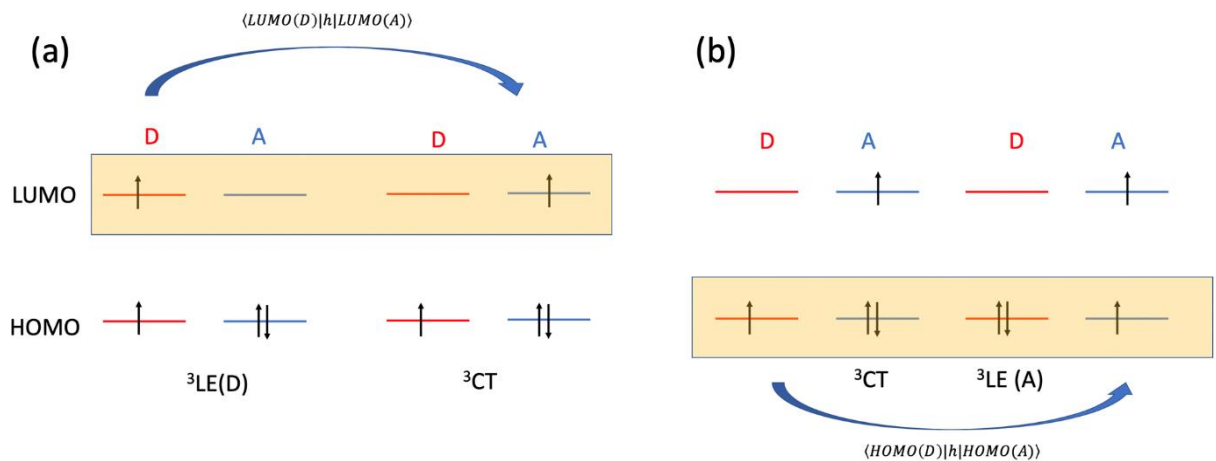

**Figure S74:** Leading electronic configurations, responsible for the coupling between **(a)** LE(D) and CT and **(b)** LE(A) and CT states.

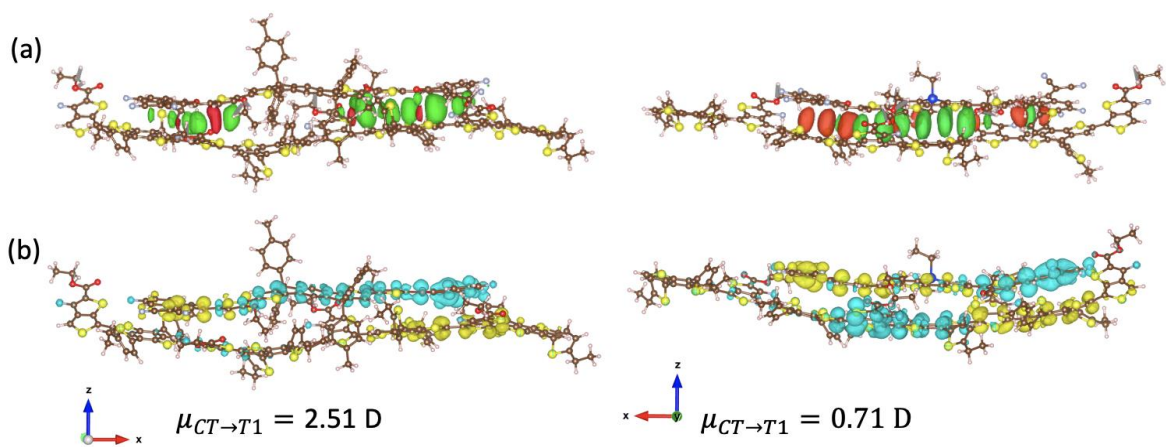

$\mu_{CT \to T1} = 2.51$ D  $\mu_{CT \to T1} = 0.71$ D

**Figure S75: (a)** Overlap between HOMOs of D and A and **(b)** transition densities for (left) PTB7-Th:IEICO-2F(G1) and (right) PTB7-Th:SiOTIC-4F(G0).